\documentclass[12pt,a4paper]{article}

\usepackage[T1]{fontenc}
\usepackage[utf8]{inputenc}
\usepackage{lmodern}
\usepackage[a4paper,left=2.45cm,right=2.45cm,top=2.45cm,textheight=660pt]{geometry}
\usepackage{amsmath,amssymb,amsthm,mathtools}
\usepackage{booktabs,tabularx,array}
\usepackage[authoryear,round]{natbib}
\usepackage{graphicx}
\usepackage{float}
\usepackage{enumitem}
\usepackage{setspace}
\usepackage{xurl}
\usepackage{xcolor}
\usepackage[colorlinks=true,linkcolor=blue!50!black,urlcolor=blue!50!black,
citecolor=blue!50!black]{hyperref}

\setlist[itemize]{leftmargin=1.45em,itemsep=0.18em,topsep=0.2em}
\setlist[enumerate,1]{label=(\alph*),leftmargin=1.75em,itemsep=0.2em,topsep=0.2em}

\newtheorem{theorem}{Theorem}
\newtheorem{proposition}{Proposition}
\newtheorem{corollary}{Corollary}
\theoremstyle{definition}

\theoremstyle{remark}

\newcommand{\E}{\mathbb E}

\newcommand{\Prob}{\mathbb P}

\newcommand{\R}{\mathbb R}
\newcommand{\dd}{\,\mathrm d}
\newcommand{\1}{\mathbf 1}
\newcommand{\TV}{d_{\mathrm{TV}}}

\newcolumntype{Y}{>{\raggedright\arraybackslash}X}

\graphicspath{{figures/}}
\title{\vspace{-1.5em}Residual Model Criticism for Fitted SDE Implementations}
\author{Shengda Zheng\\
\normalsize Independent Researcher, Hong Kong\\
\normalsize\href{mailto:clemmentin@outlook.com}{\nolinkurl{clemmentin@outlook.com}}}
\hypersetup{pdfauthor={Shengda Zheng},
  pdftitle={Residual Model Criticism for Fitted SDE Implementations}}
\date{\vspace{-1em}}
\begin{document}
\begingroup
\setstretch{1.0}
\maketitle
\begin{abstract}
\small
Stochastic differential equation (SDE) models can reproduce trajectories while missing temporal dependence. We compare residual-based scores from a held-out path with scores from paths simulated by a fitted implementation, keeping fitted parameters, simulation rules and the scoring rule fixed. We give conditions for Monte Carlo calibration and bound conditional rejection error by the gap between conditional score CDFs. Residuals use zero-noise centres and covariance approximations that track noise propagation over observation intervals; report residuals also include reporting variation. Residual weights affect sensitivity and the influence of fitting error. For a quadratic energy score in a linear example with singular true covariance, whitening with nonsingular fitted covariance matrices leaves a nonvanishing score-CDF gap as nonzero coefficient estimates approach the truth. A fixed positive-definite weighting matrix makes the gap vanish. In a controlled SIR experiment using reference paths simulated at fitted parameters, adding susceptible-state endpoints to training or downweighting a low-variance residual reduces null rejection. A separate comparison using reference paths simulated at the true parameter shows reduced sensitivity to added susceptible-coordinate noise after downweighting. An exploratory weekly-report application incorporates reporting and reconstruction into reference simulations, revealing excess correlation between squared residuals in Czechia and persistently negative residuals in Finland.
\end{abstract}


\endgroup

\section{Introduction}
\label{sec:introduction}

\begin{figure}[!ht]
\centering
\includegraphics[width=\textwidth]{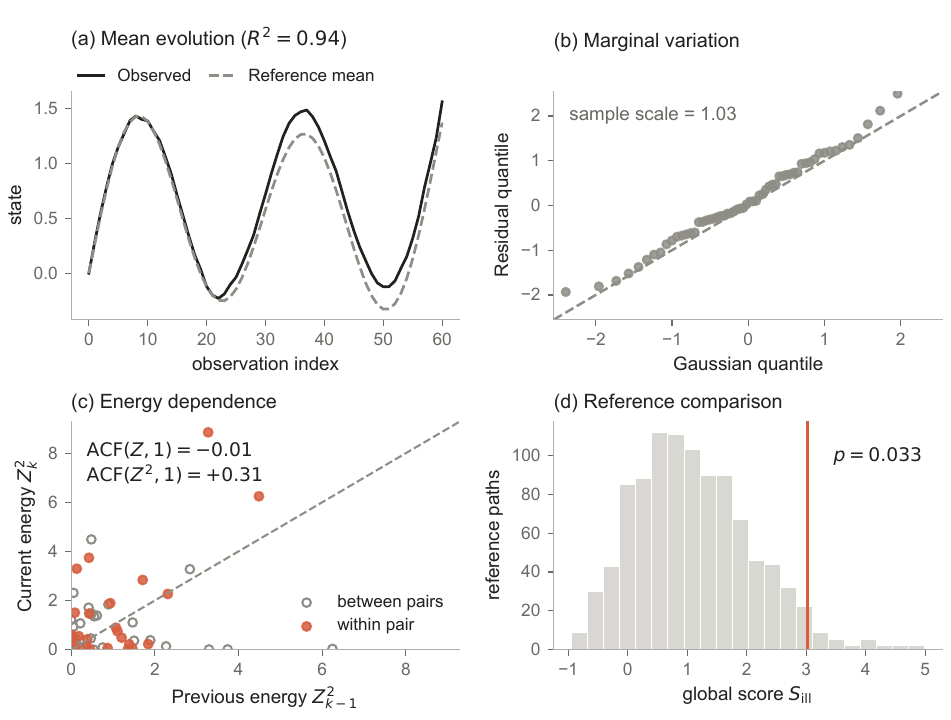}
\caption{Motivating illustration: dependence missed by marginal summaries.
Reference innovations are independent standard normals. Held-out innovations
have the same marginal distributions and zero population lag-one correlation,
but their squares are dependent within each of 30 independent pairs.
(a) A prescribed reference mean, with noise scaled to give \(R^2=0.94\).
(b) Marginal quantiles. (c) Adjacent squared innovations within and between
pairs. (d) An illustrative score takes the maximum of three features
standardised using independent pilot paths: absolute deviation of the sample
standard deviation from one, and absolute lag-one
correlations of innovations and their squares (\(p=0.033\)).
The untruncated score may be negative. The independent pilot and reference samples contain 500
and 999 paths, respectively. Supplementary Section~\ref{app:motivation-construction}
gives the generating construction and score definitions. The prescribed
curve is not an estimated SDE fit.}
\label{fig:audit-motivation}
\end{figure}

After fitting a stochastic differential equation (SDE), we assess whether
simulations from the fitted implementation reproduce selected features of
the observed variation and temporal dependence. Trajectory plots and
marginal residual summaries can miss differences in this dependence.
Figure~\ref{fig:audit-motivation}
illustrates this problem using constructed innovations around a prescribed
curve. The target and reference innovations have the same standard-normal
marginal distributions, and consecutive signed innovations have zero
population correlation in both. Their squared values nevertheless remain
dependent within pairs on the target path.
We therefore develop a residual diagnostic for comparing a held-out path with
simulations from a fitted SDE implementation. Each path is
converted into residuals, then into selected features of centring,
magnitude and temporal dependence, and finally into a global score.
The score provides an overall comparison; the individual features and
cumulative residual curves describe the departures.

The reference for this comparison is the path law induced by the complete
\emph{fitted implementation} on the finite observation grid. The fitted
coefficients, numerical solver, initialisation and feedback rules determine
this law, together with any specified observation processing. We hold these
choices and the scoring rule fixed during evaluation. The reference law
specifies what generates the comparison paths; the scoring rule specifies
which aspects of those paths we examine.

Fitting can make the reference and held-out score distributions differ
even within a correctly specified model class.
Section~\ref{sec:method} gives conditions for calibrated Monte Carlo
comparisons and bounds the error in conditional rejection probability
by the gap between the target and reference conditional score CDFs.

With the reference comparison specified, we construct residual coordinates
at the observation scale (Section~\ref{sec:residual-coordinates}). Drift can
carry noise added early in an interval into directions that receive no
instantaneous noise. We approximate observation-step covariance by
linearising the fitted solver along the path obtained with noise inputs
set to zero. For state observations, we subtract the resulting
\emph{zero-noise increments} from path increments and use the propagated
covariance to scale the residuals. For reports, the centre and scale refer
to the amount reported over the interval. These quantities are recomputed
from each path's own states and history. Noise propagation through the
drift is studied in control, filtering and hypoelliptic inference
\citep{Sontag1998,SarkkaSvensson2023,PokernStuartWiberg2009,DitlevsenSamson2019}.

Residual scaling determines how deviations within an observation interval
are measured. Feature scaling, fixed using independent pilot paths,
determines how the resulting path features enter the global score. The
maximum-based score provides an overall comparison, using a common
threshold derived from the reference scores. We also compare each raw
feature with its own reference distribution to show which features differ,
in which direction, and by how much. Cumulative curves show how departures
accumulate over time.

Residual weights determine how model departures and fitting error appear
in the score, even when the fitted reference law is held fixed. For
independent Gaussian residuals with known positive-definite covariances,
whitening maximises the standardised mean shift of quadratic scores under
local common-scale covariance departures. After fitting, however, division by a small fitted residual scale can
magnify centring error. In a linear example with singular true covariance,
the centring error and the fitted standard deviation shrink at the same
rate as nonzero coefficient estimates approach the truth. The conditional
score-CDF gap therefore remains positive under whitening. Using one fixed
positive-definite weighting matrix instead makes this gap vanish.

A controlled SIR experiment uses repeated fitting to
examine how training information and downweighting a low-variance residual
component affect rejection under the model null. We then examine how
downweighting affects sensitivity to added noise. For this comparison, we
generate reference paths at the true parameter while retaining each fitted
scoring rule. Downweighting reduces detection of the added noise
(Section~\ref{sec:fitted-stability}). Related error amplification arises in
the drift-estimation analysis of \citet{PokernStuartWiberg2009}; here the
question is how residual weights change the selected score's sensitivity
and conditional distribution gap.

The weekly-report SIR application illustrates how the diagnostic describes
unusual dependence and persistent departures in one direction
(Section~\ref{sec:sir-application}). We build the reference law around an
already fitted neural SIR implementation. To generate each reference path,
we simulate reports and repeat smoothing and state reconstruction to update
subsequent histories. For Czechia, the lag-one correlation of squared
residuals lies above its reference interval, while the mean of squared
residuals lies within its own interval.
For Finland, all nine observed reports fall below their corresponding
model centres, computed with latent noise inputs set to zero. These
exploratory, retrospective comparisons keep the fitted neural SIR model
and reporting parameters fixed.

Residual prediction tests examine structure remaining in fitted
residuals \citep{ShahBuhlmann2018,JankovaEtAl2020}, while global envelope
tests connect an overall test to a graphical description of departures
\citep{MyllymakiEtAl2017}. Our selected features and cumulative curves
serve these two roles for paths from a fixed SDE implementation.
Diffusion specification tests examine transition
distributions or the drift and diffusion functions
\citep{HongLi2005,BhardwajCorradiSwanson2008,GaoCasas2008,LopezPerezEtAl2025};
\citet{HongLi2005} also use powers of probability-integral-transform
residuals to distinguish forms of temporal dependence. We compare selected
features of the target with those of paths from the complete fitted
implementation, including its solver, feedback and observation processing.

Calibration after fitting is also central to holdout and split predictive
checks \citep{MoranBleiRanganath2024,LiHuggins2026}, classifier randomisation
\citep{JavanmardMehrabi2024} and conditioning on sufficient statistics
\citep{LinEtAl2025}. For conditional randomisation and permutation tests,
\citet{BerrettEtAl2020} bound excess type I error by total variation error
in an estimated conditional law. Our bound uses the conditional CDF gap of
the selected score, distinguishing differences in the path laws from
differences visible to that score. Proofs and complete experimental
settings are in the Supplementary Material, together with extensions that
select weights using training confidence sets. These extensions are related
to confidence-set methods for nuisance parameters \citep{BergerBoos1994}.

\section{Reference laws and Monte Carlo calibration}
\label{sec:method}

We first specify the reference law and scoring rule, then give calibration
and rejection-error results. Section~\ref{sec:residual-coordinates} gives
our residual construction.

\subsection{The fitted implementation and analysis information}
\label{sec:miniature}

Fix the fitted coefficients, solver, initialisation, conditioning history,
inputs, feedback rules and projections, together with any reporting and
reconstruction maps. These choices define the reference path law \(Q\)
on the finite observation grid.

Choose the path features to examine, such as centring, magnitude,
cumulative departures or lag dependence. Let \(S(Y)\) be a real-valued
measurable score, with larger values indicating greater discrepancy. We
apply the same rule to the target and every reference path.

Training data determine the fitted quantities. Pilot paths fix feature
transformations or other scoring choices before evaluation. Separate
reference paths \(Y^{(1)},\ldots,Y^{(M)}\) provide the comparison with the
held-out path \(Y^{held}\). The pilot paths are not included among these
\(M\) reference paths.

Let \(\mathcal A\) collect the fitted implementation, conditioning inputs,
scoring rules and any simulations used for pilot estimation or selection.
This information fixes \(Q\) and \(S\), including pilot estimates and
selection outcomes; we suppress their dependence on \(\mathcal A\).
Conditional on \(\mathcal A\), reference paths are iid from \(Q\),
independently of the target.

\subsection{Reference ranks and conditional calibration}
\label{sec:calibration}

The \emph{null hypothesis for the score} states that \(S(Y^{held})\) and
\(S(Y^{sim})\), where \(Y^{sim}\sim Q\), have the same conditional
distribution given \(\mathcal A\). Equality of the conditional path laws
implies this null, but different path laws can also give the same score
distribution.

Rank the target score against the reference scores:
\begin{equation}
p_M=\frac{1+\sum_{m=1}^M
\1\{S(Y^{(m)})\ge S(Y^{held})\}}{M+1}.
\label{eq:rank}
\end{equation}
Under the following conditions, \(p_M\) is a valid Monte Carlo p-value,
and we reject when \(p_M\le\alpha\). For descriptive comparisons, we call it the global
reference rank proportion.

\begin{samepage}
\begin{theorem}[Monte Carlo validity]
\label{thm:rank}
Conditional on \(\mathcal A\), suppose \(Y^{(1)},\ldots,Y^{(M)}\) are iid
from \(Q\), independently of \(Y^{held}\), and the reference score has a
continuous distribution. If \(S(Y^{held})\) and \(S(Y^{(1)})\) have the same
distribution,
then the Monte Carlo p-value satisfies
\begin{equation}
\Prob(p_M\le\alpha\mid\mathcal A)
=\frac{\lfloor\alpha(M+1)\rfloor}{M+1}.
\label{eq:rank-validity}
\end{equation}
\end{theorem}
\end{samepage}

Conditional exchangeability makes the descending rank uniform on
\(\{1,\ldots,M+1\}\) \citep{Dufour2006}.
Supplementary Section~\ref{app:rank-proof} gives the proof. Without
continuity, the \(\ge\) rule is conservative under conditional independence
and matching score distributions; randomised tie-breaking restores the
uniform rank. These conditions concern score distributions and require
neither exact transition covariances nor Gaussian residuals.

\subsection{Score gaps after fitting}
\label{sec:score-gaps}

Fitting can change the reference score distribution even within a correctly
specified model class. For the same scoring rule fixed by \(\mathcal A\),
the \emph{conditional score gap} \(\Delta_{\mathcal A}\) below measures
the largest absolute difference between the target and reference
conditional cumulative distribution functions (CDFs).
Under model departures, this gap can also reflect the signal that the
diagnostic is intended to detect.

\begin{samepage}
\begin{corollary}[Calibration when score distributions differ]
\label{cor:score-gap}
Put \(k_\alpha=\lfloor\alpha(M+1)\rfloor\) and
\begin{equation}
\Delta_{\mathcal A}=\sup_{t\in\R}
\left|\Prob\{S(Y^{held})\le t\mid\mathcal A\}
-\Prob\{S(Y^{sim})\le t\mid\mathcal A\}\right| .
\label{eq:score-gap-main}
\end{equation}
Under the independence and continuity assumptions of Theorem~\ref{thm:rank},
but without assuming that the two scores have the same distribution,
\begin{equation}
\left|\Prob(p_M\le\alpha\mid\mathcal A)-\frac{k_\alpha}{M+1}\right|
\le\Delta_{\mathcal A} .
\label{eq:rank-bound-main}
\end{equation}
If those assumptions hold for almost every realisation of the fitting,
selection and pilot stages, with \(M\) fixed, then
\begin{equation}
\left|\Prob(p_M\le\alpha)-\frac{k_\alpha}{M+1}\right|
\le\E[\Delta_{\mathcal A}] .
\label{eq:refit-bound-main}
\end{equation}
\end{corollary}
\end{samepage}

Conditional on the reference scores, rejection is determined by a threshold
for the target score. Replacing its conditional distribution with the
reference score distribution changes this probability by at most
\(\Delta_{\mathcal A}\). Averaging over reference draws gives
\eqref{eq:rank-bound-main}, with \(\mathcal A\) fixed. The marginal bound
\eqref{eq:refit-bound-main} also averages over fitting, selection and pilot
simulation.

At one prespecified level, rejection error can be smaller than
\(\Delta_{\mathcal A}\), which controls differences over all thresholds.
We do not estimate the score gap from a single
held-out path.

Under the same independence and continuity assumptions, for a sequence
\(\mathcal A_n\) with \(M\) fixed, \(\Delta_{\mathcal A_n}\to0\) in
probability implies that the marginal rejection probability converges to
\(k_\alpha/(M+1)\). It is sufficient that the total variation distance
between the target and reference path laws, conditional on \(\mathcal A_n\),
tends to zero in probability. Parameter consistency alone need not make
the score gap vanish (Section~\ref{sec:linear-fitted-stability}).

Choosing the score or target window after inspecting target outcomes
requires accounting for that selection, for example by using a new
independent target or reproducing selection in a justified null simulation.
Fresh reference paths alone do not establish calibration. The calibration
argument leaves the residual construction and score to be chosen.

\section{From paths to residual comparisons}
\label{sec:residual-coordinates}

We now construct residuals and scores for the reference comparison.
The global score summarises departure from the
reference simulations; contributing features and cumulative residual
curves describe how it appears in the residual sequence.

Residual scaling determines how deviations within each observation interval
are measured. Feature scaling then determines how departures in different
path features are compared when taking the global maximum.

An \emph{observation step} is an interval between observations; a
\emph{substep} is one numerical update within it. Feedback and residual
quantities are recomputed from each path's own states and history.

\subsection{Centres and residuals}
\label{sec:nonlinear-euler-coordinates}

For a path with \(K\) observation intervals, index increments by
\(k=0,\ldots,K-1\). For directly observed scalar states, subtract the
zero-noise increment calculated by the fitted implementation from the path
increment, then divide the difference by the chosen positive scale:
\begin{equation}
z_k(Y)=\frac{\Delta Y_k-\widehat m_k(Y)}{\widehat s_k(Y)},
\qquad \widehat s_k(Y)>0 .
\label{eq:scalar-coordinate}
\end{equation}
Here \(\widehat m_k\) is the zero-noise increment and \(\widehat s_k\)
is a scale for variation over the observation interval. These are choices
for centring and scaling; they need not be the mean and standard deviation
of the increment conditional on the starting state and history for that
interval.

To compute the zero-noise increment, divide the interval into \(L\)
substeps of size \(h\), with fitted drift \(\widehat b\) and noise loading
\(\widehat g\). Suppressing observation index \(k\), write one substep
with controls \(C_\ell\) as
\begin{equation}
\Psi_\ell(y,\varepsilon)
=P\{y+h\widehat b(y,C_\ell)+\sqrt h\,\widehat g(y,C_\ell)\varepsilon\},
\label{eq:microstep}
\end{equation}
where \(P\) is the state projection or identity. Starting from the current
path state, run \(\bar Y_{\ell+1}=\Psi_\ell(\bar Y_\ell,0)\). The
zero-noise increment is \(\widehat m_k=\bar Y_{k,L}-Y_k\).

For report observations, the quantity being scored is the amount reported
over the interval. In \eqref{eq:scalar-coordinate}, replace \(\Delta Y_k\)
by that report and use a centre and scale for the report. In the weekly-report
application, the centre is the expected report
at zero latent noise inputs, and the propagated scale includes both
latent variation and reporting variation. We use this centre to approximate
the report mean conditional on the starting history. If reports are smoothed and states
reconstructed to update subsequent histories, repeat those operations on
each reference path.

From these residuals, we examine means for shifts in centring, squared
values for magnitude, and lag correlations for temporal dependence.
Cumulative sums show how signed residuals and deviations in their squared
magnitudes build up over time.

\subsection{Propagating covariance over the observation step}
\label{sec:tangent-coordinates}

Noise added at an early substep changes the states from which later
substeps begin. Its effect can therefore be altered by the drift before
the next observation. We approximate observation-step covariance by
linearising the reference solver along its zero-noise path. The recursion
below tracks this propagation as new noise is added.

At the zero-noise trajectory, the state derivative
\(F_\ell=D_y\Psi_\ell(\bar Y_\ell,0)\) propagates existing perturbations,
and \(g_\ell=h^{-1/2}D_\varepsilon\Psi_\ell(\bar Y_\ell,0)\) gives the
noise loading in the linearised step. Fix the interval's starting state
and use independent innovations with mean zero and identity covariance.
The covariance of the linearised perturbations then satisfies
\begin{equation}
\Sigma_{\ell+1}^{tan}
=F_\ell\Sigma_\ell^{tan}F_\ell^\top+h g_\ell g_\ell^\top,
\qquad \Sigma_0^{tan}=0.
\label{eq:tangent-recursion}
\end{equation}
The first term propagates the existing covariance; the second adds the
covariance of the new noise. For a scalar state residual, use the
corresponding diagonal element of \(\Sigma_L^{tan}\) as an approximate
transition variance to construct the scale in \eqref{eq:scalar-coordinate}.
For a residual vector, fix a rule for retaining eigenvectors with positive
covariance eigenvalues and whiten on their span. Whitening can mix
components as well as rescale them. The squared norm of the whitened vector
weights the original residual directions according to the covariance used.
Fix any rule for zero scales before evaluation. Recompute the zero-noise
state and tangent covariance on each interval of every held-out and
reference path.

We restart this extended Kalman prediction recursion
\citep{SarkkaSvensson2023} from zero covariance on each observation interval,
without a filtering update. If controls vary with the perturbed substep
state, include their derivatives in the linearisation, directly or through
an augmented state. Hold controls determined by the interval's initial
history fixed when taking these derivatives.

For nonlinear implementations, we use this recursion as a local covariance
approximation. Supplementary Proposition~\ref{prop:tangent-first-variation}
justifies the approximation for smooth maps as the noise amplitude tends
to zero; Supplementary Section~\ref{app:sir-tangent} specifies the derivative
conventions at drift and projection kinks.

\subsection{Cumulative curves and residual features}
\label{sec:six-residual-statistics}

The following features describe the centring, magnitude and dependence
of the scalar residuals \(z_k\). Two are computed from signed residuals;
the remaining features use their squared magnitudes, called \emph{energy}.
Mean energy ignores temporal order. We therefore also use cumulative sums
and lag correlations.

Set \(E_k=z_k^2-1\) and define the cumulative curves
\[
C_z(n)=K^{-1/2}\sum_{k<n}z_k,\qquad
C_E(n)=K^{-1/2}\sum_{k<n}E_k,\qquad n=0,\ldots,K.
\]
The baseline one is fixed for target and reference paths; it need not equal
the residual second moment. Changing it can change the cumulative feature
and its ranks.
A downward step of \(C_z\) means that the increment or report at that
step is below its model centre. A downward step of \(C_E\) means that the
newly added squared residual is below one. Compare these curves with
reference curves to assess departures from the fitted implementation;
their signs alone do not establish a departure.

The maximum absolute excursion of each curve is a scalar feature; the
curve retains the sign and timing of cumulative departures. For a lag set
\(\mathcal L\) fixed before evaluation, the features are
\begin{equation}
\begin{aligned}
T_z^{mean}&=K^{-1}\sum_k z_k,&
T_z^{cum}&=\max_{n\le K}|C_z(n)|,\\
T_E^{level}&=K^{-1}\sum_kz_k^2,&
T_E^{cum}&=\max_{n\le K}|C_E(n)|,\\
T_E^{(\ell)}&=\operatorname{Corr}(E_k,E_{k-\ell}),&
\ell&\in\mathcal L.
\end{aligned}
\label{eq:ce-statistics}
\end{equation}
Lags are measured in observation intervals. Choose \(\mathcal L\) to
reflect the dependence of interest and the number of available pairs,
and use the same set on target and reference paths. Correlations on every
path follow the conventions in Supplementary Section~\ref{app:score-details}.
The weekly application uses \(\mathcal L=\{1\}\), giving five features.

\subsection{From features to scores and interpretation}
\label{sec:score-construction}
\label{sec:interpretation}

To construct the score \(S\),
let \(T_j(Y)\) denote a feature computed from the resulting residuals,
such as their mean squared magnitude. To combine \(J\) features, choose
an aggregation rule
\begin{equation}
S(Y)=s\{T_1(Y),\ldots,T_J(Y)\}.
\label{eq:statistics}
\end{equation}
To use the maximum over transformed features as the score, compute each
feature on the pilot paths and estimate its centre \(\mu_j\) and sample
standard deviation \(\tau_j>0\) from those feature values. Use these fixed
estimates during evaluation to put feature deviations on comparable scales:
\begin{equation}
R_j(Y)=d_j\!\left(\frac{T_j(Y)-\mu_j}{\tau_j}\right),
\qquad S(Y)=\max_{1\le j\le J}R_j(Y),
\label{eq:component-departures}
\end{equation}
where \(d_j(u)=[u]_+\), \([-u]_+\) or \(|u|\) for a right-tail, left-tail
or two-sided departure, respectively. Fix these tail rules before evaluation.
To construct our main score \(S_{\mathrm{CE}}\), we use the median of
each feature over the pilot paths as its centre. We apply two-sided tail rules to the signed mean, mean
energy and the correlations between squared residuals. For the two maximum
absolute excursions of the cumulative curves, we use upper-tail rules.
The score \(S_{\mathrm{CE}}\) is the maximum of these transformed
components \(R_j\).
We refer to this construction as the median-distance score. It expresses
feature departures in units of their pilot standard deviations.

\begin{figure}[H]
\centering
\includegraphics[width=\textwidth]{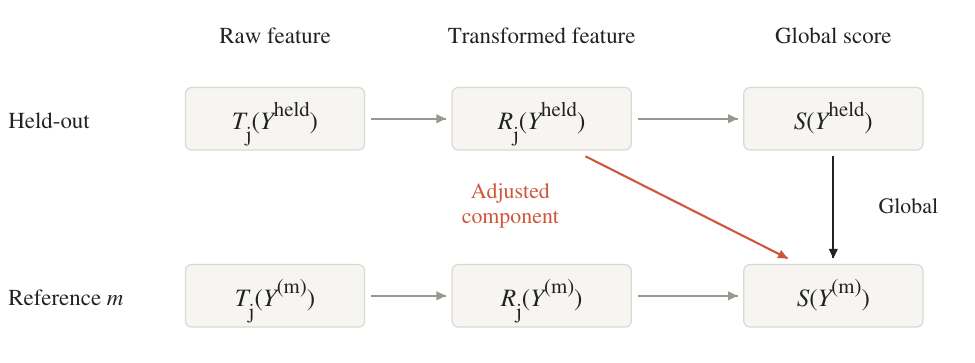}
\caption{Constructing scores and comparing them. For each path, horizontal
arrows apply the same pilot standardisation and tail rules feature by
feature, then take the maximum over \(j\).
The global comparison ranks \(S(Y^{held})\) against \(S(Y^{(m)})\);
the adjusted component comparison ranks \(R_j(Y^{held})\) against those
same reference maxima, for \(m=1,\ldots,M\).}
\label{fig:score-comparisons}
\end{figure}

The global comparison ranks the target's \(S\) against reference values of
\(S\) using \eqref{eq:rank}, excluding the pilot scores.
Figure~\ref{fig:score-comparisons} shows how this differs from comparing
an individual transformed feature with the reference maxima.

To identify contributing features, compare each transformed feature value
on the target path \(R_j(Y^{held})\) with the global score
\(S(Y^{(m)})\) on each reference path:
\begin{equation}
\widehat p_j^{max}
=\frac{1+\sum_{m=1}^M\1\{S(Y^{(m)})\ge R_j(Y^{held})\}}{M+1}.
\label{eq:component-adjusted}
\end{equation}
The smallest adjusted component rank equals \(p_M\). If the global test
rejects, these comparisons identify which features also exceed the
corresponding reference threshold. These comparisons do not control the probability
of any false component rejection for every configuration of true and
false component null hypotheses. Supplementary Section~\ref{app:score-details}
gives pointwise component comparisons and score conventions.

For interpretation, compare each raw feature \(T_j\) with its own reference
distribution. This retains the direction and magnitude of the feature
difference. A raw feature outside its reference interval need not have a
small adjusted component rank: the latter compares a transformed feature
with reference maxima over all features.

Plot \(C_z\) and \(C_E\) with pointwise reference ranges to examine the
time course. The maximum absolute excursions summarise how far the curves
move from zero; the curves themselves show how the signed residuals and
energy deviations accumulate.

Raw features and curves can describe residual patterns even without a
global rejection. Neither these descriptions nor the component comparisons
identify a unique model component as
the cause: fitting, dynamics, reporting or reconstruction can affect
the same residual feature.

An alternative transformation uses empirical pilot tail proportions to
measure how unusual a feature value is. Another choice is to use the
pilot interquartile range instead of the standard deviation as the feature
scale. Supplementary Sections~\ref{app:report-power}
and~\ref{app:report-moment-comparison} compare these choices.

\section{How residual scales affect the diagnostic}
\label{sec:finite-step-propagation}
\label{sec:fitted-stability}

The residual scale determines how departures enter the score. We use a
linear family and a single
quadratic energy score to examine noise propagation, sensitivity to
covariance departures and the effect of fitting error. The conditional
score gap from Section~\ref{sec:score-gaps} bounds the resulting error in
conditional rejection probability. A controlled SIR experiment then examines
how training information and residual weights affect rejection rates under
repeated fitting.

\subsection{Variation created by noise propagation}
\label{sec:linear-family}

In this family, noise enters the first coordinate directly. The drift in the
second coordinate depends on the first, so the same noise also affects later
increments of the second coordinate. A separate noise input may enter the
second coordinate directly. With independent Brownian motions, a fixed initial state,
\(\sigma_1>0\) and \(\sigma_2\ge0\), consider
\begin{equation}
\dd X_{1,t}=\sigma_1\dd W_{1,t},\qquad
\dd X_{2,t}=\kappa X_{1,t}\dd t+\sigma_2\dd W_{2,t}.
\label{eq:linear-family}
\end{equation}
For an observation interval of length \(H\), take \(L\) Euler substeps of size \(h=H/L\).
The zero-noise increment is \((0,\kappa HX_{1,0})^\top\). This finite-\(L\)
Euler implementation has exact transition covariance
\begin{equation}
\Sigma_{H,L}^{E}=
\begin{pmatrix}
\sigma_1^2H&\kappa\sigma_1^2H^2a_L\\
\kappa\sigma_1^2H^2a_L&\sigma_2^2H+\kappa^2\sigma_1^2H^3b_L
\end{pmatrix},
\label{eq:linear-family-euler}
\end{equation}
where \(a_L=(L-1)/(2L)\) and \(b_L=(L-1)(2L-1)/(6L^2)\).
Replacing \(a_L,b_L\) by \(1/2,1/3\) gives the continuous-time transition
covariance. Noise added to the first coordinate can affect the second-coordinate
drift in later substeps. Summing the remaining propagation times and their squares gives the
factors \(a_L\) and \(b_L\).

\begin{figure}[!ht]
\centering
\includegraphics[width=\textwidth]{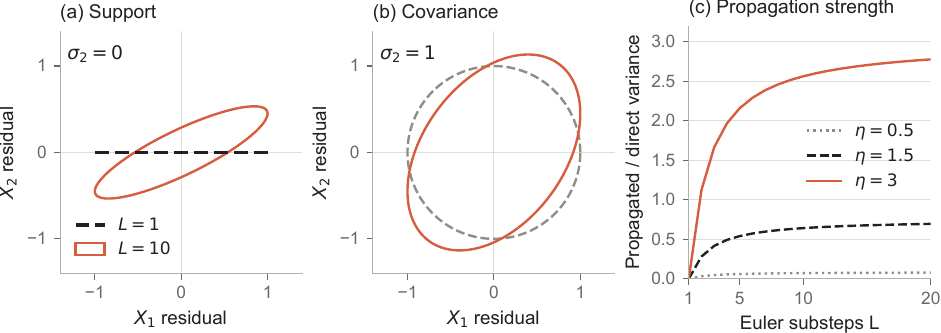}
\caption{Noise propagation changes observation-step covariance. For
\eqref{eq:linear-family}, panels (a)--(b) compare covariance contours for Euler
transitions using one substep
(\(H\operatorname{diag}(\sigma_1^2,\sigma_2^2)\)) or ten substeps at
\(H=\kappa=\sigma_1=1\). With \(\sigma_2=0\),
propagation creates a second stochastic direction; with \(\sigma_2=1\),
it changes variance and covariance in directions that already receive noise
directly. Panel (c)
shows the ratio \(\eta^2b_L\) of the variance contributed to the second
coordinate by drift propagation to the variance contributed by its direct
noise, for \(\sigma_2>0\).}
\label{fig:finite-grid-motivation}
\end{figure}

When \(\sigma_2=0\) and \(\kappa\ne0\), the Euler covariance is rank one
at \(L=1\) but full rank at \(L\ge2\). When \(\sigma_2>0\), both components
receive noise directly, so propagation changes the covariance without
changing its range. For \(\sigma_2>0\), divide the contribution of propagated
noise to the variance of the second-coordinate increment by the contribution
of its direct noise. This ratio is \(\eta^2b_L\), where
\(\eta=|\kappa|\sigma_1H/\sigma_2\).
Figure~\ref{fig:finite-grid-motivation} illustrates both effects.

Drift propagation can produce nonzero observation-step residuals in
directions that receive no instantaneous noise
(Supplementary Section~\ref{app:projected-residual-example}).
Supplementary Proposition~\ref{prop:linear-finite-support} gives the
covariance ranges for general linear systems.

\subsection{Sensitivity with known coefficients}
\label{sec:known-sensitivity}

Changing the weights in the quadratic score can improve the rejection rate only if
it changes the ordering of the scores computed on the paths. Throughout
this comparison, we hold the target and reference path laws and the
Monte Carlo rank procedure fixed. A strictly increasing
transformation applied to every target and reference score
preserves Monte Carlo ranks, including ties; this includes common positive
rescaling of a homogeneous score. Matrix weights can change the ordering
by weighting residual directions differently. We compare the local
sensitivity of these scores under a specified covariance alternative.

For a positive-definite matrix \(\Sigma\) used to standardise the residuals,
define the energy score for the linear family
\begin{equation}
T_\Sigma=\frac1{2K}\sum_{k=0}^{K-1}r_k^\top\Sigma^{-1}r_k,\qquad
r_k=X_{k+1}-\Phi_{H,L}(X_k),
\label{eq:linear-energy}
\end{equation}
where \(\Phi_{H,L}(x)\) is the state reached from \(x\) after one
observation interval under the Euler scheme with all noise inputs set to zero.

In this linear family, the residuals are independent Gaussian vectors. We compare
quadratic scores by the change in each score's mean under the alternative,
divided by its null standard deviation.

Under the common-scale alternative below, the variance of the residual
projected onto any fixed direction increases by the same proportion.
Under the null, the components of the whitened residual have equal variances.
Their squared values therefore enter the energy score on the same scale.

\begin{samepage}
\begin{proposition}[Whitening under common-scale departures]
\label{prop:common-scale-efficiency}
Let \(r_1,\ldots,r_K\) be independent zero-mean Gaussian residuals with known
covariances \(\Sigma_k\succ0\). Under local alternatives that retain zero means
and have covariances
\((1+\tau K^{-1/2})\Sigma_k\), \(\tau>0\), the statistic
\(\sum_k r_k^\top\Sigma_k^{-1}r_k\) maximises the change in its mean divided
by its null standard deviation among quadratic statistics
\(\sum_k r_k^\top W_kr_k\) with fixed symmetric weights, not all zero.
The maximising weights are \(W_k=c\Sigma_k^{-1}\) for a common \(c>0\).
Under the array conditions in Supplementary Section~\ref{app:scale-optimality},
this gives the local-efficiency comparison for upper-tail tests.
\end{proposition}
\end{samepage}
Under other local covariance alternatives, different weights may give a
larger standardised mean shift. Supplementary
Proposition~\ref{prop:quadratic-local-efficiency} gives the general result and
the conditions for the corresponding local-efficiency comparison.

For a finite-sample illustration under a common-scale departure in the
linear family, compare
\(\Sigma^{inst}=H\operatorname{diag}(\sigma_1^2,\sigma_2^2)\) and
\(\Sigma^{step}=\Sigma_{H,L}^{E}\) in \eqref{eq:linear-energy}.
At \(H=\sigma_1=\sigma_2=1\), \(\kappa=3\), \(L=10\) and \(K=40\),
we multiply both diffusion loadings by 1.25 to generate the target paths.
At the nominal 5\% level with 499 reference paths, the estimated rejection
rates are 0.7170 when residuals are standardised using \(\Sigma^{inst}\)
and 0.8784 when they are standardised using \(\Sigma^{step}\).
Subtracting the rejection rate using \(\Sigma^{inst}\) from that using
\(\Sigma^{step}\) gives a paired difference of 0.1614, with a 95\% simultaneous interval
\([0.1345,0.1877]\) over the 96 comparisons.
Supplementary Section~\ref{sec:linear-mechanism} reports the full 96-cell comparison.

\subsection{Fitting error relative to the whitening scale}
\label{sec:linear-fitted-stability}

Dividing a residual by a small scale calculated from fitted coefficients
can magnify centring error. We examine how this changes the target and
reference score distributions.

We now estimate the coupling coefficient in the same linear family. The
same estimate determines the reference path law, the zero-noise centre
and the covariance used for whitening.

Fix \(H>0\), \(L\ge2\), \(K\ge2\), the common initial state \(X_0=0\),
and known \(\sigma_1>0,\sigma_2\ge0\). Estimate \(\kappa\) from independent
training data. Generate held-out and reference paths using the
\(L\)-substep Euler implementation of \eqref{eq:linear-family} at
\(\kappa_0\) and \(\widehat\kappa\), respectively. Score both kinds of path
using \(\widehat\kappa\) in \(\Phi_{H,L}\) and \(\Sigma_{H,L}^{E}\) when
the fitted covariance is positive definite, and denote this score by
\(T_{\widehat\kappa}\). Under the Euler law at
\(\kappa\), the variance of the second-coordinate increment conditional
on the interval's starting state and first-coordinate increment is
\begin{equation}
v_\kappa=H\sigma_2^2+\kappa^2\sigma_1^2H^3c_L,
\qquad c_L=b_L-a_L^2=\frac{L^2-1}{12L^2}.
\label{eq:linear-conditional-variance}
\end{equation}

To whiten the residual vector, use the Cholesky factor of the covariance
calculated from the fitted coefficients. First, remove from the second residual the linear contribution
of the first, with the coefficient calculated under the fitted law.
The difference is the second-coordinate increment minus its fitted
conditional mean, given the interval's starting state and the
first-coordinate increment. We divide this difference by
\(\sqrt{\widehat v}\), the fitted conditional standard deviation.

\paragraph{A boundary where the covariance becomes singular.}
Consider the boundary \(\kappa_0=\sigma_2=0\).
Given the interval's starting state and first-coordinate increment,
the fitted conditional mean of the second-coordinate increment differs from
the true conditional mean by
\(\widehat\kappa H\{X_{1,k}+a_L\Delta X_{1,k}\}\).
The magnitude of this error in the conditional mean and the conditional
standard deviation calculated under the fitted law
\(\sqrt{\widehat v}=|\widehat\kappa|\sigma_1H^{3/2}\sqrt{c_L}\)
are both proportional to \(|\widehat\kappa|\). Dividing the error in the conditional
mean by this standard deviation therefore leaves a term that does not
vanish as nonzero estimates approach zero.

Conditional on \(\mathcal A\), the score \(T_{\widehat\kappa}\) computed on
the held-out path has the same distribution for every nonzero
\(\widehat\kappa\), with
\begin{equation}
\E[T_{\widehat\kappa}^{held}\mid\mathcal A]
=\frac12+\frac{K-1+2a_L^2}{4c_L},
\qquad
\E[T_{\widehat\kappa}^{ref}\mid\mathcal A]=1.
\label{eq:linear-boundary-energy}
\end{equation}
With \(H,L,K\) fixed, the conditional score gap is a strictly positive
constant along every nonzero sequence \(\widehat\kappa\to\kappa_0=0\).
If we continue to centre increments using the fitted zero-noise map but
use one matrix
\(\Sigma_*\succ0\) in \eqref{eq:linear-energy}, held constant across paths
and throughout the sequence of coupling estimates, the gap tends to zero.
At each estimate, the two weighting rules use the same target and fitted
reference laws.

\begin{figure}[!ht]
\centering
\includegraphics[width=\textwidth]{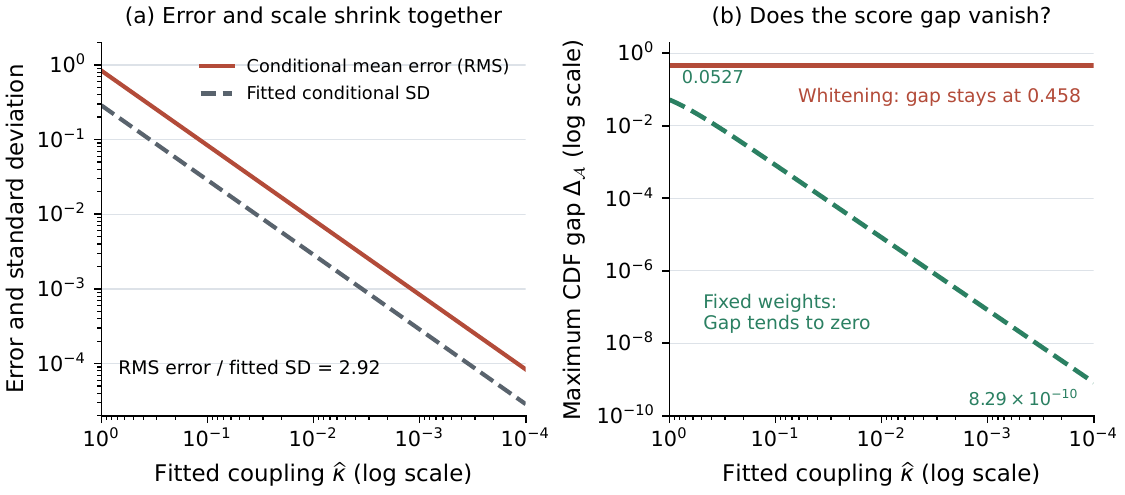}
\caption{Why the effect of fitting error can persist after whitening at the boundary
\(\kappa_0=\sigma_2=0\), with \(H=\sigma_1=1\), \(L=10\), \(K=2\)
and \(X_0=0\). (a) The root mean square error in the fitted conditional
mean and the fitted conditional standard deviation both shrink with
\(\widehat\kappa\); their ratio remains about 2.92. The mean square
averages over held-out paths and observation intervals, conditional on
the fitted value. (b) On a logarithmic vertical axis, the score-distribution
gap stays near 0.458 under fitted covariance whitening, whereas it
decreases towards zero under fixed weights \(\Sigma_*=I_2\). At each fitted
value, both rules use the same path laws and fitted zero-noise centre.
The curves are computed from the distributions derived in Supplementary
Section~\ref{app:linear-fitted-stability}. Both horizontal axes decrease
to the right and exclude zero, where fitted whitening is undefined.}
\label{fig:linear-boundary-score-gap}
\end{figure}

Figure~\ref{fig:linear-boundary-score-gap} illustrates this contrast.

The held-out and fitted reference path laws have total variation distance
one for every nonzero \(\widehat\kappa\), so a vanishing gap between the
selected score distributions does not require convergence of the path laws
in total variation.

\paragraph{Control when the conditional variance is positive.}
We now return to parameter values with positive conditional variance.
\begin{samepage}
\begin{proposition}[Score gap after fitting the linear model]
\label{prop:linear-fitted-energy-stability}
In this setting, let \(\kappa_0\) generate the held-out path and suppose
\(v_0=v_{\kappa_0}>0\) and \(\widehat v=v_{\widehat\kappa}>0\).
With \(S=T_{\widehat\kappa}\) in \eqref{eq:score-gap-main} and
\(m_K=\sup_{t\ge0}f_{\chi^2_K}(t)\),
\begin{equation}
\Delta_{\mathcal A}\le
\min\left\{1,\;Km_K\left[
\left|\frac{v_0}{\widehat v}-1\right|
+\frac{(\widehat\kappa-\kappa_0)^2\sigma_1^2H^3}{v_0}
\left(\frac{K-1}{2}+a_L^2\right)\right]\right\}.
\label{eq:linear-fitted-score-gap}
\end{equation}
\end{proposition}
\end{samepage}

The first bracketed term measures relative error in the conditional
variance. For the second term, we compare the conditional means of the
second-coordinate increment under the generating and fitted laws, at the
same starting state and first-coordinate increment. We square their
difference, average over observation intervals and over paths from the
generating law, then divide by \(v_0\). The centring error depends on the
state of the first coordinate, which accumulates earlier innovations. A small \(v_0\) need not imply
a large score gap: fitting error can decrease with \(v_0\) enough to keep
both terms in the bound small. If the bound
tends to zero in probability over repeated fits,
Corollary~\ref{cor:score-gap} implies that rejection probability, averaged
over fitting and simulation, converges to the Monte Carlo grid level for
fixed \(M\). Supplementary Section~\ref{app:linear-fitted-stability} gives
the conditional distribution and proof.

\paragraph{Training information in a controlled SIR experiment.}
A two-substep SIR experiment examines the size of fitting error relative
to the residual scale away from the singular boundary. The discrete
transition has an exact covariance, so this comparison does not depend on
a tangent approximation. We fit a constant infection rate to independent
training transitions and compare quadratic residual energies on 20-day
paths. Under the model null, targets are generated at the true coefficient,
while each fit determines the reference law and the common scoring rule
for target and reference paths.

We subtract from the susceptible residual the fitted linear contribution
of the residual for the log infectious fraction. The remaining residual
has small variance under the fitted transition law; we call it the
low-variance residual component.

At the nominal 5\% level, with 64 infectious endpoints used for training,
the whitened-energy test rejects 72.64\% of targets generated under the
null model. Adding the exact susceptible endpoints from the same training
transitions reduces this rate to 5.20\%. At the fixed initial state, we
divide the error in the fitted zero-noise centre for this low-variance
component by its fitted standard deviation. Across repeated
fits, this standardised error has root mean square \(2.498\) with
infectious-only training and \(0.1245\) with joint training.

With infectious-only training, downweighting this residual reduces the
rejection rate for null targets to 5.71\%.

\begin{figure}[H]
\centering
\includegraphics[width=\textwidth]{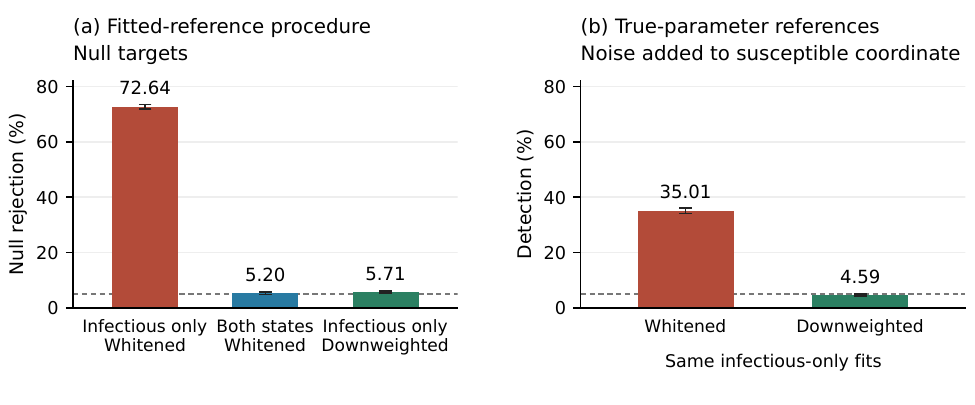}
\caption{Controlled SIR comparisons with 64 training transitions.
(a) Null rejection using fitted-parameter references: adding susceptible
endpoints changes the fit, the reference law and the scoring rule, whereas
downweighting retains the infectious-only fit.
(b) Detection of noise added to the susceptible coordinate using true-parameter references
and the same infectious-only fitted scoring rules. This ideal comparison
measures the sensitivity cost of downweighting. Bars show rates over
10,000 independent repetitions with 499 fresh references each; error bars
are pointwise 95\% Wilson intervals and dashed lines mark 5\%.}
\label{fig:sir-training-weights}
\end{figure}

To compare the sensitivity of the weighting rules at a common null level,
we generate reference paths at the true coefficient while retaining each
fitted scoring rule. The null target and reference paths then have the same
law. We add independent noise to the susceptible coordinate of the target
paths to assess sensitivity. The detection rate is 35.01\% with whitened energy and 4.59\% after
reducing the weight. Each comparison uses 10,000 complete repetitions and 499 new
reference paths per repetition; Supplementary Section~\ref{app:sir-refit}
gives the design and complete results.
Figure~\ref{fig:sir-training-weights} shows the fitted-reference and
true-reference comparisons in separate panels.

The practical issue is the size of fitting error relative to the residual
scale: additional training information reduces that error, while
lowering a residual's weight can also reduce sensitivity to model
departures. Supplementary Sections~\ref{app:nonlinear-selection}
and~\ref{app:feedback-selection} develop separate extensions for selecting
weights using training confidence sets.

\section{Criticising a SIR implementation on weekly reports}
\label{sec:empirical-studies}
\label{sec:sir-application}

We now apply the residual diagnostic to weekly epidemic reports.
Here reporting and state reconstruction are part of the reference law
built around an already fitted neural SIR implementation. The solver
generates seven-day state paths, from which a negative-binomial reporting
model generates counts. Smoothing and reconstruction provide the history
for the next simulation. We hold the fitted quantities and these rules
fixed when comparing residual features.

\subsection{Reports, reconstructed states and the reference model}

We use the archived Our World in Data snapshot and the six previously
examined country windows listed in Supplementary Table~\ref{tab:report-windows}.
In each 60-day window, nonzero case counts occur only at weekly report
dates. There are 53 reports in total: nine each for Czechia, Greece,
Finland, Norway and Sweden, and eight for Moldova. The archived daily
incidence is reproduced by rounding the trailing seven-day average of
those counts to two decimal places. We reconstruct the susceptible fraction from cumulative
incidence and the infectious fraction from incidence accumulated with an
exponential decay time of 28 days. These daily
states do not supply new observations between reports; the reports
themselves may also be dependent.

All countries share fitted parameters, but each reference path begins
from its own country's reconstructed history. In unnormalised coordinates,
let \(X=(S,Z)^\top\), where \(S\) is the susceptible fraction and
\(Z=\log I\) is the infectious log-fraction. The network receives the
normalised state \(Y=\mathsf W(X-\mu)\). The centring vector \(\mu\) and
diagonal scaling matrix \(\mathsf W\) are estimated from the training data
and held fixed during evaluation. With the fitted parameters, the equations are
\begin{equation}
\begin{aligned}
\dd S_t&=-\beta_\theta(Y_t,C_t)S_te^{Z_t}\dd t,\\
\dd Z_t&=\{\beta_\theta(Y_t,C_t)S_t-\widehat\gamma_0
-\tfrac12\sigma^2\}\dd t+\sigma\dd W_t,\\
\beta_\theta(y,c)&=[\widehat\beta_0+0.25\tanh f_{\theta,1}(y,c)]_+,
\end{aligned}
\label{eq:audited-sir-sde}
\end{equation}
Here \(f_{\theta,1}\) is the first network output. We fit one model
jointly to data from nine other countries, holding previously estimated
baseline rates fixed
and estimating the network parameters and a common scalar volatility
\(\sigma>0\). Controls \(C\) are computed from the current state and the preceding 19 states.
The solver takes ten projected Euler substeps per day and updates controls
daily. The baseline removal rate depends on the decay time used in state
reconstruction; its reciprocal therefore does not independently estimate
a biological recovery duration. Supplementary Section~\ref{app:sir-provenance}
gives the training and reconstruction details.

Starting from the reconstructed history at the preceding report date,
simulate seven days and let \(L_k\) be the nonnegative decrease in the
susceptible fraction multiplied by population. Generate a count \(B_k\) with
\begin{equation}
B_k\mid L_k\sim\mathrm{NB}\{\text{mean}=aL_k,
\ \text{variance}=aL_k+\phi(aL_k)^2\}.
\label{eq:report-law}
\end{equation}
Apply the same smoothing and reconstruction to each generated count,
then start the next seven-day prediction from the resulting history.
Each reference path therefore generates its own reports and reconstructed
states. The state is reset after each report; the latent forecast endpoint
is not continued into the next week. This construction defines the
reference law \(Q\) used below, which differs from the law obtained by
directly simulating the original SIR states.

With the network fixed, fitting \(a\) and \(\phi\) to 332 training-country
reports dated before 2021 gives \(a=1.04873\) and \(\phi=0.09670\).
The network was fitted using later records from other countries, so the
assessment remains retrospective.

\subsection{Residuals at the report scale}
\label{sec:report-residuals}

Write \(L_k(\varepsilon)\) for the seven-day simulated loss as a function of
its Gaussian inputs, conditional on the current reconstructed history.
Use \(m_{0,k}=aL_k(0)\) as the report centre and
\(v_k^{tan}=\|D_\varepsilon L_k(0)\|^2\) as the first-order variance of
the loss, including daily feedback. We compare
\begin{equation}
\begin{aligned}
d_k^{report}&=m_{0,k}+\phi m_{0,k}^2,\\
d_k^{prop}&=d_k^{report}+(1+\phi)a^2v_k^{tan},\qquad
z_k=\frac{B_k-m_{0,k}}{\sqrt{\max\{d_k,1\}}}.
\end{aligned}
\label{eq:report-scales}
\end{equation}
Here \(d_k\) is either \(d_k^{report}\) or \(d_k^{prop}\).
The total-variance decomposition for \eqref{eq:report-law} contains the
term \((1+\phi)a^2\operatorname{Var}(L_k)\), in addition to reporting
variation evaluated at the mean loss. We approximate the mean of \(L_k\)
by \(L_k(0)\) and the variance of \(L_k\) by \(v_k^{tan}\), giving
\(d_k^{prop}\). A variance floor of one squared case handles zero
local scales; we also compare a fixed floor of 100 squared cases.

Whether propagation changes the relative weights of reports depends on
its size relative to reporting variation at each history. Supplementary
Section~\ref{app:report-scale-checks} examines its size and the effect of
reporting variation.

\subsection{Residual features on the report grid}

We use \(\mathcal L=\{1\}\) in \eqref{eq:ce-statistics}, giving five
weekly features. Lag seven would leave at most two pairs in these
windows.
For each country and residual scale rule, we estimate and fix the feature
medians and standard deviations using 2,500 independent pilot paths, then
calculate the global rank against 2,500 separate reference paths.
All six windows have been inspected previously. The
following comparisons are exploratory, with no adjustment across countries.

Table~\ref{tab:report-features} gives all five features for Czechia.
The residual mean is 0.003 and mean energy is 0.635, inside
its central 95\% reference interval \([0.284,2.497]\). Energy correlation is
0.647, above its own interval \([-0.587,0.539]\). Thus a summary of average
magnitude misses temporal structure retained by the dependence feature.
The global rank based on the median-distance score is 0.0164. The transformed component for
energy correlation is the largest, so its adjusted rank against the
reference global maxima is also 0.0164.

\begin{table}[H]
\centering
\small
\begin{tabularx}{\textwidth}{@{}Yrrr@{}}
\toprule
Feature & Observed & \shortstack{95\% reference\\interval} & \shortstack{Adjusted\\rank} \\
\midrule
Signed mean \(T_z^{mean}\) & 0.003 & \([-0.574,0.703]\) & 0.9996 \\
Maximum \(\lvert C_z\rvert\), \(T_z^{cum}\) & 0.593 & \([0.379,2.316]\) & 1.0000 \\
Mean energy \(T_E^{level}\) & 0.635 & \([0.284,2.497]\) & 1.0000 \\
Maximum \(\lvert C_E\rvert\), \(T_E^{cum}\) & 1.096 & \([0.465,5.045]\) & 1.0000 \\
Energy correlation \(T_E^{(1)}\) & 0.647 & \([-0.587,0.539]\) & 0.0164 \\
\bottomrule
\end{tabularx}
\caption{Residual features for Czechia's nine weekly reports, using the
propagated scale and median-distance score. Intervals are the central
95\% ranges of each raw feature on the reference paths. Adjusted ranks
compare transformed target features with reference global maxima,
as in \eqref{eq:component-adjusted}. The smallest adjusted rank equals
the global rank, 0.0164. This is an exploratory comparison with the
fitted network and report parameters held fixed.}
\label{tab:report-features}
\end{table}

\begin{figure}[!ht]
\centering
\includegraphics[width=\textwidth]{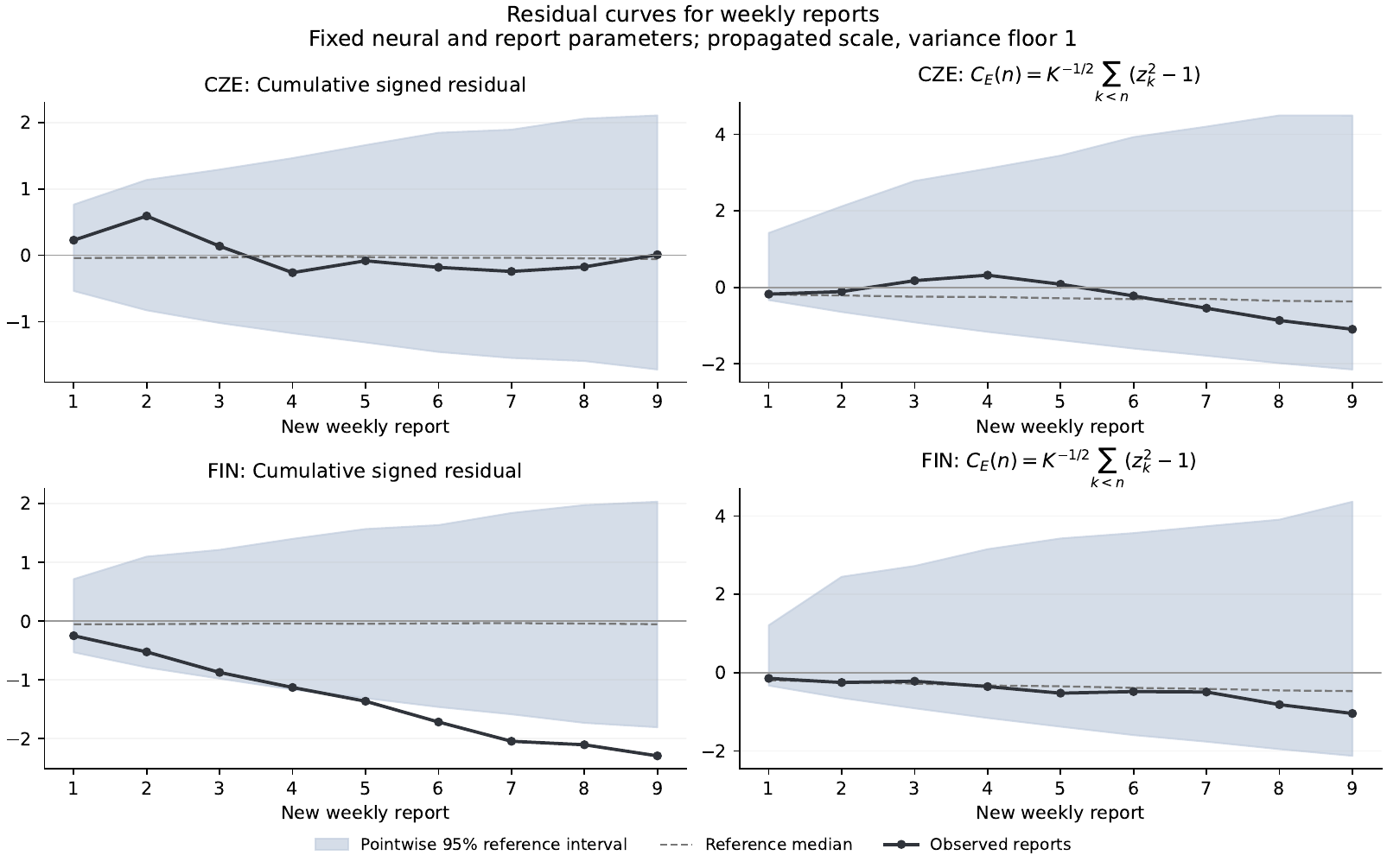}
\caption{Weekly residual curves for Czechia and Finland, chosen for
illustration after examining the six windows. Residuals are standardised
using the propagated scale in \eqref{eq:report-scales}. Shading gives pointwise 95\% ranges
from 2,500 reference paths; dashed lines are reference medians.
The fitted network and report parameters are fixed.}
\label{fig:sir-report-curves}
\end{figure}

For Finland, residuals standardised using the propagated scale have
mean \(-0.764\). The cumulative signed curve decreases at each of the
nine reports, showing that each reported count is below its model centre
computed with latent noise inputs set to zero (Figure~\ref{fig:sir-report-curves}). The global rank
based on the median-distance score is 0.0676.

Keeping the residuals and five raw features unchanged, an alternative
transformation measures each feature's extremeness through empirical
pilot tail proportions. It gives global ranks 0.0268 for Finland and
0.1004 for Czechia. The descriptive feature findings are unchanged, but whether a rank falls
below 0.05 depends on the feature transformation. Supplementary
Table~\ref{tab:report-ranks} reports both rules for all six windows.

In simulations starting from Norway's recorded history, we compare the report-only
and propagated residual scales using the five-feature median-distance
score, with the network and report parameters fixed. The propagated scale gives a higher detection rate than the report-only
scale for an infection-rate increase of 0.005 per day, but a lower rate when
SIR volatility is multiplied by 1.25
(Supplementary Section~\ref{app:report-power}).

We then examine how the local centre and variance approximations affect
the pilot feature scales. At one
selected pilot history in each of Czechia and Greece, both the seven-day
susceptible loss \(L_k(0)\) and its derivative \(D_\varepsilon L_k(0)\) are
zero, even though finite noise inputs can produce a positive loss.
In the Czech pilot, dividing report deviations by the local scales produces
some very large residuals. Their squared values inflate the pilot standard
deviations of the mean-energy and cumulative-energy features, reducing these
features' relative weight in the five-feature maximum.

On common simulated paths for Czechia, we replace the zero-input centre
\(m_{0,k}\) and propagated variance \(d_k^{prop}\) by conditional moments
estimated by simulation and recompute the pilot feature transformations.
For an infection-rate increase of 0.005 per day, the post hoc comparison
using standard-deviation feature scaling gives detection rates of 3.4\%
with the local construction and 46.8\% with simulated moments, at nominal
level 5\%. The prespecified comparison using interquartile-range feature
scaling gives 43.0\% and 42.4\%, respectively. Changing the feature scale
alone thus raises detection with the local residuals; replacing their
centre and variance approximations adds no clear gain under
interquartile-range scaling
(Supplementary Table~\ref{tab:report-moment-scaling}). These comparisons
hold the fitted law fixed. Supplementary
Section~\ref{app:report-moment-comparison} gives the conditional-moment
calculations and complete-path comparisons.

\section{Discussion}
\label{sec:discussion}

Our diagnostic combines departures in centring, magnitude and temporal
dependence in a global score, while raw features and cumulative curves
describe their direction and timing.

An approximate residual centre or covariance does not by itself invalidate
Monte Carlo calibration. Under the independence and continuity conditions
of Theorem~\ref{thm:rank}, calibration holds when the resulting target and
reference scores have the same conditional distribution.

Propagating covariance over the observation interval tracks how substep
noise affects later states through the drift. The linear results show
that this can change both the variance and the directions of residual
variation. With known coefficients, whitening maximises the standardised
mean shift of quadratic scores under the specified local common-scale
departures. In the linear family, the bound for positive conditional
variances separates relative variance error from standardised centring
error. At the singular boundary, the error in the fitted conditional mean
shrinks at the same rate as the whitening scale, so the score gap persists
as nonzero coefficient
estimates approach the truth. For the same sequence of fitted reference
laws, keeping one positive-definite weighting matrix fixed makes the score
gap tend to zero.

In the controlled SIR experiment with references simulated at fitted
parameters, adding susceptible-state endpoints to training and
downweighting the low-variance residual both lower null rejection. The
separate comparison using true-parameter references shows the sensitivity
cost of downweighting for added noise in the susceptible coordinate.

In the exploratory weekly-report analysis, the reference law includes
reporting and reconstruction. For Czechia, the correlation between squared
residuals lies above its reference interval, while mean energy lies within
its own interval. For Finland, all nine reports fall below their zero-input
report centres. In simulations for Czechia with an infection-rate increase of 0.005 per day,
replacing the local centre and variance approximations with simulation
estimates of conditional moments raises detection under standard-deviation
feature scaling but gives no clear gain under interquartile-range scaling.
Calibration after jointly repeating the neural and reporting fits has
not been evaluated.

In a separate comparison at the 53 observed report histories, a
negative-binomial model using recent weekly growth has lower mean scaled
continuous ranked probability score for the next report than the neural
SIR reporting model
(Supplementary Section~\ref{app:report-benchmarks}). The residual diagnostic
describes particular aspects of disagreement; distinguishing reporting,
reconstruction and latent epidemic causes requires further information.

\section*{Data and code availability}

The epidemiological data are from the Our World in Data COVID-19 dataset
\citep{OWIDCOVID}. The archived input snapshot is available from the
\href{https://github.com/owid/covid-19-data/blob/f0d53320b19703cf565940d8269181a8423d7032/public/data/owid-covid-data.csv}{official repository version last updated on 19 August 2024}.
Supplementary Section~\ref{app:sir-reconstruction} describes the reconstruction.
The archived data, original fitted model and earlier retained results
are publicly available at
\href{https://doi.org/10.5281/zenodo.22212995}{Zenodo (doi:10.5281/zenodo.22212995)}.
Source code is available in the GitHub repository at
\url{https://github.com/clemmentin/Residual-Model-Criticism-for-Fitted-SDE-Implementations}.
Scripts, seeds and small result tables for the additional experiments are in
\path{experiments/sir_refit_scale/} and
\path{experiments/sir_case_observation/}. Supplementary
Sections~\ref{app:sir-refit} and~\ref{app:report-reproduction} give the files
and reproduction instructions.

\begingroup
\setstretch{1.1}
\setlength{\parskip}{0.1em}
\section*{Acknowledgements}
The author thanks James Foster at the University of Bath for his continued
supervision and guidance. The author also thanks their family for their support.

\noindent\textbf{AI declaration.} The author used ChatGPT over the course of the project to assist with discussion of mathematical arguments, code development and debugging, and manuscript preparation and refinement. The author directed its use, reviewed and verified all outputs incorporated into the work, and takes full responsibility for the final content.

\noindent\textbf{Funding.} The author received no financial support for this research.

\noindent\textbf{Conflict of interest.} The author declares no conflicts of interest.

\endgroup

\clearpage
\setcounter{section}{0}
\setcounter{subsection}{0}
\setcounter{equation}{0}
\setcounter{figure}{0}
\setcounter{table}{0}
\setcounter{theorem}{0}
\setcounter{proposition}{0}
\setcounter{corollary}{0}
\setcounter{definition}{0}
\setcounter{example}{0}
\setcounter{remark}{0}
\renewcommand{\thesection}{S\arabic{section}}
\renewcommand{\theHsection}{S\arabic{section}}
\renewcommand{\theequation}{S\arabic{equation}}
\renewcommand{\theHequation}{S\arabic{equation}}
\renewcommand{\thefigure}{S\arabic{figure}}
\renewcommand{\theHfigure}{S\arabic{figure}}
\renewcommand{\thetable}{S\arabic{table}}
\renewcommand{\theHtable}{S\arabic{table}}
\renewcommand{\thetheorem}{S\arabic{theorem}}
\renewcommand{\theHtheorem}{S\arabic{theorem}}
\renewcommand{\theproposition}{S\arabic{proposition}}
\renewcommand{\theHproposition}{S\arabic{proposition}}
\renewcommand{\thecorollary}{S\arabic{corollary}}
\renewcommand{\theHcorollary}{S\arabic{corollary}}
\renewcommand{\thedefinition}{S\arabic{definition}}
\renewcommand{\theHdefinition}{S\arabic{definition}}
\renewcommand{\theexample}{S\arabic{example}}
\renewcommand{\theHexample}{S\arabic{example}}
\renewcommand{\theremark}{S\arabic{remark}}
\renewcommand{\theHremark}{S\arabic{remark}}
\pdfbookmark[0]{Supplementary Material}{supplement}
\section*{Supplementary Material}

\noindent This supplement gives proofs, experimental specifications and
protocol records for the main paper. Unprefixed equation, theorem, figure,
table and section numbers refer to the main paper; supplementary numbers
begin with S.

The table links topics in the main paper to proofs, constructions and
related experiments in this supplement.

\begin{center}
\small
\setstretch{1.1}
\begin{tabularx}{\textwidth}{@{}>{\raggedright\arraybackslash}p{0.23\textwidth}YY@{}}
\toprule
Topic & Proofs or constructions & Related experiments \\
\midrule
Calibration and scoring (Sections~\ref{sec:method}--\ref{sec:residual-coordinates}) &
Score conventions: \ref{app:score-details}; calibration proofs: \ref{app:rank-proof}, including the Euler bound in \ref{app:smooth-score-bound} &
--- \\
Propagation and sensitivity (Sections~\ref{sec:residual-coordinates}--\ref{sec:fitted-stability}) &
Covariance: \ref{app:projected-residual-example}--\ref{app:tangent-proof}; local efficiency: \ref{app:scale-optimality} &
Known-coefficient comparison: \ref{sec:linear-mechanism}; residual weights: \ref{app:coordinate-ablation} \\
Fitting error (Section~\ref{sec:fitted-stability}) &
Linear score distributions: \ref{app:linear-fitted-stability} &
Neural fits: \ref{sec:refit-size}--\ref{app:ce-conditional}; controlled SIR fits: \ref{app:sir-refit} \\
Weight selection (Section~\ref{sec:fitted-stability}) &
Selection guarantees: \ref{app:nonlinear-selection}--\ref{app:feedback-selection} &
Nonlinear comparisons: \ref{app:nonlinear-computation}; feedback repetitions: \ref{app:feedback-selection}; retrospective exclusion: \ref{app:joint-feedback} \\
SIR fit and analysis history (Section~\ref{sec:sir-application}) &
Neural fit and reconstruction: \ref{app:sir-provenance} &
Earlier state comparisons: \ref{app:sir-legacy-results} \\
Weekly report analysis (Section~\ref{sec:sir-application}) &
Report model: \ref{app:report-observation} &
Scale and sensitivity comparisons: \ref{app:report-scale-checks}--\ref{app:report-moment-comparison}; predictive benchmarks: \ref{app:report-benchmarks} \\
\bottomrule
\end{tabularx}
\end{center}

\section{Score conventions}
\label{app:score-details}

For the transformed components in \eqref{eq:component-departures}, the
pointwise Monte Carlo p-value is
\begin{equation}
\widehat p_j^{pt}
=\frac{1+\sum_{m=1}^M\1\{R_j(Y^{(m)})\ge R_j(Y^{held})\}}{M+1}.
\label{eq:component-pointwise}
\end{equation}
To compute the max-adjusted value in \eqref{eq:component-adjusted}, we
compare each transformed observed feature value with the maximum over the
transformed feature values on each reference path. Any transformed component \(R_j\) attaining
the observed maximum has the smallest max-adjusted p-value, which equals
the global p-value. We use these
adjusted values to identify features
contributing to a global rejection
\citep{MyllymakiEtAl2017,MrkvickaMyllymakiHahn2017,ShahBuhlmann2018};
pointwise values have no multiplicity adjustment. We estimate feature
centres and scales from the feature values computed on independent pilot
paths. The pilot scores are excluded from the Monte Carlo rank used for evaluation.

We apply the same correlation convention to the target and every pilot and
reference path. We calculate each lagged segment's standard deviation with
divisor equal to the number of valid pairs. We set the correlation to zero
if there are fewer than two pairs or either standard deviation is below
\(10^{-12}\).
The calculation stops if any component, score or pilot scale is non-finite,
or if any pilot scale is at most \(10^{-10}\).

For the state-path centring--energy analyses, take \(\mathcal L=\{1,7\}\) in
\eqref{eq:ce-statistics}, giving six features. The weekly-report
analyses take \(\mathcal L=\{1\}\), giving five. The same lag set is
used on all target, pilot and reference paths in each comparison.

We also compare centring and energy as groups. For a group \(G\),
containing the centring or energy components retained on the chosen grid,
its \emph{reference rank proportion} uses the plus-one rule in
\eqref{eq:rank} with \(S\) replaced by \(\max_{j\in G}R_j\) on both
target and reference paths. Unlike the adjusted component comparison,
this uses the same group on both sides.

\subsection*{Construction of the introductory illustration}
\phantomsection\label{app:motivation-construction}
Figure~\ref{fig:audit-motivation} uses 30 independent pairs
\((U_i,B_i\{0.5U_i+\sqrt{0.75}V_i\})\), with independent standard-normal
\(U_i,V_i\) and independent signs \(B_i\) taking values \(-1,1\) with equal
probability. The signs make the population lag-one covariance of the signed
innovations zero. The shared \(U_i\) makes the squares within each pair
dependent.
Concatenate the pairs to form the innovations \(z_1,\ldots,z_{60}\).
Let \(T_1=|s_z-1|\), \(T_2=|\widehat\rho_1(z)|\), and
\(T_3=|\widehat\rho_1(z^2)|\), where \(s_z\) uses divisor 59 and each
correlation centres its two lagged segments separately. The illustration uses
\[
S_{\rm ill}=\max_{j=1,2,3}\frac{T_j-\mu_j}{\tau_j},
\]
with pilot medians \(\mu_j\) and sample standard deviations \(\tau_j\).
There is no positive-part transformation after standardisation, unlike the
right-tail rule in \eqref{eq:component-departures}. The held-out, pilot and
reference samples use NumPy generator seeds 1001, 20260717 and 1, respectively.
The plotted path is \(x_n=m_n+a\sum_{k=1}^n z_k\), with
\(m_n=0.95\sin(n/4.7)+0.45\sin(n/9.2)+0.018n\) and \(0\le n\le60\).
We choose the scale \(a\) so that
\(1-\sum_n(x_n-m_n)^2/\sum_n(x_n-\bar x)^2=0.94\).
This descriptive illustration uses a prescribed curve, not an estimated
SDE fit. The function \texttt{make\_motivation\_figure} in
\path{docs/draft/make_paper1_core_figures.py} reproduces it.

\section{Calibration proofs}
\label{app:rank-proof}

\paragraph{Proof of Theorem~\ref{thm:rank}.}
Conditional on \(\mathcal A\), the \(M+1\) independent scores have a common
distribution and are exchangeable. Continuity excludes ties, so the descending
rank of the held-out score has a discrete uniform conditional distribution
on \(\{1,\ldots,M+1\}\), giving \eqref{eq:rank-validity}. With ties,
the test using the deterministic \(\ge\) comparison is conservative;
randomised tie-breaking restores the discrete uniform conditional rank
distribution.

\paragraph{Proof of Corollary~\ref{cor:score-gap}.}
Condition on \(\mathcal A\) and fix the reference scores. Rejection is then
a right-tail threshold event in the held-out score, or an empty or certain
event at boundary levels. Replacing the conditional held-out score
distribution by the conditional reference score distribution changes the
rejection probability by at most \(\Delta_{\mathcal A}\) in
\eqref{eq:score-gap-main}. After this replacement, the rank has a discrete uniform distribution
when we average over the reference scores conditional on \(\mathcal A\). This
proves \eqref{eq:rank-bound-main}. Averaging over fitting, selection and
pilot simulation then gives \eqref{eq:refit-bound-main}. Positive and negative conditional rejection-probability errors can cancel
when we average over fits. Rejection rates over repeated fits therefore
do not estimate
\(\E\Delta_{\mathcal A}\). Related work accounts for parameter estimation
in diffusion specification tests \citep{CorradiSwanson2005} and studies
the calibration of p-values under composite null models
\citep{BayarriBerger2000,RobinsVanDerVaartVentura2000}.

\paragraph{The error at a specified level.}
To relate the score CDF gap to rejection error at one specified level,
fix the analysis information \(\mathcal A\).
Write \(F\) for the CDF of the target score and \(F_0\) for the CDF of a
score computed on a path from the fitted reference path law, both
conditional on \(\mathcal A\).
Put \(\alpha_M=k_\alpha/(M+1)\) and
\(g(u)=\Prob\{\operatorname{Bin}(M,1-u)\le k_\alpha-1\}\).
Here \(g(u)\) is the rejection probability, averaged over the fresh reference
scores, when the held-out score is at percentile \(u\) of the conditional
reference-score distribution. For continuous \(F_0\) and \(1\le k_\alpha\le M\),
\begin{equation}
\Prob(p_M\le\alpha\mid\mathcal A)-\alpha_M
=-\int_{\R}\{F(t)-F_0(t)\}\,\dd[g\{F_0(t)\}].
\label{eq:level-cdf-integral}
\end{equation}
The integration measure has mass one. CDF differences contribute little
where this measure places little weight, and positive and negative
differences can cancel. Thus the error depends on a weighted integral of
CDF differences.

Conditional on \(\mathcal A\) and a held-out score \(t\), the number of
reference scores that exceed \(t\) is binomial. The test therefore rejects
with probability
\(\phi(t)=g\{F_0(t)\}\). The integrals of \(\phi\) under \(F\) and \(F_0\)
differ by
\[
\int\phi\,\dd(F-F_0)=-\int(F-F_0)\,\dd\phi,
\]
by Stieltjes integration by parts: \(\phi\) is continuous and the boundary
terms vanish. Moreover,
\[
g'(u)=\frac{u^{M-k_\alpha}(1-u)^{k_\alpha-1}}
{\mathrm B(M-k_\alpha+1,k_\alpha)}.
\]
For the reference quantile function \(Q_0\), define
\(H(u)=F\{Q_0(u)\}-u\). Changing to reference quantiles gives
\begin{equation}
\Prob(p_M\le\alpha\mid\mathcal A)-\alpha_M
=-\E H(U),\qquad U\sim\operatorname{Beta}(M-k_\alpha+1,k_\alpha).
\label{eq:level-beta-integral}
\end{equation}
Thus \(\E|H(U)|\) bounds the absolute conditional rejection error at the
specified level without cancellation between positive and negative CDF
differences.

\paragraph{From path-law convergence to calibration.}
Let \(P_{\mathcal A}\) and \(Q_{\mathcal A}\) be the held-out and reference
path laws conditional on \(\mathcal A\). The scoring rule is fixed and measurable
under this conditioning, so
\begin{equation}
\Delta_{\mathcal A}\le\TV(P_{\mathcal A},Q_{\mathcal A}).
\label{eq:tv-bound}
\end{equation}
For a sequence of analysis configurations \(\mathcal A_n\) with \(M\) fixed,
\(\Delta_{\mathcal A_n}\to0\) in probability implies
\(\E\Delta_{\mathcal A_n}\to0\) because the gap is bounded by one.
Corollary~\ref{cor:score-gap} then gives marginal rejection probability
converging to \(k_\alpha/(M+1)\). Path-law convergence in total variation
is sufficient. These conditional laws must include any outcome-based
selection used to obtain the held-out path.

\subsection{Euler score bound: statement, constants and proof}
\label{app:smooth-score-bound}
\label{sec:euler-stability}

We bound the conditional score gap for nonlinear Euler implementations by
coupling paths generated with the true and fitted coefficients and applying
the same scoring rule, fixed after fitting, to both paths. We first bound
the expected absolute difference between their scores using bounds on the
weight matrices and their variation with the state. We then impose a condition
on the probability mass that the reference-score distribution assigns to
short intervals. Together, these two bounds give a bound on the conditional
score gap.

Condition on the analysis information \(\mathcal A\), which fixes the
coefficients and scoring rule; all expectations, probabilities and norms
in this subsection use this conditioning unless further conditioning is stated.
Couple the two Euler
implementations using the same iid standard-normal vectors \(Z_j\):
\[
X_{j+1}=X_j+h b_0(X_j)+\sqrt h G_0Z_j,\qquad
\widehat X_{j+1}=\widehat X_j+h\widehat b(\widehat X_j)
+\sqrt h\widehat GZ_j.
\]
Both paths start from the same state. Each of the \(K\) observation
intervals has \(L\) substeps. Use the fitted
zero-noise map \(\widehat\Phi\) to define
\[
r_k=X_{(k+1)L}-\widehat\Phi(X_{kL}),\qquad
S(X)=K^{-1}\sum_{k=0}^{K-1}r_k^\top W(X_{kL})r_k.
\]
Assume the fitted drift is globally Lipschitz, the drift error
\(b_0-\widehat b\) is uniformly bounded, and the difference
\(G_0-\widehat G\) between the constant diffusion matrices has finite
Frobenius norm. Assume also that \(W\) is
symmetric positive semidefinite, globally bounded and Lipschitz in operator
norm. The bound \(D_S\) in \eqref{eq:smooth-score-perturbation} accounts for
changes in the residuals and in the weights evaluated along the two paths.

Choose bounds \(d_b,d_g\) on coefficient error and \(L_b\) on the fitted
drift's Lipschitz constant such that
\[
\sup_x\|b_0(x)-\widehat b(x)\|\le d_b,\qquad
\|G_0-\widehat G\|_F\le d_g,\qquad
\operatorname{Lip}(\widehat b)\le L_b.
\]
The symmetric positive semidefinite weight function \(W\) has operator norm
at most \(w\) and Lipschitz constant \(L_W\) in that norm.
Put \(B=1+hL_b\). The substep index runs from \(j=0\) to \(KL\), and
the observation index from \(k=0\) to \(K-1\). Define the deterministic bounds
\begin{equation}
\begin{aligned}
D_0&=0,&D_{j+1}&=BD_j+hd_b+3^{1/4}\sqrt h\,d_g,\\
R_k&=D_{(k+1)L}+B^LD_{kL},&
R&=3^{1/4}\sqrt h\|\widehat G\|_F\sum_{j=0}^{L-1}B^j.&&
\end{aligned}
\label{eq:smooth-score-constants}
\end{equation}
Set
\begin{equation}
D_S=\frac1K\sum_{k=0}^{K-1}
\{wR_k(2R+R_k)+L_WD_{kL}R^2\}.
\label{eq:smooth-score-perturbation}
\end{equation}

The quantities \(D_j,R_k,R\) bound the \(L^4\) norms of the coupled path
differences, residual differences and reference residuals,
respectively. The first term in \(D_S\) bounds the effect of changing the
residuals at fixed weights. The second bounds the effect of evaluating the
weights at different states along the two paths.

A small expected absolute score difference need not imply a small CDF gap
when the reference-score distribution assigns substantial probability to a
short interval. To bound this probability, we condition on all Gaussian
innovations except one coordinate of the final innovation. We require a
positive lower bound on the coefficient of its square in the reference score.

\begin{samepage}
\begin{proposition}[Score gap for Euler implementations]
\label{prop:smooth-score-stability}
Under these assumptions, \(\E|S(X)-S(\widehat X)|\le D_S\).
Suppose a column \(\widehat g\) of \(\widehat G\) satisfies
\[
\frac hK\widehat g^\top W(\widehat X_{(K-1)L})\widehat g
\ge a_*>0\qquad\text{almost surely}.
\]
Then the conditional score gap satisfies
\begin{equation}
\Delta_{\mathcal A}\le
\min\left\{1,\left(1+\frac2{\sqrt\pi}\right)
\left(\frac{D_S}{a_*}\right)^{1/3}\right\}.
\label{eq:smooth-score-gap}
\end{equation}
\end{proposition}
\end{samepage}

Here \(D_S\) bounds the conditional expected absolute score difference,
and \(a_*\) gives a bound on the probability mass that the reference-score
distribution assigns to a short interval. We obtain the cube-root rate by
choosing the interval width to balance two probability bounds: one for a
large difference between the coupled scores and one for a reference score
lying near a threshold.

\begin{proof}[Proof of Proposition~\ref{prop:smooth-score-stability}]
Minkowski's inequality and
\(\E\|GZ\|^4\le3\|G\|_F^4\) give
\(\|X_j-\widehat X_j\|_{L^4}\le D_j\).
Since \(\widehat\Phi\) is \(B^L\)-Lipschitz, the residual differences have
\(L^4\) norm at most \(R_k\). Comparing a stochastic observation step with
the fitted zero-noise step gives \(\|\widehat r_k\|_{L^4}\le R\) for the
reference residual. Decompose the difference between corresponding score summands:
\[
\begin{aligned}
&r_k^\top W(X_{kL})r_k-\widehat r_k^\top
W(\widehat X_{kL})\widehat r_k\\
&\quad=(r_k-\widehat r_k)^\top W(X_{kL})(r_k+\widehat r_k)\\
&\qquad+\widehat r_k^\top\{W(X_{kL})-W(\widehat X_{kL})\}\widehat r_k.
\end{aligned}
\]
Cauchy--Schwarz bounds the expected absolute values of the two terms on the right
by the corresponding terms in
\eqref{eq:smooth-score-perturbation}.

Condition on all reference innovations except the coordinate of the final
innovation multiplying \(\widehat g\). The reference score is then
\(aZ^2+bZ+c\), where \(Z\sim N(0,1)\) and \(a\ge a_*\).
Completing the square shows that the inverse image of an interval of length
\(2\varepsilon\) has total length at most \(2\sqrt{2\varepsilon/a_*}\).
The standard-normal density is at most \(1/\sqrt{2\pi}\), hence
\[
\sup_t\Prob\{|S(\widehat X)-t|\le\varepsilon\}
\le\frac2{\sqrt\pi}\sqrt{\varepsilon/a_*}.
\]
For the coupled scores constructed above, event inclusion and Markov's inequality give
\[
\Delta_{\mathcal A}\le D_S/\varepsilon+
\sup_t\Prob\{|S(\widehat X)-t|\le\varepsilon\}.
\]
Taking \(\varepsilon=D_S^{2/3}a_*^{1/3}\) proves the result; if \(D_S=0\),
the two scores agree almost surely.
\end{proof}

\paragraph{Weights obtained by flooring covariance eigenvalues.}
For a bounded Lipschitz covariance estimate, replacing eigenvalues below
a positive floor by that floor before inversion gives bounded Lipschitz
weights. The resulting bounds limit the contribution of fitting error to
the score perturbation. Changing these weights can also change test sensitivity.
The constants below show how the floor enters the bound.

For a positive semidefinite covariance estimate \(\widehat\Sigma(x)\) with operator norm at most
\(s\) and Frobenius Lipschitz constant \(c_\Sigma\), flooring its eigenvalues
at \(\tau>0\) before inversion gives
\[
w\le\tau^{-1},\qquad L_W\le c_\Sigma\tau^{-2},\qquad
a_*=\frac{h\|\widehat g\|^2}{K\max(s,\tau)}
\]
for any nonzero diffusion column \(\widehat g\).

To obtain the Lipschitz bound, express the Frobenius norm of the matrix
difference in the two eigenvector bases and use the scalar Lipschitz
constant of \(u\mapsto1/\max(u,\tau)\).
Also \(W(x)\succeq I/\max(s,\tau)\), giving the stated value of \(a_*\).
The global propagation and cube-root bounds can be conservative.

The bound on the conditional score gap after fitting depends on coefficient
error, residual scales and the probability mass that the reference-score
distribution assigns to short intervals.

\section[Propagation, score distributions and weight selection]{Propagation, score distributions and\\ weight selection}
\label{app:finite-step-geometry}

\subsection{Propagation and a zero-variance restriction}\label{app:projected-residual-example}
For \(\dd X_1=\dd W,\ \dd X_2=X_1\dd t\), stochastic Fubini gives
\[
X_{2,H}-X_{2,0}-HX_{1,0}=\int_0^H(H-s)\dd W_s,
\qquad \operatorname{Var}(X_{2,H}-X_{2,0}-HX_{1,0})=H^3/3.
\]
Thus the increment of a coordinate over an observation interval can have
positive variance even when no noise enters that coordinate directly. A Monte Carlo test can remain calibrated when
its score uses residuals projected onto a direction with no direct noise.
In the damped linear system
\(A=\left(\begin{smallmatrix}-0.4&1\\0&-0.7\end{smallmatrix}\right)\),
\(G=(0,1)^\top\), with unit observation spacing, a test based on the squared norms of
residuals projected orthogonally to the instantaneous noise loading rejected
at rate \(0.048\). The check uses exact linear transitions and subtracts
\(e^A X_k\) from the next state to form each residual. The target and
reference scores therefore have the same distribution. The test compares
the held-out score with reference scores; it does not treat every nonzero
projected residual as impossible. The check used 60
increments, 5,000 replicates and 250 reference paths per replicate, with
Monte Carlo grid rejection level \(12/251=0.04781\). The script
\path{experiments/finite_step_geometry_synthetic.py} uses seed 20260710;
results are in \path{cache/summaries/finite_step_geometry_synthetic/}.

\subsection{Linear covariance range}

For the linear additive system \(\dd X_t=AX_t\dd t+G\dd W_t\), put
\(F_h=I+hA\). The continuous-time and Euler transition covariances are
\begin{equation}
\Sigma_H^{ex}=\int_0^H e^{Au}GG^\top e^{A^\top u}\dd u,\qquad
\Sigma_{H,L}^{E}=h\sum_{j=0}^{L-1}F_h^jGG^\top(F_h^\top)^j.
\label{eq:linear-covariance}
\end{equation}
The following standard controllability result gives their ranges.

\begin{samepage}
\begin{proposition}[Linear covariance range]
\label{prop:linear-finite-support}
Let \(H>0\), \(L\ge1\), \(h=H/L\), and define
\[
\mathcal R_q=\operatorname{span}_{0\le j<q}\operatorname{Range}(A^jG).
\]
In state dimension \(d\),
\begin{equation}
\operatorname{Range}(\Sigma_H^{ex})=\mathcal R_d,\qquad
\operatorname{Range}(\Sigma_{H,L}^{E})=\mathcal R_L.
\label{eq:linear-range}
\end{equation}
\end{proposition}
\end{samepage}

\begin{proof}
For the continuous-time transition covariance,
\(v^\top\Sigma_H^{ex}v=\int_0^H\|G^\top e^{A^\top u}v\|^2\dd u\).
It vanishes exactly when \(G^\top e^{A^\top u}v=0\) throughout the interval.
Derivatives at zero and Cayley--Hamilton then give
\(\operatorname{Null}(\Sigma_H^{ex})=\mathcal R_d^\perp\).
The Euler covariance has range
\(\operatorname{span}_{j<L}\operatorname{Range}(F_h^jG)\).
The identity \(F_h^jG=\sum_{r=0}^j\binom jr h^rA^rG\) gives a triangular
change of generators with nonzero diagonal. The span is therefore
\(\mathcal R_L\).
\end{proof}

\subsection{Covariance calculations for the two-driver family}

The independent Brownian drivers in \eqref{eq:linear-family} give the exact
continuous-time transition covariance over an interval of length \(H\):

\begin{equation}
\Sigma_H^{ex}=\begin{pmatrix}
\sigma_1^2H & \kappa\sigma_1^2H^2/2\\
\kappa\sigma_1^2H^2/2 & \sigma_2^2H+\kappa^2\sigma_1^2H^3/3
\end{pmatrix}.
\label{eq:linear-family-exact}
\end{equation}

The covariance formulas follow from \eqref{eq:linear-covariance}.
In the Euler scheme, the noise propagated from the first driver into the
second component is
\[
\kappa\sigma_1h^{3/2}\sum_{\ell=0}^{L-1}(L-1-\ell)\xi_{1,\ell},
\]
and the second component may also receive its own direct noise. Summing
the propagation weights and their squares gives \(a_L\) and \(b_L\); the corresponding
continuous integrals
give \eqref{eq:linear-family-exact}.

The statements about covariance rank follow from the determinant of
\eqref{eq:linear-family-euler}. The limits \(a_L\to1/2\), \(b_L\to1/3\)
recover \eqref{eq:linear-family-exact}. The general range characterisation
is the standard finite-horizon controllability result \citep{Sontag1998}.

\goodbreak
\begin{samepage}
\subsection{Small-noise tangent covariance: statement and proof}
\label{app:tangent-proof}

\begin{proposition}[Small-noise covariance of a smooth observation-step map]
\label{prop:tangent-first-variation}
Fix \(L,h\), the implementation and its conditioning inputs, and an initial
state \(y_0\). Let \(\Psi_\ell:\R^d\times\R^r\to\R^d\) be \(C^2\), with
globally bounded first and second derivatives. For independent
\(\xi_\ell\sim N(0,I_r)\), set
\[
Y_0^{(\epsilon)}=\bar Y_0=y_0,\qquad
Y_{\ell+1}^{(\epsilon)}
=\Psi_\ell(Y_\ell^{(\epsilon)},\epsilon\xi_\ell),\qquad
\bar Y_{\ell+1}=\Psi_\ell(\bar Y_\ell,0).
\]
Use the state and noise derivatives \(F_\ell,g_\ell\) defined before
\eqref{eq:tangent-recursion}. With \(U_0=0\) and
\(U_{\ell+1}=F_\ell U_\ell+\sqrt h\,g_\ell\xi_\ell\),
\begin{equation}
\E\left\|\frac{Y_L^{(\epsilon)}-\bar Y_L}{\epsilon}-U_L\right\|^2
=O(\epsilon^2),\qquad
\epsilon^{-2}\operatorname{Cov}(Y_L^{(\epsilon)})
\longrightarrow\Sigma_L^{tan}.
\label{eq:small-noise}
\end{equation}
\end{proposition}
\end{samepage}

\begin{proof}[Proof of Proposition~\ref{prop:tangent-first-variation}]
Let \(B_\ell\) be the derivative of \(\Psi_\ell\) with respect to its noise
input, evaluated at \((\bar Y_\ell,0)\), so \(B_\ell=\sqrt h\,g_\ell\).
All constants below may depend on the fixed maps and \(L\), but not on
\(0<\epsilon\leq1\).  Put
\(\delta_\ell=Y_\ell^{(\epsilon)}-\bar Y_\ell\).
The bounded first derivatives imply
\[
\|\delta_{\ell+1}\|\leq C(\|\delta_\ell\|+\epsilon\|\xi_\ell\|).
\]
Since \(\delta_0=0\) and the Gaussian innovations have finite fourth moments,
induction over the fixed \(L\) substeps gives
\(\E\|\delta_\ell\|^4=O(\epsilon^4)\).
Taylor expansion jointly in state and noise input yields
\[
\delta_{\ell+1}=F_\ell\delta_\ell+\epsilon B_\ell\xi_\ell+R_\ell,
\qquad
\|R_\ell\|\leq C(\|\delta_\ell\|+\epsilon\|\xi_\ell\|)^2.
\]
The bounded second derivatives and the preceding fourth-moment bound imply
\(\E\|R_\ell\|^2=O(\epsilon^4)\).
For \(e_\ell=\delta_\ell/\epsilon-U_\ell\),
\[
e_0=0,\qquad e_{\ell+1}=F_\ell e_\ell+R_\ell/\epsilon.
\]
A second induction gives \(\E\|e_L\|^2=O(\epsilon^2)\).
Thus \(\delta_L/\epsilon\to U_L\) in \(L^2\); Cauchy--Schwarz gives
convergence of means and second-moment matrices, hence of covariances.
Because \(\bar Y_L\) is deterministic,
\(\epsilon^{-2}\operatorname{Cov}(Y_L^{(\epsilon)})
=\operatorname{Cov}(\delta_L/\epsilon)\).
Finally, \(F_\ell,B_\ell\) are deterministic once the implementation,
conditioning inputs and initial state have been fixed. Since \(U_\ell\)
is independent of the fresh \(\xi_\ell\),
\[
\operatorname{Cov}(U_{\ell+1})
=F_\ell\operatorname{Cov}(U_\ell)F_\ell^\top+B_\ell B_\ell^\top.
\]
Using \(B_\ell=\sqrt h\,g_\ell\) proves the stated recursion.
\end{proof}

Restore the observation index \(k\), and set
\(\Pi_{k,\ell}=F_{k,L-1}\cdots F_{k,\ell+1}\) and \(\Pi_{k,L-1}=I\).
Expanding \eqref{eq:tangent-recursion} gives
\begin{equation}
\widehat\Sigma_k^{tan}
=h\sum_{\ell=0}^{L-1}
\Pi_{k,\ell}g_{k,\ell}g_{k,\ell}^\top\Pi_{k,\ell}^\top.
\label{eq:tangent-sum}
\end{equation}
The propagated loadings need not be parallel, even with one Brownian driver.
Let the columns of \(U_r\) be orthonormal covariance eigenvectors retained
by a prespecified rule, and let \(\Lambda_r\) be the diagonal matrix of
their positive eigenvalues. If
\(\delta_k\) is the residual under the linearised recursion, with covariance
\(\widehat\Sigma_k^{tan}\), then
\(\Lambda_r^{-1/2}U_r^\top\delta_k\) has identity covariance. The earlier
daily-state SIR score standardises the scalar residual in
\eqref{eq:sir-coordinate} using its marginal variance. The report application uses
\eqref{eq:report-scales}.

\subsection{Score distributions after fitting the linear model}
\label{app:linear-fitted-stability}

Gaussian norm comparison can bound differences between score distributions
directly \citep{GoetzeEtAl2019}. For our linear family, a conditional
chi-square argument gives the explicit constant in
Proposition~\ref{prop:linear-fitted-energy-stability}.

Use the Euler implementation of \eqref{eq:linear-family} with
\(H>0\), \(L\ge2\), \(K\ge2\), \(X_0=0\), and known
\(\sigma_1>0,\sigma_2\ge0\). Estimate \(\kappa\) from independent
training data. The held-out path is generated at \(\kappa_0\) and the
reference path at \(\widehat\kappa\); both use the fitted zero-noise
centre and covariance in \eqref{eq:linear-energy}. Denote this score by
\(T_{\widehat\kappa}\), and define \(v_\kappa\) by
\eqref{eq:linear-conditional-variance}.

\begin{proof}[Proof of Proposition~\ref{prop:linear-fitted-energy-stability}]
The held-out transitions follow the \(L\)-substep Euler law at \(\kappa_0\).
Put \(q=\sigma_1^2H\) and condition on the analysis information \(\mathcal A\).
Conditional on \(X_k\), Gaussian regression of \(\Delta X_{2,k}\) on
\(\Delta X_{1,k}\)
gives the exact representation
\begin{equation}
\begin{aligned}
\Delta X_{1,k}&=\sqrt q\,\xi_k,\\
\Delta X_{2,k}&=\kappa_0 H\{X_{1,k}+a_L\Delta X_{1,k}\}
+\sqrt{v_0}\,\zeta_k,
\end{aligned}
\label{eq:fitted-linear-increments}
\end{equation}
where all \(\xi_k,\zeta_k\) are independent standard normals. The variance
\(v_0\) is the Schur complement of the first diagonal entry in
\eqref{eq:linear-family-euler}. Subtract the fitted zero-noise increment from the
held-out increment, then apply the inverse of the Cholesky factor of the fitted
covariance to the resulting residual to obtain
\[
z_{1,k}=\xi_k,\qquad
z_{2,k}=\frac{(\kappa_0-\widehat\kappa)H
\{X_{1,k}+a_L\sqrt q\,\xi_k\}+\sqrt{v_0}\,\zeta_k}
{\sqrt{\widehat v}}.
\]
Let \(\mathsf A_K\) be the \(K\times K\) matrix with diagonal \(a_L\), ones
strictly below the diagonal and zeros above it. Since \(X_{1,0}=0\),
\(X_{1,k}=\sqrt q\sum_{j<k}\xi_j\), and hence
\begin{equation}
\begin{aligned}
2KT_{\widehat\kappa}^{held}
&\overset d=\|\xi\|^2+\|\beta\mathsf A_K\xi+\rho\zeta\|^2,\\
\beta&=\frac{(\kappa_0-\widehat\kappa)\sigma_1H^{3/2}}
{\sqrt{\widehat v}},\qquad
\rho^2=\frac{v_0}{\widehat v},
\end{aligned}
\label{eq:fitted-linear-energy-law}
\end{equation}
with independent \(\xi,\zeta\sim N(0,I_K)\).
On the reference path, the two residual coordinates whitened using the Cholesky factor of the fitted covariance are independent
standard normals at every step, so \(2KT_{\widehat\kappa}^{ref}\sim\chi^2_{2K}\).
The held-out residuals need not be independent across observation steps
after we subtract the fitted zero-noise increments and whiten using the covariance calculated from the fitted parameters. In \eqref{eq:fitted-linear-energy-law}, the term
\(\beta\mathsf A_K\xi\) depends both on the error in estimating \(\kappa\)
and on first-coordinate innovations from the current and earlier steps.

First compare \(U+\rho^2V\) with \(U+V\), where \(U=\|\xi\|^2\) and
\(V=\|\zeta\|^2\) are independent \(\chi^2_K\). Given \(V\), the density
bound \(m_K\) for \(U\) bounds their CDF difference by
\(m_K|\rho^2-1|V\). Averaging gives \(m_KK|\rho^2-1|\).

Next condition on \(\xi\). The second term in
\eqref{eq:fitted-linear-energy-law}, divided by \(\rho^2\), has a noncentral
\(\chi^2_K\) distribution with noncentrality
\(\ell=\|\beta\mathsf A_K\xi\|^2/\rho^2\).
Its Poisson mixture representation is \(V+W_N\), where
\(N\sim\operatorname{Poisson}(\ell/2)\), \(W_N\mid N\sim\chi^2_{2N}\),
and \(V\) is independent of \((N,W_N)\); \(W_0=0\).
The representation follows by expanding the noncentral chi-square moment
generating function in \(\ell\). The variable \(\rho^2V\) has density bounded
by \(m_K/\rho^2\), and the added nonnegative variable \(\rho^2W_N\) has
mean \(\rho^2\ell\) conditional on \(\xi\). Adding this variable changes the
conditional CDF by at most \(m_K\ell\). Averaging over \(\xi\) and combining the two comparisons
gives
\[
\Delta_{\mathcal A}\le
m_K\left\{K|\rho^2-1|+
\frac{\beta^2}{\rho^2}\|\mathsf A_K\|_F^2\right\}.
\]
Now \(\|\mathsf A_K\|_F^2=K\{(K-1)/2+a_L^2\}\), and substitution of
\(\beta,\rho\), together with \(\Delta_{\mathcal A}\le1\), proves
\eqref{eq:linear-fitted-score-gap}. The constant is finite for \(K\ge2\):
\(m_2=1/2\), and \(m_K=f_{\chi^2_K}(K-2)\) for \(K>2\).
\end{proof}

\paragraph{Comparing weighting rules at the boundary.}
At \(\kappa_0=\sigma_2=0\) and \(\widehat\kappa\ne0\), the representation
\eqref{eq:fitted-linear-energy-law} still holds, now with \(\rho=0\) and
\(\beta=-\operatorname{sgn}(\widehat\kappa)/\sqrt{c_L}\). Thus
\begin{equation}
2KT_{\widehat\kappa}^{held}
\overset d=\|\xi\|^2+c_L^{-1}\|\mathsf A_K\xi\|^2,
\qquad
2KT_{\widehat\kappa}^{ref}\sim\chi^2_{2K}.
\label{eq:fitted-linear-boundary-law}
\end{equation}
The held-out score distribution is therefore the same for every nonzero
\(\widehat\kappa\). Taking expectations gives
\eqref{eq:linear-boundary-energy}. For \(K\ge2,L\ge2\) the two means differ,
so the score gap is a strictly positive constant along any nonzero
sequence \(\widehat\kappa\to0\).

Use one matrix \(\Sigma_*\succ0\), held constant across paths and throughout
the sequence \(\widehat\kappa\to0\). The residual formed by subtracting the fitted
zero-noise increment from a held-out increment is
\((\sqrt q\xi_k,-\widehat\kappa H\sqrt q\sum_{j<k}\xi_j)^\top\).
The reference residual is
\((\sqrt q\xi'_k,\widehat\kappa Ha_L\sqrt q\xi'_k+
\sqrt{\widehat v}\zeta'_k)^\top\), with independent standard normal
\(\xi',\zeta'\). As \(\widehat\kappa\to0\), both stacked residual laws
converge to the same law supported on the first coordinate. The energies
computed with these fixed weights converge in distribution to
\[
\frac{q(\Sigma_*^{-1})_{11}}{2K}\,\chi^2_K.
\]
This limit has a continuous CDF, so both CDFs converge uniformly and the
score gap tends to zero. The conclusion also holds in probability for
random fits converging to zero because the conditional gap is a function
of \(\widehat\kappa\). This argument fixes \(K,H,L\) and the
positive-definite weights.
At this boundary the second coordinate is deterministic under the true law
and has a nondegenerate Gaussian increment under every nonzero fitted coupling.
Hence \(\TV(Q_{\widehat\kappa},Q_0)=1\) for every \(\widehat\kappa\ne0\),
even though the fixed-weight score gap vanishes. Thus convergence of the path laws in total variation is sufficient for
the score gap to vanish, but this example shows that it is not necessary.

For Figure~\ref{fig:linear-boundary-score-gap}, we take
\(H=\sigma_1=1\), \(L=10\), \(K=2\) and \(\Sigma_*=I_2\).
In panel (a), the fitted conditional mean error is
\(e_k=\widehat\kappa H(X_{1,k}+a_L\Delta X_{1,k})\). Under the held-out
law at the boundary,
\[
\left\{\frac1K\sum_{k=0}^{K-1}\E[e_k^2\mid\mathcal A]\right\}^{1/2}
=|\widehat\kappa|\sigma_1H^{3/2}
\sqrt{\frac{K-1}{2}+a_L^2}.
\]
Dividing by the fitted conditional standard deviation
\(\sqrt{\widehat v}=|\widehat\kappa|\sigma_1H^{3/2}\sqrt{c_L}\)
gives \(\sqrt{281/33}\simeq2.918\). This averages over paths and intervals
at a fixed fitted value, not over repeated fits.
For panel (b), the score representations above reduce to sums of two
weighted chi-square variables. We evaluate the target CDFs by one-dimensional angular
quadrature and the reference CDFs by explicit formulas. A combined
linear and logarithmic threshold grid locates candidate extrema of both
signed CDF differences, which are then refined by numerical optimisation.
The upper search endpoint bounds both omitted tail probabilities by
\(10^{-13}\). Doubling the quadrature order from 64 to 128 and each
threshold grid from 1,025 to 2,049 points changes the reported gaps by
less than \(10^{-12}\). The script
\path{experiments/linear_boundary/plot_score_gap.py} retains the curve,
settings and numerical checks; it does not simulate paths or repeat fitting.

\paragraph{A consistent estimator with a persistent score gap at the boundary.}
Take \(n\) independent training transitions, each starting from zero and using
the same \(H,L\), independently of the held-out path. Put
\(V_j=Ha_L\Delta X_{1,j}\) and define the least-squares fit
\(\widehat\kappa_n^{LS}=\sum_{j=1}^nV_j\Delta X_{2,j}/\sum_{j=1}^nV_j^2\).
Under \(\sigma_2=0\), Gaussian regression gives
\[
\widehat\kappa_n^{LS}\mid(V_1,\ldots,V_n)
\sim N\!\left(\kappa_0,\frac{v_0}{\sum_{j=1}^nV_j^2}\right),
\qquad
\frac1n\sum_{j=1}^nV_j^2\longrightarrow H^2a_L^2q>0
\quad\text{almost surely}.
\]
Given \(V_1,\ldots,V_n\), the estimator's variance is proportional to \(v_0\).
Thus a small conditional transition variance alone does not imply instability.
The least-squares estimator is consistent. At \(\kappa_0=0\), all
second-coordinate training increments vanish, and the least-squares estimate
is exactly zero. The modified estimator
\(\widetilde\kappa_n=\max\{\widehat\kappa_n^{LS},n^{-1}\}\) is therefore
consistent for every \(\kappa_0\ge0\), but its strictly positive estimates
produce the persistent score gap in \eqref{eq:fitted-linear-boundary-law}
when the true coupling is zero.
At \(\kappa_0=0\), scoring only the first, stochastic coordinate with the
unmodified zero estimate gives matching held-out and reference score
distributions.

A nonvanishing score gap need not imply nonvanishing rejection error at a
specified level: the latter depends on the weighted CDF difference in
\eqref{eq:level-cdf-integral}.

\subsection{Selecting weights in the nonlinear coupling model}
\label{app:nonlinear-selection}
\label{sec:nonlinear-weights}
\label{sec:nonlinear-weight-comparison}

We extend the main-paper diagnostic by using confidence sets constructed from
training data to select residual weights in two specified models. For each fit,
we hold the fitted reference path law fixed and vary the parameter used to
generate target paths over the confidence set constructed from training data.
Using independent simulations, we compute simultaneous upper bounds on
rejection-probability error for the tests defined by the candidate weights. We then evaluate the
selected score on a new independent target path and fresh paths from the fitted law.

Both constructions use model-specific bounds rather than the general
Euler bound in Proposition~\ref{prop:smooth-score-stability}. In the first
model, we compare bounding the discrepancy between score CDFs with bounding
rejection error at one specified level. In the second, the parameter changes both path
coordinates through feedback.

The first model is

\begin{equation}
\dd X_t=-X_t\,\dd t+\dd W_t,\qquad
\dd Y_t=\kappa\tanh(X_t)\,\dd t,\qquad \kappa>0.
\label{eq:nonlinear-coupling}
\end{equation}
Only \(\kappa\) is unknown. Each unit observation interval uses two Euler
substeps. Training consists of \(n\) independent transitions from \((0,0)\).
Held-out and reference paths start there and run for \(K\) observation steps. We generate
the held-out path at the unknown \(\kappa\) and reference paths at the
training estimate \(\widehat\kappa\). We construct this estimate using a
statistic whose distribution at the generating parameter is exactly chi-square.
Inverting that statistic gives a confidence interval \([\kappa_-,\kappa_+]\)
with coverage \(1-\gamma\). The corresponding
relative-error interval is
\(I=[\kappa_-/\widehat\kappa-1,\kappa_+/\widehat\kappa-1]\).
The formulas are given below.

As in confidence-set methods for nuisance parameters \citep{BergerBoos1994},
the training interval restricts the parameter set used in selection.

\paragraph{Confidence interval from training.}
The training \(Y\) increments after two substeps from \((0,0)\) satisfy
\(D_i=\kappa h\tanh(\sqrt h Z_i)\), with \(h=1/2\) and independent
standard-normal \(Z_i\). For \(u>\max_i|D_i|/h\), define
\begin{equation}
T_n(u)=h^{-1}\sum_{i=1}^n
\operatorname{atanh}^2\!\left(\frac{D_i}{hu}\right),
\qquad T_n(\widehat\kappa)=n.
\label{eq:coupling-fit}
\end{equation}
Then \(T_n(\kappa)=\sum_i Z_i^2\). The interval endpoints solve
\begin{equation}
T_n(\kappa_-)=\chi^2_{n,1-\gamma/2},\qquad
T_n(\kappa_+)=\chi^2_{n,\gamma/2},
\label{eq:coupling-interval-roots}
\end{equation}
where \(\chi^2_{n,p}\) is the \(p\)-quantile of \(\chi_n^2\).
Almost surely, \(T_n\) decreases continuously from infinity to zero on
\(u>\max_i|D_i|/h\). The solutions in \eqref{eq:coupling-interval-roots}
are therefore unique and give
\(\Prob_\kappa\{\kappa\in[\kappa_-,\kappa_+]\}=1-\gamma\).
Multiplying all \(D_i\) by a positive constant multiplies the fit and both
endpoints by that constant.

To precompute bounds, suppose the two quantiles in
\eqref{eq:coupling-interval-roots} lie on opposite sides of \(n\). Then
\begin{equation}
I\subseteq I_n^*=
\left[\sqrt{\frac{n}{\chi^2_{n,1-\gamma/2}}}-1,
      \sqrt{\frac{n}{\chi^2_{n,\gamma/2}}}-1\right].
\label{eq:coupling-outer-interval}
\end{equation}
Indeed, convexity of \(\operatorname{atanh}\) on \([0,1)\) gives
\(T_n(u)\le(\widehat\kappa/u)^2n\) for \(u\ge\widehat\kappa\), and
the reversed inequality for \(u\le\widehat\kappa\).
Substitution at the two endpoints proves the inclusion.
All reported training sizes satisfy this quantile condition.

\paragraph{Tangent weights and the score polynomial.}
Subtract the fitted zero-noise increments from the path increments to
obtain \(R_{x,k},R_{y,k}\). The first residual has known variance \(q=5/8\).
In the tangent approximation, regressing the second residual on the first
gives coefficient \(\widehat s\) and remaining variance \(\widehat v\);
these need not be exact for the nonlinear residuals. Use \(\ell\ge0\) to
control the weight on the adjusted second residual in the score
\begin{equation}
S_\ell=\frac1K\sum_{k=0}^{K-1}\left\{
\frac{R_{x,k}^2}{q}+
\frac{\{R_{y,k}-\widehat s(X_k)R_{x,k}\}^2}
{\widehat v(X_k)+\widehat\kappa^2\ell}\right\}.
\label{eq:nonlinear-weighted-energy}
\end{equation}
Full tangent whitening uses \(\ell=0\); increasing \(\ell\) reduces the
weight on the adjusted second residual. The functions \(\widehat s\) and
\(\widehat v\) are fixed after fitting, but their values are recomputed at
the states along each path.

To assess a candidate weighting rule throughout \(I\), we keep the fitted
scoring rule and reference law fixed and vary the parameter generating the
target path. With identical Gaussian innovations at each parameter value,
changing \(\kappa\) leaves the \(X\) path unchanged. The resulting score is
quadratic in \(d=\kappa/\widehat\kappa-1\):
\begin{equation}
S_\ell(d)=S_\ell(0)+2dA_\ell+d^2B_\ell,
\qquad B_\ell\ge0.
\label{eq:coupling-score-quadratic}
\end{equation}

Set \(m(x)=\{\tanh x+\tanh(x/2)\}/2\) and
\(c(x)=1-\tanh^2(x/2)\). The residuals and tangent quantities in
\eqref{eq:nonlinear-weighted-energy} are
\[
\begin{aligned}
R_{x,k}&=X_{k+1}-X_k/4,&
R_{y,k}&=Y_{k+1}-Y_k-\widehat\kappa m(X_k),\\
q&=5/8,&
\widehat s(x)&=\widehat\kappa c(x)/5,\qquad
\widehat v(x)=\widehat\kappa^2c(x)^2/10.
\end{aligned}
\]
Put \(a=1-h=1/2\). A transition from \((x,y)\) obeys
\[
X_{mid}=ax+\sqrt h Z_1,\quad
X'=aX_{mid}+\sqrt h Z_2,\quad
Y'-y=\kappa J,\qquad J=h\{\tanh x+\tanh X_{mid}\}.
\]
The first innovation has tangent loading
\((a\sqrt h,\widehat\kappa h\sqrt h\,c(x))^\top\), and the second has
loading \((\sqrt h,0)^\top\). These loadings give \(q\), \(\widehat s\)
and \(\widehat v\) above. The regression coefficient \(\widehat s\) and
remaining variance \(\widehat v\) come from the tangent approximation;
the fitted zero-noise \(Y\) increment is \(\widehat\kappa m(x)\).

Write \(\bar s=\widehat s/\widehat\kappa\),
\(\bar v=\widehat v/\widehat\kappa^2\), and
\(e_k=J_k-m(X_k)-\bar s(X_k)R_{x,k}\). For
\(\kappa=\widehat\kappa(1+d)\), the adjusted second residual in the energy is
\(\widehat\kappa(e_k+dJ_k)\). Thus
\begin{equation}
A_\ell=\frac1K\sum_k\frac{e_kJ_k}{\bar v(X_k)+\ell},\qquad
B_\ell=\frac1K\sum_k\frac{J_k^2}{\bar v(X_k)+\ell},
\label{eq:coupling-polynomial-coefficients}
\end{equation}
This proves \eqref{eq:coupling-score-quadratic}. For weight selection, we can
simulate the polynomial's coefficients at coupling one because their joint
distribution does not depend on \(\widehat\kappa\).
Conditional on the other innovations, the score is a quadratic function of
the final \(X\) innovation with a positive quadratic coefficient. Its
distribution is therefore continuous.

\paragraph{Bounds over a continuous parameter interval.}
For prespecified \(\alpha,M\), put \(\alpha_M=k_\alpha/(M+1)\).
Define the largest CDF discrepancy and rejection error at this level over \(I\):
\begin{equation}
\begin{aligned}
D_\ell(I)&=\sup_{d\in I}\sup_t|F_{\ell,d}(t)-F_{\ell,0}(t)|,\\
R_\ell(I)&=\sup_{d\in I}|r_{\ell,d}-\alpha_M|,
\end{aligned}
\label{eq:interval-errors}
\end{equation}
Here \(F_{\ell,d}\) is the target score CDF at relative error \(d\),
conditional on the training information fixing the fitted law and score.
The test uses \(M\) fresh paths from the fitted reference law and has rejection
probability \(r_{\ell,d}\). At \(d=0\), both path laws are the fitted law.

For this model, we can calculate the score's minimum and maximum on each
cell of a finite cover of \(I\). We use independent paths for selection to construct CDF bands that,
conditional on training, hold jointly across parameter cells and candidate
weights. These bands yield upper bounds
\(\widehat B_{\mathrm{CDF}}(\ell)\) for \(D_\ell(I)\) and
\(\widehat B_\alpha(\ell)\) for \(R_\ell(I)\), as follows.

Fix a finite candidate set \(\mathcal L\). Cover \(I\) by finitely many closed
intervals \(I_c=[a_c,b_c]\), chosen using training information alone, or
use a deterministic partition of \(I_n^*\) and retain every cell intersecting
\(I\). For each of \(N\) independent paths for weight selection define
\[
L_{\ell,c}=\min_{d\in I_c}S_\ell(d),\qquad
U_{\ell,c}=\max_{d\in I_c}S_\ell(d).
\]
For \(B_\ell>0\), the minimum occurs at the projection of
\(-A_\ell/B_\ell\) onto \(I_c\), and the maximum at an endpoint.
If \(B_\ell=0\), also \(A_\ell=0\), so the score is constant in \(d\).
For every \(d\in I_c\),
\(F_{U_{\ell,c}}\le F_{\ell,d}\le F_{L_{\ell,c}}\).

Let \(N_F\) bound the total number of reference and envelope CDFs used,
across all candidates and all training sizes used for precomputation, and put
\begin{equation}
b_N=\sqrt{\frac{\log(2N_F/\eta)}{2N}}.
\label{eq:coupling-dkw}
\end{equation}
The Dvoretzky--Kiefer--Wolfowitz--Massart inequality
\citep{Massart1990,Reeve2024} and a union bound give simultaneous CDF errors
at most \(b_N\) with probability at least \(1-\eta\), conditional on training.
Within the sample used to estimate any one CDF, the paths are independent.
The random inputs generating that sample may be reused across candidates
and parameter cells.

Write \(\widehat F_{\ell,0}\), \(\widehat F_{L_{\ell,c}}\), and
\(\widehat F_{U_{\ell,c}}\) for the empirical CDFs. An upper bound for
\(D_\ell(I)\) on this event is
\begin{equation}
\widehat B_{\mathrm{CDF}}(\ell)=
\min\left\{1,\max_c\max_{V\in\{L_{\ell,c},U_{\ell,c}\}}
\|\widehat F_V-\widehat F_{\ell,0}\|_\infty+2b_N\right\}.
\label{eq:coupling-cdf-bound}
\end{equation}
Here and below the maximum is over the retained covering cells.

For the specified level, use the function \(g\) in
\eqref{eq:level-cdf-integral} and let \(\mathbb P_N\) denote the average over the selection simulations. Define
\begin{equation}
\begin{aligned}
\underline r_{\ell,c}
 &=\left[\mathbb P_Ng\{(\widehat F_{\ell,0}(L_{\ell,c})-b_N)_+\}
       -b_N\right]_+,\\
\overline r_{\ell,c}
 &=\min\left[1,\mathbb P_Ng\{\min(\widehat F_{\ell,0}(U_{\ell,c})+b_N,1)\}
       +b_N\right],\\
\widehat B_\alpha(\ell)
 &=\max_c\max\{\alpha_M-\underline r_{\ell,c},
                         \overline r_{\ell,c}-\alpha_M\}.
\end{aligned}
\label{eq:coupling-level-bound}
\end{equation}
\begin{samepage}
\begin{proposition}[Weight selection after fitting]
\label{prop:nonlinear-selection}
In \eqref{eq:nonlinear-coupling}, fix a finite ordered set of candidates
\(\ell\ge0\) and a tolerance \(\epsilon\). The training interval has
noncoverage probability at most \(\gamma\). Conditional on training, the
simultaneous simulation bounds above fail with probability at most \(\eta\).
Use selection innovations independent of training and evaluation. Select the smallest
candidate for which the chosen bound is at most \(\epsilon\).
With probability at least \(1-\gamma-\eta\) over training and selection simulations,
the CDF rule gives \(\Delta_{\mathcal A}\le\epsilon\); the level rule gives
\(|\Prob(p_M\le\alpha\mid\mathcal A)-\alpha_M|\le\epsilon\).
If no candidate qualifies, use only \(K^{-1}\sum_kR_{x,k}^2/q\).
Both procedures satisfy
\[
|\Prob(p_M\le\alpha)-\alpha_M|\le\epsilon+\gamma+\eta.
\]
\end{proposition}
\end{samepage}

\paragraph{Proof of Proposition~\ref{prop:nonlinear-selection}.}
The CDF ordering and the simultaneous bands prove
\(D_\ell(I)\le\widehat B_{\mathrm{CDF}}(\ell)\) for every candidate.
For the level bound, monotonicity of \(g\) first gives
\[
\E g\{F_{\ell,0}(L_{\ell,c})\}
\le r_{\ell,d}\le
\E g\{F_{\ell,0}(U_{\ell,c})\},\qquad d\in I_c.
\]
Replace \(F_{\ell,0}\) by its bands. A CDF error at most \(b_N\) bounds
\(|\mathbb P_Nf-\E f|\) by \(b_N\) simultaneously for all nondecreasing
\(f:\R\to[0,1]\), by integration against a positive Stieltjes measure of
mass at most one. The bound therefore applies to functions built from
\(\widehat F_{\ell,0}\), even when the selection simulations used to estimate
that CDF are also used to construct the envelope CDFs. Applied to \(L_{\ell,c}\)
and \(U_{\ell,c}\), it gives the two rejection bounds and
\(R_\ell(I)\le\widehat B_\alpha(\ell)\).

With probability at least \(1-\gamma-\eta\), the training interval contains
the generating parameter and all selection bounds hold. The bounds then
hold for whichever candidate is selected. The fallback statistic has
distribution \(\chi_K^2/K\) under both laws because its
\(R_{x,k}/\sqrt q\) are independent standard normals.
Corollary~\ref{cor:score-gap} controls rejection error for the CDF rule;
the level rule controls it directly. On the complementary event, rejection
error is at most one. Averaging gives \(\epsilon+\gamma+\eta\).

The rule retains the most weight on the adjusted second residual among
candidates that meet the chosen error bound. The fallback score omits that residual and has law \(\chi_K^2/K\) for
every \(\kappa\). A test based on this score cannot detect changes confined
to \(Y\).

\paragraph{Response to added noise in \(Y\).}
Add independent Brownian noise of loading \(\omega\) to the \(Y\) equation
only for the test path. Over an observation step, this noise contributes
an increment independent of the original innovations, with mean zero and
variance \(\omega^2\). Conditional on the training information, the expected score
\(S_\ell\) increases by
\[
\omega^2\E\left[\frac1K\sum_k
\frac1{\widehat v(X_k)+\widehat\kappa^2\ell}\right].
\]
This change in the expected score decreases with \(\ell\) and vanishes for the
first-coordinate fallback. The power calculations in
Section~\ref{app:nonlinear-computation} measure how the
added noise affects rank-based rejection.

\subsection{Selecting weights when the parameter changes the path}
\label{app:feedback-selection}
\label{sec:feedback-weights}

In the preceding model, varying \(\kappa\) with fixed innovations leaves
the \(X\) path unchanged. We now consider a model with feedback between
the coordinates, using four Euler substeps per unit observation interval:
\begin{equation}
\begin{aligned}
X^+&=0.75X+0.125\tanh Y+0.5Z,\\
Y^+&=0.875Y+0.25\kappa\tanh X,
\end{aligned}
\label{eq:feedback-step}
\end{equation}
with independent standard-normal \(Z\), zero initial state and
\(\kappa\in[0.2,3]\) the only unknown. Changing \(\kappa\) now changes both
coordinates and subsequent residuals. We obtain a confidence set \(C\)
by inverting tests that compare the observed training statistic with
simulated statistics using ranks. We then hold \(Q\) and each candidate
scoring rule fixed and propagate interval bounds through the solver.
These bounds enclose scores at every parameter value in \(C\).
We give the fitting rule, five candidate weights and bounds below.

Each path runs for \(K=20\) observations.
Write \(P_\kappa\) for its path law and \(Q=P_{\widehat\kappa}\) for the
reference law induced by the fitted implementation. The prespecified alternative adds
\(0.1\kappa Z_y\) to each \(Y\) substep only for the future test path,
independently of the \(X\) innovations. This corresponds to Brownian
loading \(0.2\kappa\) at substep size \(1/4\). Throughout this experiment,
the training data are generated under the unmodified model.

We select the first candidate whose rejection-error upper bound is at most
\(\epsilon\) and whose power lower bound is at least \(\beta\), throughout
\(C\). If no candidate qualifies, we do not reject. If \(C\) misses the generating
parameter with probability at most \(\gamma\), and the simultaneous bounds
fail with probability at most \(\eta\) conditional on training and weight
preparation, then
\begin{equation}
\Prob(\text{reject under the model})
\le\alpha_M+\epsilon+\gamma+\eta.
\label{eq:feedback-level}
\end{equation}
The rejection probability in this bound is averaged over training,
weight preparation and selection.
Under the future-path alternative, marginal power is at least
\mbox{\(\beta[\Prob(\text{a candidate is selected})-\gamma-\eta]_+\)}
(Proposition~\ref{prop:feedback-selection}). It therefore depends on the
probability of selecting a candidate.

\paragraph{Training by simulation.}
For \(n\) independent first observations, use
\(T(D)=n^{-1}\sum_iY_{i,1}^2\). At each \(\kappa\in[0.2,3]\), simulate
\(B\) independent training datasets and obtain
\(T_1(\kappa),\ldots,T_B(\kappa)\). Their Gaussian inputs are shared across
parameter values and independent of the observed training data. Retain
\begin{equation}
C^*=\{\kappa:T_{(j)}(\kappa)\le T(D)
                       \le T_{(B+1-j)}(\kappa)\}.
\label{eq:feedback-training-set}
\end{equation}
At the generating parameter, these \(B+1\) statistics are exchangeable
and have continuous distributions, so
\(\Prob_\kappa(\kappa\notin C^*)=2j/(B+1)\).
We take \(B=2047\), \(j=2\), hence \(\gamma=2/1024\).
We compute the estimate by solving
\(T_{(1024)}(\widehat\kappa)=T(D)\) with bracketed root finding. If \(T_{(1024)}(\kappa)-T(D)\) has the same sign at both endpoints,
choose the endpoint with the smaller absolute difference. The estimate
sets the reference law \(Q\) and enters the residual and weight
constructions below; coverage of \(C^*\) follows from the rank inversion,
not from this point-estimation rule.

To retain coverage between parameter grid points, we bound each
\(T_b(\kappa)\) over a whole interval. Increasing an input cannot decrease
an order statistic, so these bounds also bound the two order statistics
in \eqref{eq:feedback-training-set}. We remove an interval only when the
bounds show that every point fails the rank condition. Otherwise, we
retain or bisect it, retaining unresolved intervals of width at most
\(0.001\). The hull \(C\) of the retained intervals therefore contains
\(C^*\). If no intervals remain, we select no candidate and do not reject.

\paragraph{Scores and bounds over continuous parameter intervals.}
Let \(R_k=U_{k+1}-\Phi_{\widehat\kappa}(U_k)\), where
\(U_k=(X_k,Y_k)^\top\) and \(\Phi_{\widehat\kappa}\) applies four fitted
zero-noise updates. The score is \(S_W=K^{-1}\sum_kR_k^\top W R_k\).
Before selection, we simulate 1024 independent \(Q\) paths and average the
fitted tangent covariances over paths and observation steps to obtain
\(\bar C\).
The five candidates, in order, are
\[
W_{\rm raw}=q_0^{-1}I_2,\qquad
W_\ell=(\bar C+\widehat\kappa^2\ell e_y e_y^\top)^{-1},
\quad \ell\in\{0,0.01,0.03,0.1\},
\]
where \(q_0=\tfrac14\sum_{a=0}^3(3/4)^{2a}\) and \(e_y=(0,1)^\top\).
These matrices remain fixed on all target and reference paths. The bounds
below use the actual simulated scores and do not require \(\bar C\)
to equal an exact transition covariance.

For fixed Gaussian inputs, differentiation of \eqref{eq:feedback-step}
with respect to \(\kappa\) gives
\[
\begin{aligned}
\dot X^+&=0.75\dot X+0.125\operatorname{sech}^2(Y)\dot Y,\\
\dot Y^+&=0.875\dot Y+0.25\tanh X
                  +0.25\kappa\operatorname{sech}^2(X)\dot X.
\end{aligned}
\]
Under the departure, add \(0.1Z_y\) to the update for \(\dot Y\).
Starting from zero derivatives, propagate interval bounds for states and
derivatives on each parameter cell. Differentiate the fitted zero-noise
centre with respect to \(\kappa\) through the starting state supplied by each
path, while holding \(\widehat\kappa\) fixed.
Because the weight matrices are fixed and positive definite, this gives
\(\dot S_W=2K^{-1}\sum_k\dot R_k^\top W R_k\).
Induction through the finite recursions proves that the intervals contain
the states, derivatives and scores at every parameter in the cell.

On a cell with midpoint \(m_c\) and half-width \(d_c\), let
\([a_c,b_c]\) enclose the score derivative. The mean value theorem gives
\begin{equation}
S_W(\kappa)\in
[S_W(m_c)-d_c\max(|a_c|,|b_c|),\,
 S_W(m_c)+d_c\max(|a_c|,|b_c|)].
\label{eq:feedback-envelope}
\end{equation}
Intersect this interval with the direct interval evaluation of the score
and with \([0,\infty)\). Apply the same calculation to \(T_b\) to construct
the training confidence set. The code advances the stochastic states
separately and computes residuals by subtracting the fitted zero-noise
update equations from the stochastic update equations. This subtraction
preserves cancellation between their terms. The code uses double precision, outward rounding
for elementary interval operations and guards for transcendental evaluation
and final bounds. The inclusion proof is in real arithmetic; the numerical
checks are not a formal certification of all floating-point operations.

\paragraph{Paired rejection bounds.}
Condition on the observed and simulated training datasets and the
simulations used to prepare the weights. The confidence set \(C\), fitted
reference law \(Q\), and \(J=5\) candidate score functions are then fixed. Cover \(C\) by \(D\) cells before drawing selection simulations.
For candidate \(j\) and cell
\(c\), write \(L_{jc},U_{jc}\) for its null score bounds and
\(L^1_{jc}\) for its departure-score lower bound, all driven by the same
Gaussian input. Each of \(N\) independent selection repetitions also
draws a fresh group of \(M\) independent \(Q\) reference paths. Groups
may be shared across candidates and cells within a repetition.

Let \(T_j\) be the \((M-k_\alpha+1)\)th reference order statistic in
ascending order, where \(k_\alpha=\lfloor\alpha(M+1)\rfloor\).
For \(1\le k_\alpha\le M\), rejection occurs when the
test score exceeds \(T_j\). Generate a path from \(Q\) using the same Gaussian input as the score
bounds, independently of the reference group. Compute its score and let
\(I_0\) be the corresponding rejection indicator. Define \(I_L,I_U,I_1\)
analogously using \(L_{jc},U_{jc},L^1_{jc}\), respectively, with the same
reference group. The score computed from a path with law \(Q\) has a continuous distribution:
conditional on all other innovations, the energy is a strictly convex
quadratic function of the final \(X\) innovation. Thus \(\E I_0=\alpha_M\).
Let \(r_j(\kappa)\) and \(\pi_j(\kappa)\) denote the rejection probabilities
under the null and the specified alternative, respectively, conditional on
the same training and weight-preparation information. Monotonicity gives, for every
\(\kappa\) in cell \(c\),
\begin{equation}
\E(I_L-I_0)\le r_j(\kappa)-\alpha_M\le\E(I_U-I_0),
\qquad \pi_j(\kappa)\ge\alpha_M+\E(I_1-I_0).
\label{eq:feedback-paired-order}
\end{equation}

For each difference \(I-I_0\), count the two events \((I,I_0)=(1,0)\)
and \((0,1)\). Its expectation is their probability difference. Each
one-sided difference bound uses two Clopper--Pearson tails, each with
error probability \(\eta/(6J)\). Write
\(\underline d_{L,jc},\overline d_{U,jc},\underline d_{1,jc}\)
for the lower, upper and lower difference bounds. Set
\begin{equation}
\widehat B_j=\max\{0,\max_c\overline d_{U,jc},
                            -\min_c\underline d_{L,jc}\},\qquad
\widehat b_j=[\alpha_M+\min_c\underline d_{1,jc}]_+.
\label{eq:feedback-selection-bounds}
\end{equation}
The final bounds hold simultaneously for all candidates with conditional
probability at least \(1-\eta\). To see why, put
\(a_{jc}=\E(I_L-I_0)\) and fix a cell \(c_*\) minimising \(a_{jc}\).
This cell depends on training and weight preparation, but not on the
selection simulations.
Then
\[
\Prob\{\min_c\underline d_{L,jc}>\min_c a_{jc}\}
\le\Prob\{\underline d_{L,jc_*}>a_{jc_*}\}\le\frac{\eta}{3J}.
\]
The upper bound on a maximum and the power lower bound follow in the same
way. A union bound over three extrema and \(J\) candidates proves that
\(\widehat B_j\) bounds \(\sup_C|r_j-\alpha_M|\) and
\(\widehat b_j\) bounds \(\inf_C\pi_j\) from below.
Individual cell bounds are not claimed to hold jointly. Selection repetitions
are independent; within each repetition, candidates and cells share inputs.

\begin{samepage}
\begin{proposition}[Weight selection using score bounds throughout a confidence set]
\label{prop:feedback-selection}
Suppose the training confidence set covers the generating parameter with
probability at least \(1-\gamma\), and the envelopes bound the scores throughout
that set. Suppose also that the candidate bounds from independent selection
repetitions hold simultaneously with probability at least \(1-\eta\),
conditional on training and weight preparation. Select the first candidate satisfying
\(\widehat B_j\le\epsilon\) and \(\widehat b_j\ge\beta\); if none
qualifies, do not reject. With fresh final test and reference paths,
\eqref{eq:feedback-level} holds. If the training data are still generated under the unmodified model and
only the future test path is generated under the specified alternative, then
\begin{equation}
\Prob(\text{reject under the departure})
\ge\beta[\Prob(\text{a candidate is selected})-\gamma-\eta]_+.
\label{eq:feedback-power}
\end{equation}
\end{proposition}
\end{samepage}

\begin{proof}
With probability at least \(1-\gamma-\eta\), the confidence set contains
the generating parameter and all selection bounds hold. On this event,
any selected candidate has conditional rejection probability at most
\(\alpha_M+\epsilon\) under the null and at least \(\beta\) under the
specified alternative.
When no candidate is selected, the rejection probability is zero. Average
these bounds, using rejection probability at most one on the complementary
event. For the power bound, subtract at most \(\gamma+\eta\) from the
probability that a candidate is selected and use nonnegativity.
\end{proof}

The set obtained by inverting the training tests has the stated coverage, and
its outer approximation retains that coverage.
Equation~\eqref{eq:feedback-envelope} bounds the scores, while
\eqref{eq:feedback-paired-order}--\eqref{eq:feedback-selection-bounds}
give the simulation bounds. If training data also came from an alternative,
we would need a separate argument for coverage of the training confidence set.

\paragraph{Complete repetitions and reproduction.}
Before generating this batch, we specified 16 independent repetitions at
\(\kappa=1\), \(n=1024\). Each repetition regenerated the training data, the \(B=2047\)
simulated training datasets, the weight-preparation simulations and the
selection simulations, using \(N=2048\),
\(M=499\), \(D=16\), \(\alpha_M=0.024\), \(\eta=0.0025\),
\(\epsilon=0.02\) and \(\beta=0.12\).
Selection used \(1{,}021{,}952\) reference paths per fit.
We chose these settings using an earlier fit, before generating this batch.
Seeds and commands are in the experiment README. Each final test used 499
fresh references; null and departure tests were paired within a repetition.

\begin{figure}[!htb]
\centering
\includegraphics[width=\textwidth]{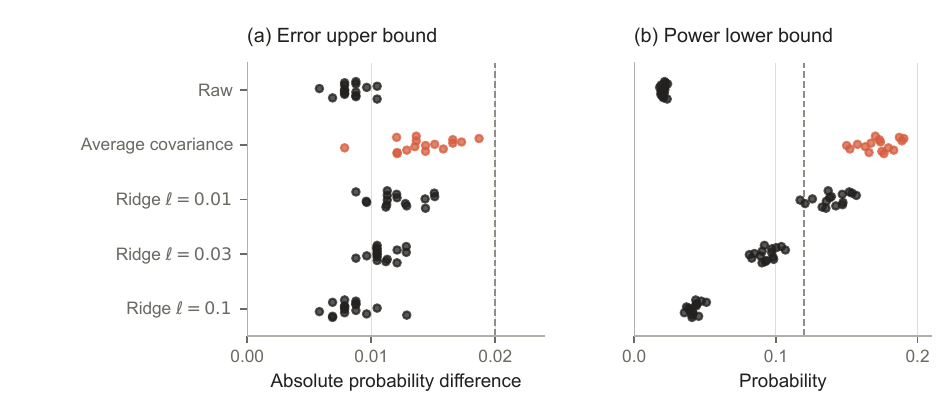}
\caption{Weight selection in the feedback model. Each point gives a bound
over the full training confidence set \(C\) in one of 16 independent repetitions.
Vertical offsets separate repetitions. Dashed lines mark the error
tolerance \(0.02\) and required power \(0.12\). Orange marks the candidate
selected in every repetition. Within each repetition, the bounds hold
simultaneously across the five candidates with at least 99.75\% conditional
simulation confidence. A candidate that does not qualify need not violate
either requirement.}
\label{fig:feedback-weight-selection}
\end{figure}

In all 16 repetitions, the selection rule chose the average-covariance weight
(Figure~\ref{fig:feedback-weight-selection}). Ridge \(0.01\) also qualified
in 15 repetitions, but appears later in the candidate order. The final tests
rejected in 1/16 repetitions under the model and 4/16 under the departure.
Separate exact 95\% binomial intervals for the corresponding rejection
probabilities are \([0.00158,0.30233]\) and \([0.07266,0.52378]\), rounded outwards.
The null frequency is imprecise. Proposition~\ref{prop:feedback-selection}
gives the upper bound on the overall rejection probability
\[
0.024+0.02+2/1024+0.0025=0.048453125<0.05
\]
which is 4.85\% when rounded to two decimal places as a percentage. An outward-rounded
four-decimal version of the 95\% interval under the alternative is
\([0.0726,0.5238]\). The power interval concerns the specified generating parameter and the
alternative applied only to the future test path. All repetitions
are included.

The directory \path{experiments/feedback_weight_selection/} contains the
source, retained per-fit settings, binomial counts and final ranks. Its
\path{README.md} gives commands and dependencies. The script
\path{four_step.py} runs one full repetition; \path{summarize.py}
recomputes the reported counts and intervals from \path{results/}.
Saved checks compare the computed score with that obtained from a direct
Euler simulation and check whether the intervals contain training statistics
and scores on finite parameter grids. These checks assess the implementation; the argument
above proves inclusion over continuous parameter intervals.

\subsection{Excluding candidates in a retrospective analysis}
\label{app:joint-feedback}

In the retrospective analysis below, simulation bounds show that none of
the five candidates meets both requirements throughout the set of
\((\kappa,\sigma)\) values retained in this analysis. Here the diffusion
loading is also unknown, and the fitted reference path law remains fixed during candidate comparisons.

Replace \(0.5Z\) in \eqref{eq:feedback-step} by \(0.5\sigma Z\), with
\(\theta=(\kappa,\sigma)\in[0.2,3]\times[0.5,1.5]\).
We retrospectively analyse one dataset of \(n=1024\) independent
four-observation paths generated at \((1,1)\).
For \(t=1,\ldots,4\) and \(D_{it}=Y_{it}-0.875^4Y_{i,t-1}\), the training summaries are
\[
T_X=\frac1{4n}\sum_{i,t}X_{it}^2,\qquad
T_Y(\kappa)=\frac1n\sum_{i,t}
\{D_{it}^2-0.6\kappa\tanh(X_{i,t-1})D_{it}\}.
\]
At each parameter value, we compute both summaries on each of 2,047
independent simulated datasets. Include the parameter in \(C_{\rm joint}\)
when both observed statistics have ascending ranks in \(\{2,\ldots,2047\}\).
If this rule is fixed independently of the observed training data,
exchangeability and a union bound give coverage at least \(1-4/2048\).
Here we use the realised set to exclude candidates retrospectively and do
not assert repeated-sampling coverage for it. The simulation bounds below
are conditional on that set and the prepared weights.

We estimate the parameters by matching the second moments of the first
\(X\) and \(Y\) observations. Let \(m\) contain the two observed second
moments, and let \(\bar m(\theta)\) contain their averages over the 2,047
simulated datasets. We minimise
\[
\sum_{a=1}^2\left\{\frac{\bar m_a(\theta)-m_a}
{\max(m_a,10^{-8})}\right\}^2
\]
over the stated rectangle, starting L-BFGS-B at the midpoint. The resulting
fit is \(\widehat\theta\approx(0.97734,0.99720)\); the function
\texttt{fit\_moments} in \path{experiments/feedback_weight_selection/joint_point_check.py}
specifies the numerical calculation. We prepare and fix the five weights
using 1,024 independent paths from the fitted reference law, as in
Section~\ref{app:feedback-selection}, replacing \(q_0\) by
\(\widehat\sigma^2q_0\) and the ridge term by
\(\widehat\kappa^2\widehat\sigma^2\ell\).
We retain \(M=499\), \(\alpha_M=0.024\), tolerance \(0.02\) and required
power \(0.12\). The departure adds independent \(Y\) noise with loading
\(0.2\kappa\), leaving the \(X\) loading \(\sigma\) unchanged. The error
tolerance requires null rejection probability to be at most \(0.044\).
Two parameters in \(C_{\rm joint}\) are
\[
\theta_+\approx(0.96508,1.03720),\qquad
\theta_-\approx(0.98960,0.95720),
\]
with training ranks \((2,2024)\) and \((2041,2024)\), respectively.

\begin{table}[htbp]
\centering
\small
\begin{tabular}{@{}lll@{}}
\toprule
Weight & Probability bound & Requirement \\
\midrule
Raw & \(\pi(\theta_-)\le0.01928\) & \(\pi\ge0.12\) \\
Average covariance & \(r(\theta_+)\ge0.04780\) & \(r\le0.044\) \\
Ridge \(0.01\) & \(r(\theta_+)\ge0.04736\) & \(r\le0.044\) \\
Ridge \(0.03\) & \(\pi(\theta_-)\le0.09979\) & \(\pi\ge0.12\) \\
Ridge \(0.1\) & \(\pi(\theta_-)\le0.04041\) & \(\pi\ge0.12\) \\
\bottomrule
\end{tabular}
\caption{Each candidate fails at a parameter in the realised set \(C_{\rm joint}\).
\(r\) is null rejection probability and \(\pi\) is power for the specified
alternative, using reference paths from the law induced by the fixed fitted implementation. Bounds are rounded outwards
and hold jointly with at least 99.5\% simulation confidence, conditional
on training and weight preparation.}
\label{tab:joint-feedback-exclusion}
\end{table}

The first independent check used 16,384 test/reference groups to bound
null rejection and power at both points for all five weights. A second
check used 32,768 groups to bound null rejection at \(\theta_+\) for the
average-covariance and ridge-0.01 weights. Each group used 499 fresh
reference paths. Each check's family of Clopper--Pearson bounds has failure
probability at most \(0.0025\). Together they give the joint guarantee in
Table~\ref{tab:joint-feedback-exclusion}.

For each of these five candidates, a violation at one parameter in the
realised set \(C_{\rm joint}\) suffices to rule it out; bounds over the entire
set are unnecessary. Settings, tail allocations
and seeds are in the README;
saved ranks and probability bounds are reproduced by
\path{experiments/feedback_weight_selection/joint_point_check.py}.

\section{Linear experiment and sensitivity calculation}
\label{app:linear-mechanism}

\subsection{Sensitivity with known coefficients}
\label{sec:linear-mechanism}

For independent Gaussian residuals with known covariances,
Proposition~\ref{prop:common-scale-efficiency} identifies weights that
maximise the standardised mean shift under common-scale covariance
departures. We now compare finite-sample rejection rates under
these departures and under changes to only one noise loading. We compare
\(\Sigma^{inst}=H\operatorname{diag}(\sigma_1^2,\sigma_2^2)\) with
\(\Sigma^{step}=\Sigma_{H,L}^{E}\) using the upper-tail energy score
\eqref{eq:linear-energy} and the common Euler zero-noise map.
We compute both scores on each reference and held-out draw, with no fitting.
For each weighting rule, we rank the held-out score against reference
scores computed using that rule.

The prespecified design uses
\(H=\sigma_1=\sigma_2=1\), \(K=40\),
\(\kappa\in\{0,0.5,1.5,3\}\), \(L\in\{1,2,5,10\}\), and six alternatives:
multiply either loading or both by \(1.10\) or \(1.25\), retaining the drift.
For each cell, we run 5,000 replicates, each with 499 independent reference
paths. We generate references under the Euler law at the unmodified parameter
values for that setting. Since \(\sigma_2>0\),
both covariance matrices are positive definite and we use their full inverses.
We give the exact simulation method and simultaneous interval construction below.

\begin{figure}[!htbp]
\centering
\includegraphics[width=\textwidth]{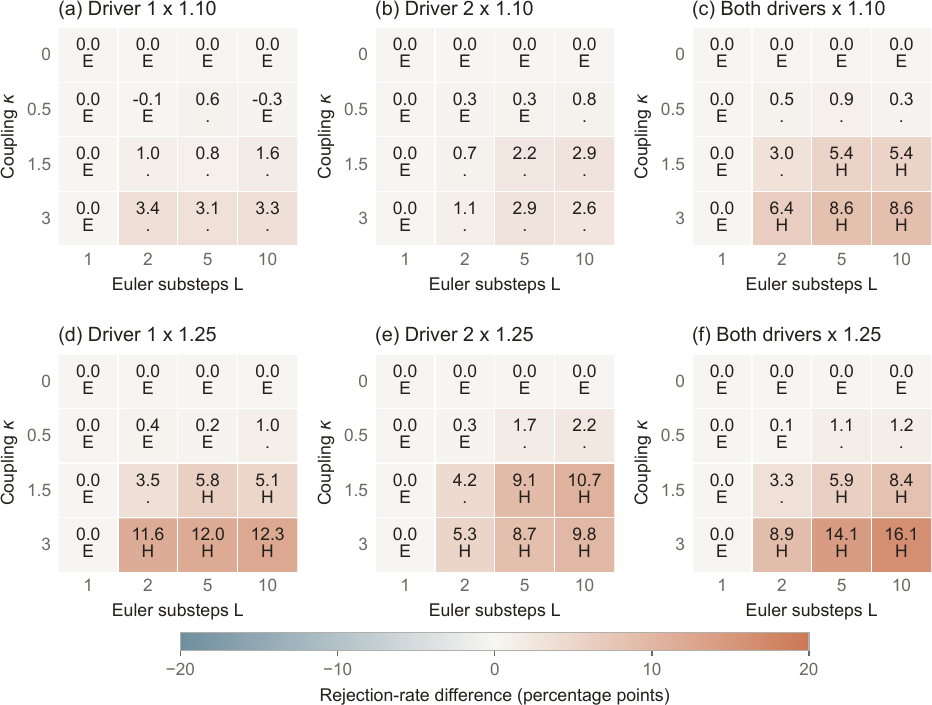}
\caption{All 96 prespecified comparisons of observation-step and instantaneous
standardisation. Each entry gives the rejection rate with observation-step
weights minus that with instantaneous weights, in percentage points. Columns give \(L\);
rows give \(\kappa\) (here \(\eta=|\kappa|\)). E denotes equivalence within \(\pm2\) percentage
points, H an interval wholly above +2 percentage points and a dot an unresolved comparison,
using 95\% simultaneous paired-difference intervals over all cells.
At \(\kappa=0\) or \(L=1\), Monte Carlo ranks are equal pathwise.}
\label{fig:linear-coordinate-sensitivity}
\end{figure}

Across the two standardisations, all 32 estimated null rejection rates lie
between 4.34\% and 5.50\%; their simultaneous binomial intervals include 5\%.

Of the 96 non-null comparisons in Figure~\ref{fig:linear-coordinate-sensitivity},
42 have identical scores and ranks pathwise because \(\kappa=0\) or \(L=1\).
Using the simultaneous intervals for paired differences, we classify these
and eight other comparisons as equivalent within two percentage points.
The test using observation-step weights has materially higher rejection
rates in 20 comparisons; 26 remain unresolved. None has a materially lower rate under
this classification.

At \(\kappa=3,L=10\), multiplying both loadings by 1.25 gives rejection
rates \(0.7170\) with instantaneous weights and \(0.8784\) with
observation-step weights, a paired difference \(0.1614\) with
simultaneous interval \([0.1345,0.1877]\). This finite-sample benefit of
reweighting occurs in the common-scale direction of
Proposition~\ref{prop:common-scale-efficiency}, with both tests using the
same Monte Carlo ranking procedure.

\paragraph{Exact simulation.}
We recorded all settings internally before simulation. Under the null, the
held-out and reference paths have the same law. For covariance \(CC^\top\), Bartlett's construction
with independent \(A\sim\chi^2_K\), \(B\sim N(0,1)\) and
\(D\sim\chi^2_{K-1}\) gives
\[
W=\begin{pmatrix}A&\sqrt A\,B\\\sqrt A\,B&B^2+D\end{pmatrix},
\qquad T_\Sigma=\frac{1}{2K}\operatorname{tr}(C^\top\Sigma^{-1}C\,W).
\]
This construction samples the exact joint distribution of the quadratic
scores of Gaussian residuals. In each of 5,000 outer replicates, we draw 499
independent reference scatter matrices and one independent held-out scatter
matrix. We use the same draws for both weighting rules and the same underlying
held-out scatter matrix for all alternatives. We compute each test's
p-value from the Monte Carlo rank of its score.

NumPy PCG64 streams use SeedSequence entropy \((2026082807,g,s)\), with
setting index \(g\), reference stream \(s=0\), held-out stream \(s=1\), and
blocks of 100 replicates. Float64 covariance factorisations use no
regularisation: a non-positive pivot stops the calculation, and condition numbers above
\(10^{10}\) are flagged. The power grid excludes \(\sigma_2=0\). At
\(\kappa=0\) or \(L=1\), scores agree to \(10^{-12}\) and ranks agree
pathwise. Settings are in
\path{docs/results/linear_geometry_design.json}; the 112-row
\path{summary.csv} and 560,000-row \path{trials.csv} are in
\path{cache/summaries/linear_geometry_mechanism/frozen/}.

\paragraph{Paired differences and uncertainty.}
For each non-null cell,
\(D=\Prob(p_{step}\le0.05)-\Prob(p_{inst}\le0.05)=p_{10}-p_{01}\), where
\(p_{10}\) is the probability that only the test using step weights rejects,
and \(p_{01}\) is the probability that only the test using instantaneous
weights rejects.
We use Clopper--Pearson limits with each of four tails at
\(0.05/(4C)\), \(C=96\), to construct simultaneous difference intervals
\([\ell_{10}-u_{01},u_{10}-\ell_{01}]\) with coverage at least 95\% by the
union bound. Pointwise intervals use \(C=1\). We treat the 32 null rejection
rates as a separate simultaneous family and do not pool distinct strengths.
Equivalence requires the full simultaneous interval to lie inside \([-0.02,0.02]\);
a materially higher or lower rate requires it to lie wholly above 0.02 or below \(-0.02\).
All other comparisons are unresolved.

\paragraph{Scope of scalar rescaling.}
Multiplying all residuals by the same positive constant,
\(z'_k=cz_k\), multiplies total energy by its square and leaves the energy
ranks unchanged. If we apply the same rescaling to the residuals on every target, pilot and reference
path, five pilot-standardised centring--energy components also remain
unchanged, provided the pilot scales are non-degenerate and the numerical
thresholds are inactive.
The transformed component based on \(T_E^{cum}\), the maximum absolute
excursion of \(C_E(n)\), need not remain unchanged, because
\[
\sum_{k=0}^{n-1}\{(cz_k)^2-1\}
=c^2\sum_{k=0}^{n-1}(z_k^2-1)+(c^2-1)n.
\]
The experiment therefore concerns the specified quadratic energy of the
full residual vector; it does not compare the SIR score \(S_{\mathrm{CE}}\), which takes a maximum over transformed features.

\subsection{Local efficiency of quadratic residual statistics}
\label{app:scale-optimality}

The next result includes the common-scale conclusion of
Proposition~\ref{prop:common-scale-efficiency} and also covers covariance
departures in other directions. For each \(K\), let
\(r_1,\ldots,r_K\in\R^d\) be independent, with
\(r_k\sim N(0,\Sigma_k)\), \(\Sigma_k\succ0\), under the null and
\(r_k\sim N(0,\Sigma_k+K^{-1/2}\Delta_k)\) under a local alternative, where
\(\Delta_k\) is symmetric and the alternative covariance is positive definite.

\begin{proposition}[Local efficiency of quadratic residual statistics]
\label{prop:quadratic-local-efficiency}
For fixed symmetric weights \(W_k\), not all zero, define
\begin{equation}
\begin{aligned}
Z_K(W)&=\frac{\sum_{k=1}^K
\{r_k^\top W_kr_k-\operatorname{tr}(W_k\Sigma_k)\}}
{[2\sum_{k=1}^K\operatorname{tr}\{(W_k\Sigma_k)^2\}]^{1/2}},\\
\Lambda_K(W,\Delta)&=
\frac{\sum_{k=1}^K\operatorname{tr}(W_k\Delta_k)}
{[2K\sum_{k=1}^K\operatorname{tr}\{(W_k\Sigma_k)^2\}]^{1/2}}.
\end{aligned}
\label{eq:quadratic-local-shift}
\end{equation}
Under the boundedness and no-dominant-term conditions in
\eqref{eq:quadratic-array-conditions}, if \(\Lambda_K(W,\Delta)\to\Lambda\),
then \(Z_K(W)\) converges to \(N(0,1)\) under the null and to
\(N(\Lambda,1)\) under the local alternative. Put
\[
A_k=\Sigma_k^{1/2}W_k\Sigma_k^{1/2},\qquad
B_k=\Sigma_k^{-1/2}\Delta_k\Sigma_k^{-1/2}.
\]
If \(\sum_k\lVert B_k\rVert_F^2>0\), the relative squared local efficacy is
\begin{equation}
\rho_K^2(W,\Delta)=
\frac{\{\sum_k\langle A_k,B_k\rangle_F\}^2}
{(\sum_k\lVert A_k\rVert_F^2)(\sum_k\lVert B_k\rVert_F^2)}
\leq1.
\label{eq:quadratic-angle-efficiency}
\end{equation}
Equality holds exactly when \(A_k=cB_k\) for every \(k\), for a common
nonzero \(c\), equivalently
\(W_k=c\Sigma_k^{-1}\Delta_k\Sigma_k^{-1}\). In particular, for a
common-scale departure \(\Delta_k=\tau\Sigma_k\), \(\tau>0\), observation-step
whitening \(W_k=\Sigma_k^{-1}\) is locally optimal.
\end{proposition}

The equality criterion concerns squared efficacy: for an upper-tail test
against a covariance increase, the weights must give a positive mean shift
(\(c>0\) in the common-scale case).
The squared cosine measures how well the transformed weights \(A_k\) align
with the transformed covariance changes \(B_k\). For common-scale departures, inverse-covariance weights make these two
arrays proportional. Under other covariance departures, maximising the
standardised mean shift may require different, possibly indefinite,
weights. Thus no energy score is optimal for every covariance departure.

\begin{proof}[Proof of Proposition~\ref{prop:quadratic-local-efficiency}]
Suppress triangular-array dependence on \(K\), and use the \(A_k,B_k\) in the
proposition.  Assume
\begin{equation}
\sup_{K,k}\lVert B_k\rVert_{op}<\infty,\qquad
0<\sum_{k=1}^K\lVert A_k\rVert_F^2<\infty,\qquad
\frac{\max_{k\leq K}\lVert A_k\rVert_{op}}
{(\sum_{k=1}^K\lVert A_k\rVert_F^2)^{1/2}}\longrightarrow0.
\label{eq:quadratic-array-conditions}
\end{equation}
The first condition ensures positive-definite local covariances for large
\(K\); the last prevents one quadratic term from dominating the sum.

Under the null, \(r_k=\Sigma_k^{1/2}z_k\) with independent
\(z_k\sim N(0,I_d)\), so the centred quadratic form is
\(z_k^\top A_kz_k-\operatorname{tr}(A_k)\), with variance
\(2\lVert A_k\rVert_F^2\).  Diagonalisation and
\eqref{eq:quadratic-array-conditions} give the Lindeberg condition for the
resulting weighted centred \(\chi_1^2\) variables.  Under the local alternative,
\(\Sigma_k^{-1/2}r_k\) has covariance \(I_d+K^{-1/2}B_k\); the corresponding quadratic form's mean
changes by \(K^{-1/2}\operatorname{tr}(A_kB_k)\), while the total variance is
\(2\sum_k\lVert A_k\rVert_F^2\{1+O(K^{-1/2})\}\).  The normal limit therefore holds under the local alternative with the
mean shift in \eqref{eq:quadratic-local-shift}.

Finally,
\[
\operatorname{tr}(W_k\Delta_k)=\langle A_k,B_k\rangle_F,
\qquad
\operatorname{tr}\{(W_k\Sigma_k)^2\}=\lVert A_k\rVert_F^2.
\]
Cauchy--Schwarz in the direct sum of the matrix spaces proves
\eqref{eq:quadratic-angle-efficiency}.  Equality requires
\(A_k=cB_k\) for a common nonzero \(c\), equivalently
\(W_k=c\Sigma_k^{-1}\Delta_k\Sigma_k^{-1}\).  When
\(\Delta_k=\tau\Sigma_k\) with \(\tau>0\), choosing \(c>0\) gives the
positive weighting \(W_k\propto\Sigma_k^{-1}\) for the upper-tail test.
\end{proof}

The common-scale conclusion also has an exact finite-sample form.  If \(c>1\),
\(r_k\) are independent \(N(0,\Sigma_k)\) under \(P_1\) and independent
\(N(0,c^2\Sigma_k)\) under \(P_c\), then
\begin{equation}
\log\frac{\mathrm dP_c}{\mathrm dP_1}
=-Kd\log c+\frac{1-c^{-2}}{2}
\sum_{k=1}^K r_k^\top\Sigma_k^{-1}r_k .
\label{eq:common-scale-likelihood-ratio}
\end{equation}
Thus the Gaussian likelihood ratio increases with observation-step energy
for this specified alternative.

\paragraph{Two-driver specialisation.}
Let \(\Sigma_0=\Sigma^{step}\succ0\), let \(\Sigma\succ0\) be a candidate
standardising covariance, and write \(\lambda_1,\lambda_2\) for the eigenvalues
of \(\Sigma^{-1}\Sigma_0\).  Under the Euler null,
\begin{equation}
2K T_\Sigma\ \overset{d}{=}\
\lambda_1\chi_K^2+\lambda_2\chi_K^2 .
\label{eq:weighted-energy-law}
\end{equation}
The two chi-square variables on the right are independent.
For local alternatives \(c_K^2=1+\tau/\sqrt K\) with \(\tau>0\), centre the energy
based on \(\Sigma\) by its null mean and divide by its null standard deviation.
The resulting statistic has limiting mean shift
\[
\tau\,
\frac{\lambda_1+\lambda_2}
{\sqrt{2(\lambda_1^2+\lambda_2^2)}}.
\]
The shift is maximised when \(\lambda_1=\lambda_2\), which determines
\(\Sigma\) up to a positive scale: \(\Sigma\propto\Sigma_0\). This also
follows from Proposition~\ref{prop:quadratic-local-efficiency}.  When
\(H=\sigma_1=\sigma_2=1\), the relative Pitman efficiency is
\begin{equation}
\operatorname{ARE}(\mathrm{step}:\mathrm{inst})
=1+\frac{\eta^4b_L^2+4\eta^2a_L^2}
{(2+\eta^2b_L)^2}.
\label{eq:step-inst-are}
\end{equation}
It equals one when \(\eta=0\) or \(L=1\), is at least one, and tends to two
as \(\eta\to\infty\) for \(L>1\).  This common-scale result is for the
quadratic energy of the full Gaussian residual vector. It does not compare
power under single-loading alternatives or establish an ordering for the
score \(S_{\mathrm{CE}}\) built from scalar residual features.

\section{Experiments on fitting, calibration and sensitivity}
\label{app:benchmark}

The first three experiments use different generators and scores to answer
different calibration questions. Section~\ref{sec:refit-size} examines rejection
rates averaged over complete fits; Section~\ref{app:ce-conditional} compares
score distributions conditional on each fit, prefix and pilot sample.
Section~\ref{app:coordinate-ablation} generates null target paths from the
fitted law to compare residual weights and reference calibration, without
introducing a discrepancy between the target and fitted path laws.

\subsection{Marginal calibration over repeated neural fits}
\label{sec:refit-size}
This experiment repeats fitting and network-width selection before each
held-out comparison. The scalar generator is
\[
Y_{t+1}=Y_t+b(Y_t,C_t)+0.42\varepsilon_t,
\qquad Y_0\sim N(0,0.7^2),\quad
\varepsilon_t\overset{iid}{\sim}N(0,1).
\]
The controls use the available history up to time $t$.
Let $a_t=\max(0,t-7)$ and $\Delta Y_j=Y_{j+1}-Y_j$.
For $t>0$, the latest increment is $\Delta Y_{t-1}$ and the increment
mean is $\overline\Delta_t=(Y_t-Y_{a_t})/(t-a_t)$; both are zero at $t=0$.
The level mean is
$\overline Y_t=(t-a_t+1)^{-1}\sum_{j=a_t}^tY_j$.
Together with the period-20 sine and cosine phases, these define the drift
\[
\begin{aligned}
b(y,C_t)={}&-0.22y+0.12\tanh(1.5y)+0.18\overline\Delta_t
+0.04\overline Y_t\\
&+0.06\sin(2\pi t/20)-0.025\cos(2\pi t/20)+0.02\Delta Y_{t-1}.
\end{aligned}
\]

Each replicate contains 32 training paths, eight selection paths and one
held-out path, each with 96 transitions. We fit a Gaussian transition model
with a one-hidden-layer tanh drift network and constant noise amplitude
$\sigma_\theta=\exp(\nu)+10^{-5}$. Inputs are standardised using the
training data. Each candidate receives 100 full-batch Adam updates with
learning rate 0.003 and an $L^2$ weight penalty of $10^{-4}$; width 8 or 16
is selected by the selection-set Gaussian negative log-likelihood.
The first seven transitions use the shorter available history defined
above and are excluded from fitting and scoring. The $n=89$ scoring
residuals are
\[
Z_t=\frac{Y_{t+1}-Y_t-\widehat b(Y_t,C_t)}{\widehat\sigma},
\qquad t=7,\ldots,95.
\]
Pilot and reference paths continue from the held-out path's first eight
states, using the fitted transition model and recomputing controls along
each path.

For each of 5,000 fits (seed 20260710), 500 independent pilot paths fix the
feature means and sample standard deviations; 999 independent reference
paths supply the final Monte Carlo rank. The six features are
\[
|\overline Z|,\quad |s_Z-1|,\quad |\widehat\rho_1(Z)|,\quad
|\widehat\rho_7(Z)|,\quad
\sup_z|\widehat F_Z(z)-\Phi(z)|,\quad
n(n+2)\sum_{h=1}^{10}\frac{\widehat\rho_h(Z)^2}{n-h}.
\]
Here $s_Z$ is the sample standard deviation with divisor $n-1$,
$\widehat F_Z$ is the empirical residual
CDF, $\Phi$ is the standard-normal CDF, and $\widehat\rho_h$ is the
correlation of the two overlapping lagged segments, each centred by its
own sample mean. The score is the untruncated maximum of the six features
after subtracting their pilot means and dividing by their pilot sample
standard deviations. It differs from the score \(S_{\mathrm{CE}}\) defined in Section~\ref{sec:score-construction}.
We use each fit's Monte Carlo p-value directly, without a second rank
calibration.

The rejection rates at nominal levels 0.010, 0.025, 0.050 and 0.100 are
0.0094, 0.0256, 0.0516 and 0.1000, respectively. At 5\%, the pointwise
95\% Wilson interval is $[0.0458,0.0581]$. These rates average over complete
fits and do not establish calibration conditional on a realised fit.
The separate test using the residual-scale component rejects in
$285/5000=0.0570$ of fits at 5\% (Wilson interval $[0.0509,0.0638]$), a small
excess in this descriptive comparison unadjusted across six components.
The full-refit calibration section of \path{REPRODUCE.md} gives the
simulation command and the command for rebuilding these summaries from
the retained results.

\subsection{Score distributions after two neural fitting procedures}
\label{app:ce-conditional}
\label{sec:ce-conditional}
We now compare score distributions conditional on each fit, prefix and
pilot sample. The two-dimensional generator has noise loading
$(0,0.42)^\top$, unit observation spacing and ten Euler substeps per
interval, with no state projection. Its drift is
\begin{align*}
b_{0,1}(x,c)&=-0.44\tanh(0.5x_1)+\tanh(0.5x_2)
 +0.18\tanh(c_3)\\
&\qquad+0.06\tanh(c_7)-0.025\tanh(c_8),\\
b_{0,2}(x,c)&=0.24\tanh(0.5x_1)-0.60\tanh(0.5x_2)
 +0.18\tanh(c_4)+0.04\tanh(c_6)\\
&\qquad+0.06\tanh(c_7)+0.02\tanh(c_2).
\end{align*}
At observation time $k\ge7$, the eight controls are, in order, the latest
increment in each coordinate, their seven-increment means, the eight-state
means, and $\sin(2\pi k/20),\cos(2\pi k/20)$. Thus the means in coordinate
$i$ are $(X_{i,k}-X_{i,k-7})/7$ and
$8^{-1}\sum_{j=k-7}^{k}X_{i,j}$. Controls are held fixed within each
interval and recomputed along each path. A width-16 tanh network with an
affine output and inputs $(x,c)$ standardised from training data can
represent this drift exactly. This representation does not ensure that
either surrogate fitting loss recovers the generating drift and diffusion.

We generate 12 independent groups of 192 training paths, each with 112
transitions and initial state $N(0,0.7^2I_2)$. The first seven transitions
use zero controls and are excluded from fitting; each smaller training set
contains the first 48 paths. Within each group and training size, the two
fits share the data, input standardisation, initialisation and simulation
streams. Both estimate the drift and noise amplitude
$\sigma_\theta=\exp(\nu)+10^{-5}$ with 3,000 full-batch Adam updates at
learning rate 0.003. The loss averages
\[
\frac{r_2^2}{2v_\theta}+\frac12\log v_\theta+\frac12\|r\|^2
\]
over training transitions and adds
$10^{-4}(\|W_1\|_F^2+\|W_2\|_F^2)$, where $W_1,W_2$ are the hidden
and output weight matrices. Here $r$ is the observed increment minus the
solver's zero-noise increment. The losses differ only through $v_\theta$:
plug-in variance $\sigma_\theta^2$ or tangent variance
$\max\{\Sigma^{\rm tan}_{22},10^{-10}\}$, with gradients through the
drift Jacobian, covariance recursion and variance floor. We initialise
hidden and output weights independently from zero-mean Gaussian
distributions with SDs 0.15 and 0.10, set biases to zero, and set
$\nu=\log(0.45)$. Thus this comparison changes the fitted parameters as
well as the resulting diagnostic; neither loss reproduces the SIR
training loss.

Each group supplies an independent eight-state prefix from the generating
law. For each fit, 2,500 fitted-law pilot continuations fix feature medians
and sample SDs. Both diagnostics divide second-coordinate residuals by
their tangent standard deviations and use the six features in
\eqref{eq:ce-statistics} with
$\mathcal L=\{1,7\}$, and the tail rules in
Section~\ref{sec:score-construction}. Residual centres, tangent variances
and pilot standardisers are computed separately for each fit.
Conditional on the fit, prefix and pilot, independent samples of
$N=8192$ continuations from each of the generating and fitted laws give
empirical score CDFs $\widehat F$ and $\widehat G$. Each continuation
spans 60 observation intervals. The oracle comparison uses the generating
parameters. We report
$\widehat\Delta=\sup_s|\widehat F(s)-\widehat G(s)|$ and estimate the
conditional rejection probability for $M=2500$ fresh references by
\[
\widehat r=\frac1N\sum_{i=1}^N h\{\widehat G(S_i^-)\},\qquad
h(u)=\Pr\{\operatorname{Bin}(2500,1-u)\le124\},\quad S_i\sim F.
\]
The left limit implements the upper-tail rank convention. This estimate
averages over a fresh reference sample; matching continuous score
distributions give rejection probability $125/2501$.

Simultaneous 95\% Dvoretzky--Kiefer--Wolfowitz--Massart envelopes
\citep{Reeve2024} cover both CDFs across 60 configurations: 12 oracle
comparisons and 24 fits per loss. The paired runs share oracle arrays,
which are counted once. The envelope radius is
$\epsilon=\{\log(4\cdot60/0.05)/(2N)\}^{1/2}$, giving gap bounds
$[\widehat\Delta-2\epsilon,\widehat\Delta+2\epsilon]\cap[0,1]$ with
untruncated half-width 0.0455. For rejection bounds, we replace
$\widehat G$ in the average by $(\widehat G-\epsilon)\vee0$ and
$(\widehat G+\epsilon)\wedge1$, then subtract or add $\epsilon$ and
truncate to $[0,1]$. Monotonicity of $h$ and the envelope for $F$ justify
these bounds. Subtracting opposite endpoints gives bounds on paired
differences (tangent minus plug-in). These bounds describe simulation
error with fits, prefixes and pilots fixed, rather than variability over
new training data.

\begin{table}[H]
\centering
\small
\begin{tabular}{@{}llrr@{}}
\toprule
Training paths & \shortstack{Variance used\\in loss} & Estimated gap: median (range) & \shortstack{Mean estimated\\rejection probability} \\
\midrule
Oracle & --- & 0.011 (0.009--0.019) & 0.0504 \\
48 & Plug-in & 0.433 (0.222--0.637) & 0.4454 \\
48 & Tangent & 0.104 (0.065--0.229) & 0.1330 \\
192 & Plug-in & 0.232 (0.160--0.369) & 0.2373 \\
192 & Tangent & 0.025 (0.011--0.050) & 0.0607 \\
\bottomrule
\end{tabular}
\caption{Diagnostics after paired refitting. Each row summarises 12
conditional comparisons with the fit, prefix and pilot fixed. The oracle's
population score gap is zero. The last column averages the 12 conditional
rejection estimates; the Monte Carlo target is $125/2501$.}
\label{tab:ce-conditional}
\end{table}

\begin{figure}[!ht]
\centering
\includegraphics[width=\textwidth]{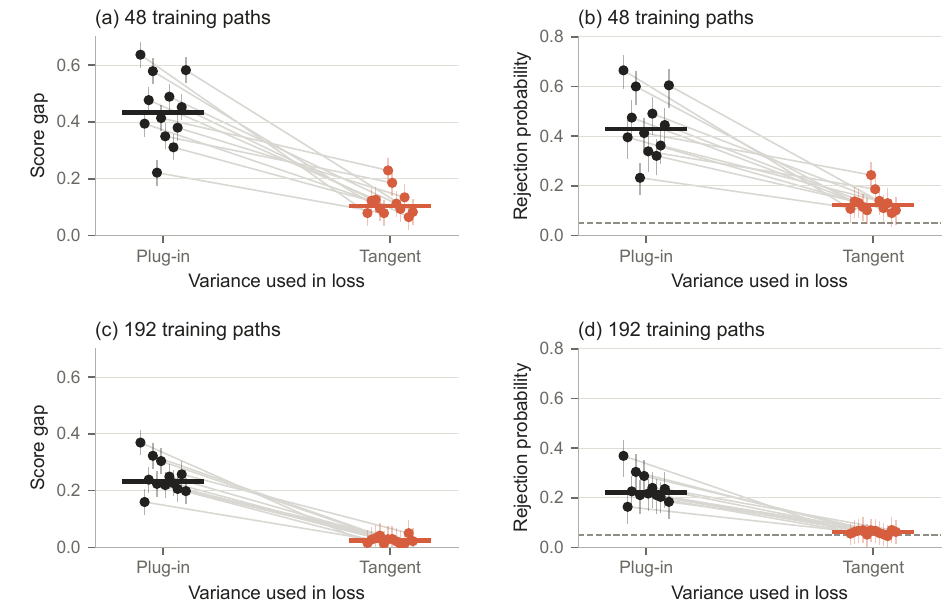}
\caption{Paired neural fits evaluated with the same residual features and aggregation rule.
Lines join results from fits to the same training data; thick marks show medians.
Vertical intervals give simultaneous 95\% Monte Carlo error bounds over
60 analysis configurations, including the 12 oracle comparisons in Table~\ref{tab:ce-conditional}. Dashed lines in panels (b) and (d) mark the Monte Carlo target
\(125/2501\).}
\label{fig:ce-training-variance}
\end{figure}

Tangent-variance fitting reduces the score gap in all 24 pairs, with all
paired intervals below zero (Figure~\ref{fig:ce-training-variance}). It
also reduces estimated conditional rejection in every pair; the
corresponding intervals are below zero in 12 of 12 pairs at 48 paths and
11 of 12 at 192 paths. Substantial mismatch remains at 48 paths: the
tangent-loss mean rejection estimate is 0.1330, and eight of its 12 lower
bounds exceed $125/2501$. At 192 paths, no such lower bound exceeds the
target, but this does not establish conditional calibration.
Table~\ref{tab:ce-conditional} summarises these conditional estimates.
The Monte Carlo bounds condition on the realised fits, prefixes and pilots
and quantify only simulation error. The master seed is 2026090907.
The retained summaries and commands for score-level reanalysis are
described in \path{output/neural_training_comparison/README.md};
\path{REPRODUCE.md} gives the fitting commands for both losses.

\subsection{Residual weights and reference calibration}
\label{app:coordinate-ablation}
\label{sec:coordinate-ablation}
Here the null held-out, pilot and reference continuations all come from
the fitted implementation. This removes target--reference path-law
mismatch and compares three ways to weight and calibrate the full
residual vector. Method A uses a pooled pilot covariance and Monte Carlo
ranks; B uses a transition-specific tangent covariance and a chi-square
tail; C uses B's score with Monte Carlo ranks.

The two-dimensional generator has loading $(0,0.42)^\top$, unit observation
spacing and 20 Euler substeps. It uses the history and phase controls
defined in Section~\ref{app:ce-conditional}, held fixed within each
interval and recomputed along each path. With seven-increment means
$\overline\Delta_{i,k}$, eight-state means $\overline x_{i,k}$ and
phases $s_k=\sin(2\pi k/20)$ and $c_k=\cos(2\pi k/20)$, its drift is
\[
\begin{aligned}
b_1(x,C_k)&=-0.22x_1+\kappa f(x_2)+0.18\overline\Delta_{1,k}
+0.06s_k-0.025c_k,\\
b_2(x,C_k)&=-0.30x_2+0.12\tanh(1.5x_1)+0.18\overline\Delta_{2,k}
+0.04\overline x_{2,k}+0.06s_k.
\end{aligned}
\]
Weak and mid coupling use $f(x)=x$ with $\kappa=0.5,2$; nonlinear
coupling uses $\kappa=2$ and $f(x)=\tanh(0.6x)$.
Paths start from $N(0,0.7^2I_2)$. We fit a width-eight tanh drift network
with an affine output and a constant second-coordinate noise amplitude
$\sigma_\theta=\exp(\nu)+10^{-5}$ to 48 paths of 112 transitions,
excluding the first seven transitions with zero controls. Inputs are
standardised using the training data. Fitting uses 3,000 full-batch Adam
updates at learning rate 0.003 and an $L^2$ weight penalty of $10^{-4}$,
with no architecture selection. The loss adds the second-coordinate
Gaussian negative log-likelihood, using plug-in variance, to the mean
squared drift residual across both coordinates with unit weight.

An independent generating-law path supplies a seven-transition prefix.
Given that prefix and the fit, we simulate 100 pilot and 200 independent
reference continuations of $K=105$ transitions. All methods subtract the
solver's zero-noise increment to obtain $r_k$ and use
\[
E_j=\frac{1}{2K}\sum_{k=0}^{K-1}\|L_{j,k}^{-1}r_k\|^2,
\qquad E_B=E_C.
\]
For A, we compute the centred covariance from all $100K$ pilot residuals,
add $10^{-10}I_2$ to that covariance, and then compute its Cholesky factor
$L_{A,k}$. Scoring does not subtract the pilot mean. For B and C, the
tangent factor is recomputed
from each path's current state and controls. All squared Cholesky pivots
are floored at $10^{-10}$. A and C use plus-one ranks against their own
200 reference scores; B uses
$p_B=\Pr\{\chi^2_{210}\ge210E_C^{\rm held}\}$. A's pilot residuals already
include propagation over the observation interval, so A versus C changes
standardisation, pilot estimation and regularisation together.

Departures retain the fitted drift and prefix. Targeted departures add an
independent noise channel through loading
$\operatorname{diag}(\delta,\widehat\sigma)$; uniform departures use
$(0,\lambda\widehat\sigma)^\top$. The two $(\delta,\lambda)$ strengths are
$(0.020,1.06)$ and $(0.030,1.08)$ for weak coupling,
$(0.120,1.05)$ and $(0.160,1.10)$ for mid coupling, and
$(0.050,1.05)$ and $(0.060,1.10)$ for nonlinear coupling.
Each strength uses 500 refits, with master seeds 20260828--20260833 in
the order just listed. Within a replicate, A and C share the fit,
held-out path and references. Null rates pool the 1,000 refits per
setting; sensitivity differences average the paired differences at the
two strengths.

\begin{table}[H]
\centering
\small
\setlength{\tabcolsep}{5pt}
\begin{tabular}{@{}lrrrcc@{}}
\toprule
& \multicolumn{3}{c}{Null rejection (\%)}
& \multicolumn{2}{c}{C minus A (percentage points)} \\
\cmidrule(lr){2-4}\cmidrule(l){5-6}
Setting & A & B & C & Targeted & Uniform \\
\midrule
Weak linear & 4.9 & 7.6 & 4.5 & $4.0\ [1.9,6.1]$ & $3.5\ [1.5,5.5]$ \\
Mid linear & 4.0 & 3.5 & 3.9 & $0.0\ [-1.8,1.8]$ & $1.3\ [-0.1,2.8]$ \\
Nonlinear & 4.7 & 5.5 & 4.5 & $3.6\ [1.3,5.9]$ & $1.9\ [-0.2,4.1]$ \\
\bottomrule
\end{tabular}
\caption{Fitted-law calibration and sensitivity at nominal 5\%. Brackets
give pointwise 95\% intervals from 20,000 paired bootstrap resamples
stratified by departure strength (seed 2026082803). The tie-free rank
target for A and C is $10/201$.}
\label{tab:coordinate-ablation}
\end{table}

Under weak coupling, B rejects 7.6\% of null paths (Wilson 95\% interval
6.11--9.41\%). Its adjusted $p=0.0033$ comes from two-sided exact
binomial tests with Holm adjustment over the nine setting-by-method null
comparisons, a family chosen retrospectively. Their targets are 0.05 for
B and $10/201$ for A and C. Neither Monte Carlo method differs detectably
from its target in that family. The weak-coupling tangent energy has
pooled mean 1.012 and SD 0.164, compared with 1 and 0.0976 for
$\chi^2_{210}/210$: a mean near one does not suffice for chi-square
calibration. C's sensitivity gains over A are modest and depend on the
departure; the average gain is zero for the mid targeted departure
(Table~\ref{tab:coordinate-ablation}).

The strengths were chosen after 32-fit exploratory probes whose seeds
were not retained, so independence from the final training streams cannot
be verified. The nonlinear setting changes both coupling shape and slope
at zero (from 2 to 1.2), and targeted noise can affect the second-coordinate
path through drift and feedback. These full-vector, upper-tail energy
comparisons do not directly assess the scalar SIR score or its negative
cumulative-energy excursions. Saved trial CSVs in
\path{cache/summaries/coordinate_ablation/} support the summaries;
\path{REPRODUCE.md} gives the simulation and summary commands.
Reference banks, per-step covariances and fitted models were not retained.

\subsection{Nonlinear selection and interval verification}
\label{app:nonlinear-computation}

We compare the two error requirements in
Section~\ref{sec:nonlinear-weight-comparison} using 64 fits
of \eqref{eq:nonlinear-coupling} at each \(n\in\{64,1024,16384\}\), with
\(K=20\), \(M=499\), \(\alpha=0.05\), \(\epsilon=0.02\) and
\(\gamma=\eta=0.0025\). We generate the training data and independent held-out
null paths under the unmodified model, and reference paths under the fitted
implementation law. Table~\ref{tab:nonlinear-weights}
compares whitening, the two selection rules and fixed weights \(W=I_2/q\).

\paragraph{Paired design.}
The candidates are
\(\mathcal L=\{0,0.001,0.003,0.01,0.03,0.1,0.3,1,3,10\}\).
We partition each \(I_n^*\) into eight equal cells and use 262,144 independent
paths for weight selection. Two envelope CDFs per cell and one reference
CDF for each candidate and training size give \(N_F=(2\times8+1)\times10\times3=510\) CDFs in
\eqref{eq:coupling-dkw}. For each fit, the rules retain the cells that intersect
its interval \(I\). With probability at least 99.5\% over training and selection simulations,
the target and reference distributions of the selected score satisfy
the conditional CDF bound under the CDF rule; under the level rule, the
test using the selected score satisfies the conditional bound on
rejection-probability error. The marginal rejection error bound is 0.025.

\begin{table}[!ht]
\centering
\small
\begin{tabular}{@{}rlrrr@{}}
\toprule
\(n\) & Weights & \shortstack{Max. gap\\across fits} & \shortstack{Null rejection (\%)\\with fitted references} & \shortstack{Power (\%)\\against added \(Y\) noise} \\
\midrule
64 & Whitening & 0.2206 & 4.99--9.95 & 31.03 \\
   & CDF rule & 0.0036 & 5.00--5.04 & 5.34 \\
   & Level rule & 0.0224 & 4.97--5.36 & 6.44 \\
   & Fixed & 0.0007 & 5.00--5.01 & 5.24 \\
\addlinespace
1024 & Whitening & 0.0128 & 4.99--5.06 & 30.10 \\
     & CDF rule & 0.0048 & 4.97--5.04 & 9.08 \\
     & Level rule & 0.0128 & 4.99--5.06 & 30.10 \\
     & Fixed & 0.0003 & 5.00--5.00 & 5.24 \\
\addlinespace
16384 & Whitening / both rules & 0.0049 & 4.99--5.01 & 30.10 \\
      & Fixed & 0.0002 & 5.00--5.00 & 5.24 \\
\bottomrule
\end{tabular}
\caption{Paired weight comparison at \(\kappa=0.1\). We report the maximum
estimated gap and the range of estimated conditional rejection probabilities
across 64 fits, evaluated at the generating parameter. These are simulation
estimates, not confidence bounds. Power is the mean estimated detection rate
after adding independent Brownian noise with loading \(0.2\kappa\) to \(Y\).
We calibrate this power comparison at the same target level using a separate
null sample from the generating model.}
\label{tab:nonlinear-weights}
\end{table}

For each evaluation law, we simulate 32,768 paths. Within each fit, all
weighting methods use the same fitted parameters and random inputs. We draw five
independent sets of innovations to generate reference paths from the fitted
law, null paths from the generating law, a second sample of null paths from
the generating law for power calibration, paths with a 15\% increase in the
\(X\) loading, and paths with added independent Brownian noise in \(Y\), with loading \(0.2\kappa\).
The fixed-weight score is \(K^{-1}\sum_k(R_{x,k}^2+R_{y,k}^2)/q\).
Scaling the training draws between \(\kappa=0.1,1\) leaves the whitening and
selected-score results unchanged by equivariance. The fixed-weight score
distributions still depend on \(\kappa\).

The five sets of innovations are independent across purposes. To estimate
the score gap, we use each innovation sequence in the null sample to generate
a paired path under the generating law and under the fitted law, then
compare the empirical CDFs of their scores. For rejection calculations, we
estimate the fitted-reference CDF from the independent reference bank.
We estimate rejection probability by
\(\mathbb P_Ng\{\widehat F_{\ell,0}(S_{\ell,d})\}\), integrating out the
\(M\) reference draws. For the control-variate correction, we evaluate the
matched score at \(d=0\) using the same null innovations and the same
independent fitted-reference CDF. We subtract that rejection estimate and add
\(\alpha_M\). We save the estimates before and after this correction. We also
save empirical score gaps computed from independent samples under the
generating and fitted path laws. To estimate power, we use the empirical CDF
\(\widehat F_{\rm cal}\) from the second null sample generated under the
true model and take the average
\(N^{-1}\sum_i g\{\widehat F_{\rm cal}(S_i^{alt}-)\}\), with \(g\) from
\eqref{eq:level-cdf-integral}. The power calculations for increased \(X\)
loading and added \(Y\) noise both integrate out
the rank randomness from \(M\) fresh reference scores, but calibrate against true-model null
scores. This compares tests based on the selected scores at a common level;
the final procedure in Proposition~\ref{prop:nonlinear-selection} instead
uses fitted-reference scores. These are simulation estimates over 64 fits
at each training size. We reuse the five sets of evaluation
innovations across fits, training sizes, coupling values and weighting
methods, evaluating each score with its corresponding fitted parameter and
weights.

\paragraph{Distribution matching and a specified level.}
The two selection rules can choose different weights while the fitted
reference path law remains fixed.
At \(n=1024\), the CDF rule selects \(\ell=0.3\)
for every fit, while the level rule retains whitening. We independently check
the first fit's interval \(I\approx[-0.0129486,0.0296339]\) and obtain,
with at least \(99.75\%\) simulation confidence,
\begin{equation}
0.02127\le D_0(I)\le0.02853,\qquad R_0(I)\le0.00453.
\label{eq:nonlinear-separation}
\end{equation}

On the simultaneous confidence event for these simulation bounds, the
scoring rule using full whitening has a supremum CDF discrepancy
\(D_0(I)>0.02\), so it violates the CDF tolerance on this interval.
The corresponding test nevertheless satisfies the specified-level
rejection-error requirement, since \(R_0(I)\le0.00453<0.02\). The lower bound on \(D_0(I)\) is obtained at the
upper endpoint of \(I\). It is not evaluated at the generating parameter
used in Table~\ref{tab:nonlinear-weights}.

\begin{samepage}
For full whitening at the upper endpoint of this interval, the empirical
CDF difference is largest near the 36th percentile of the reference score
distribution. The integration measure in \eqref{eq:level-cdf-integral}
instead concentrates near the 95th percentile of the reference-score
distribution.
The largest empirical CDF difference therefore occurs away from the region
that receives most weight in the rejection calculation.\par
\end{samepage}

\begin{figure}[!ht]
\centering
\includegraphics[width=\textwidth]{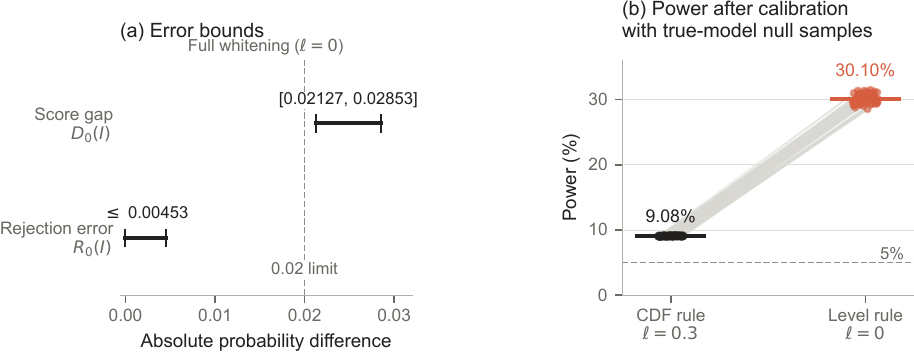}
\caption{Two calibration requirements and a separate sensitivity comparison,
\(n=1024\).
(a) Segments give lower and upper bounds on the worst CDF and rejection-probability errors for the scoring rule using full
whitening over one confidence interval obtained from training. With that
interval held fixed, the bounds have at least 99.75\% simultaneous simulation
confidence. The dashed line marks the tolerance 0.02.
(b) Paired power comparisons across 64 fits against added independent
\(Y\)-noise of loading \(0.2\kappa\), using the original selection bounds.
We calibrate each test at the nominal 5\% level using an independent null
sample from the generating law. Thin lines join the two selection rules'
results for each fit; thick marks give means.}
\label{fig:nonlinear-weights}
\end{figure}

The precision of the bounds also affects weight selection. On the interval in panel (a),
tighter bounds admit \(\ell=0.1\) rather than \(\ell=0.3\); the original
bounds therefore lead to more downweighting than the CDF tolerance requires
on this interval.
Panel (b) reports sensitivity under the original bounds; it does not quantify
the sensitivity change from using \(\ell=0.1\) on this interval.

At \(n=64\), the CDF rule selects \(\ell=10\) for 61 fits and \(3\) for
three; the level rule selects \(1\) and \(0.3\), respectively. Tests based on the scores selected by either rule have lower power
against added \(Y\)-noise than the test using full whitening (Table~\ref{tab:nonlinear-weights}). At
\(n=16384\), both retain whitening. Of the 192
confidence intervals constructed from training data, one excludes the true parameter; all fits are retained. At \(n=64\),
27 empirical score gaps under full whitening exceed 0.02, but only one
estimated deviation in rejection probability does. After reducing the second-coordinate weight, the tests have much greater
estimated power against increased \(X\) loading than against added \(Y\) noise
(Table~\ref{tab:nonlinear-strong-power}). Power against added \(Y\) noise with fixed weights at \(\kappa=1\) is 7.26\%, 7.27\%
and 7.27\% for the three training sizes.

\begin{table}[htbp]
\centering
\small
\begin{tabular}{@{}rrrrrr@{}}
\toprule
& \multicolumn{5}{c}{Power for increased \(X\) loading (\%)} \\
\cmidrule(l){2-6}
\(n\) & Whitening & CDF rule & Level rule & Fixed, \(\kappa=0.1\) & Fixed, \(\kappa=1\) \\
\midrule
64 & 28.20 & 25.17 & 25.86 & 25.10 & 26.14 \\
1024 & 27.98 & 27.16 & 27.98 & 25.10 & 26.15 \\
16384 & 27.97 & 27.97 & 27.97 & 25.10 & 26.15 \\
\bottomrule
\end{tabular}
\caption{Mean estimated detection rates over 64 fits after increasing the \(X\) loading
by 15\%, with the true-model null calibration used in
Table~\ref{tab:nonlinear-weights}. Changing \(X\) also changes the nonlinear
drift input to \(Y\). The whitening, CDF-rule and level-rule results apply to both coupling values.}
\label{tab:nonlinear-strong-power}
\end{table}

\paragraph{Independent verification on one training confidence interval.}
For the first \(n=1024\) fit in this experiment,
\(\widehat\kappa/\kappa\approx1.0027266\) and the relative error at the generating parameter
is approximately \(-0.00271916\). We verify the bounds using nested samples
of \(N=262144,2097152\) generated with seed 2026092141, independently of
the exploratory simulations generated with seed 2026092140. We check both
endpoints of the unchanged \(I\) for
\(\ell\in\{0,0.001,0.003,0.01,0.03,0.1,0.3\}\).
We compare covers formed from the original four intersecting cells, those
cells truncated to \(I\), and 32 equal cells within \(I\). The total CDF count is
\(N_F=2\times7\{3+2(4+4+32)\}=1162\).
Thus \eqref{eq:coupling-dkw}, with \(\eta=0.0025\), gives simultaneous
coverage for all bounds, including the displayed endpoint choice.

Figure~\ref{fig:nonlinear-weights}a displays the simultaneous bounds in
\eqref{eq:nonlinear-separation}.
For an endpoint \(d\in I\), a lower bound for \(D_\ell(I)\) is
\[
\left[\|\widehat F_{\ell,d}-\widehat F_{\ell,0}\|_\infty-2b_N\right]_+.
\]
At \(N=2097152\), \(b_N=0.0018101021\). For \(\ell=0\), the upper
endpoint's empirical gap is 0.0248937607, giving the lower bound in
\eqref{eq:nonlinear-separation}. The 32-cell calculation gives an upper
bound 0.02852923 for \(D_0(I)\) and bounds the rejection probability within
\([0.04626,0.05453]\). The generating relative error, approximately \(-0.00271916\), lies in
\(I\). The score-CDF gap upper bound and rejection-probability interval
therefore also apply at the generating parameter, with the same simultaneous
simulation confidence of at least \(99.75\%\).

The 32-cell upper bound at \(\ell=0.1\) is 0.01804; the lower and upper
bounds for \(D_\ell(I)\) at \(\ell=0.01,0.03\) straddle 0.02.
\path{experiments/nonlinear_weight_selection/separation.md} records the full
comparison of partitions and simulation precision. This comparison does not
identify the smallest ridge parameter that satisfies the CDF tolerance.

\paragraph{Location of the CDF difference.}
For \(M=499\) and \(\alpha=0.05\), the random reference percentile
\(U\) in \eqref{eq:level-beta-integral} has distribution
\(\operatorname{Beta}(475,25)\), with
central 95\% interval \([0.92928,0.96732]\). At the upper endpoint of the fixed
interval \(I\), the empirical CDF difference for \(\ell=0\) is largest at a
score of approximately 1.52267.
The reference CDF at that score is bounded by \([0.35794,0.36157]\).

\paragraph{Files and reproduction.}
The scripts and saved results are in
\path{experiments/nonlinear_weight_selection/}.
Its \path{README.md} and \path{separation.md} give run commands, dependencies
and file descriptions; saved settings retain seeds and full-precision values.
The scripts \path{coupling.py} and \path{verify_separation.py} reproduce the
comparison and validation. The rules were developed during this study;
the notes retain the interval verification and precision comparisons.
We construct Figure~\ref{fig:nonlinear-weights} from the saved
\mbox{\path{point_bounds.csv}}, \path{bound_refinement.csv} and
\path{coupling/results.csv}, using
\path{experiments/make_manuscript_figures.py} with \texttt{--additions-only}.

\section{Repeated fitting in a controlled SIR model}
\label{app:sir-refit}

This experiment supports the discussion in Section~\ref{sec:linear-fitted-stability}. It uses a constant-rate
SIR implementation with exact discrete-transition moments. It does not
refit the neural network or reconstruct states from case reports.

\subsection{Transition, covariance and training information}

The state is the susceptible fraction \(S\) and infectious log-fraction
\(Z=\log I\). Take \(\beta_0=0.2\), \(\gamma=0.1\), \(\sigma=0.05\),
\(S_0=0.99\), \(I_0=0.001\), an observation step \(H=1\) day, and two
Euler substeps of length \(h=1/2\). Conditional on the current state,
\begin{align*}
S_1&=S-h\beta SI,&
Z_1&=Z+h(\beta S-\gamma-\sigma^2/2)+\sigma\sqrt h\,\varepsilon_1,\\
S_2&=S_1-h\beta S_1e^{Z_1},&
Z_2&=Z_1+h(\beta S_1-\gamma-\sigma^2/2)+\sigma\sqrt h\,\varepsilon_2,
\end{align*}
with independent standard normal inputs. These Euler transitions define
the generating implementation; they are not exact transitions of the
continuous-time SIR diffusion.

Let \(\omega^2=\sigma^2h\) and
\(A=h\beta S_1I\exp\{h(\beta S-\gamma-\sigma^2/2)\}\).
The zero-noise endpoints are
\(f_S=S_1-A\) and
\(f_Z=Z+h\beta(S+S_1)-H\gamma-H\sigma^2/2\).
Writing \(r_S=S_2-f_S\), \(r_Z=Z_2-f_Z\), the exact moments give
\begin{equation}
\begin{aligned}
\operatorname{Var}(Z_2)&=2\omega^2=\sigma^2H,\\
c=\frac{\operatorname{Cov}(S_2,Z_2)}{\operatorname{Var}(Z_2)}
&=-\tfrac12 A e^{\omega^2/2},\\
v=\operatorname{Var}(S_2-cZ_2)
&=A^2e^{\omega^2}\{e^{\omega^2}-1-\omega^2/2\}.
\end{aligned}
\label{eq:sir-refit-moments}
\end{equation}
The transition is not jointly Gaussian: \(v\) is a linear residual
variance, not a general conditional variance given \(Z_2\).
The rank calculation does not assume Gaussian whitened residuals.

Only \(\beta\) is estimated. The first training design observes \(n\)
independent one-day infectious log-increments from the common initial state.
Their distribution is exactly Gaussian with variance \(\sigma^2H\) and mean
\[
\mu(\beta)=H\beta S_0-h^2\beta^2S_0I_0-H\gamma-H\sigma^2/2.
\]
We invert the sample mean under this expression to obtain the maximum
likelihood estimate, constrained to \([0.001,1]\). No retained fit reaches
a boundary. We use \(n=64\) and \(n=16{,}384\), with known-parameter
comparisons as a separate case.

The joint-training comparison adds the susceptible endpoint from each of
the same 64 transitions. If the original infectious innovation is \(u\),
draw an independent standard normal \(w\) and set
\(\varepsilon_1=(u+w)/\sqrt2\), \(\varepsilon_2=(u-w)/\sqrt2\).
This preserves every infectious training observation. For a candidate
\(\beta\), put \(q=S_1(\beta)-S_2>0\). The two innovations are recovered by
\[
\varepsilon_1=\frac{\log\{q/A(\beta)\}}{\omega},\qquad
\varepsilon_2=\frac{Z_2-Z-\mu(\beta)}{\omega}-\varepsilon_1.
\]
The absolute inverse Jacobian is \(1/(\omega^2q)\). We minimise
\(\sum_i\{(\varepsilon_{1i}^2+\varepsilon_{2i}^2)/2+\log q_i\}\)
over the same parameter range and admissible support. Thus joint training
uses the exact non-Gaussian density rather than a Gaussian approximation.

\subsection{Weights, targets and reference generation}

The fitted parameter determines the reference transitions, zero-noise
centres and covariance. For each fresh \(K=20\)-day path we compare:
\begin{align*}
T_{fixed}&=\frac1{2K}\sum_k
\frac{r_{Z,k}^2+(r_{S,k}/I_0)^2}{\sigma^2H},\\
T_{white}&=\frac1{2K}\sum_k
\left\{\frac{r_{Z,k}^2}{\sigma^2H}+
\frac{(r_{S,k}-c_k r_{Z,k})^2}{v_k}\right\},\\
T_{reduced}&=\frac1{2K}\sum_k
\left\{\frac{r_{Z,k}^2}{\sigma^2H}+
\frac{(r_{S,k}-c_k r_{Z,k})^2}{v_k+4K\widehat V_\beta d_k^2}\right\},\\
T_I&=\frac1{2K}\sum_k\frac{r_{Z,k}^2}{\sigma^2H},
\end{align*}
where \(\widehat V_\beta=\sigma^2H/\{n\mu'(\widehat\beta)^2\}\)
and \(d_k=\partial_\beta f_S-c_k\partial_\beta f_Z\), holding the current
state and \(c_k\) fixed in the derivative. The factor four was fixed before
the pilot run. This rule estimates the effect of shared training error;
it is not the confidence-set selection procedure proved elsewhere in
this supplement and has no asserted 5\% error guarantee. When \(\beta\)
is known, \(\widehat V_\beta=0\), so reduced weighting equals whitening.

Training always follows the null model. Future alternatives either
multiply \(\sigma\) by 1.2, with its It\^o drift correction, or add
independent mean-zero Gaussian noise to \(S\) after each complete day,
with standard deviation \(\sqrt{v_k}\) under the generating null model at
that day's starting state. The second alternative doubles the current
step's linear residual variance and then propagates the changed state.
It is a specified discrete-transition departure, not an identified
epidemic mechanism.

Each of 10,000 independent repetitions regenerates training data, the fit,
target and 499 references. We use the plus-one upper-tail rank with threshold
0.05. Methods share random inputs within a repetition. The main comparison
generates references from the fitted parameter. A separate comparison uses
499 independent references from the true null while keeping the same fitted
scoring rules. Conditional on training, this second construction compares scores
at a common Monte Carlo level, but requires the true parameter and is not
the fitted-reference procedure.

\begin{table}[!htbp]
\centering\small
\begin{tabularx}{\textwidth}{@{}Ylrrr@{}}
\toprule
Training & Weight & Null & Volatility & Added noise \\
\midrule
Known & Fixed & 4.83 & 35.94 & 5.65 \\
Known & Whitening & 5.08 & 52.43 & 57.82 \\
Known & Infectious only & 4.84 & 35.51 & 4.84 \\
64 infectious & Fixed & 16.97 & 49.98 & 18.52 \\
64 infectious & Whitening & 72.64 & 89.45 & 90.27 \\
64 infectious & Reduced & 5.71 & 36.48 & 5.80 \\
64 infectious & Infectious only & 5.32 & 35.61 & 5.32 \\
16,384 infectious & Fixed & 5.22 & 35.14 & 6.36 \\
16,384 infectious & Whitening & 5.78 & 53.44 & 59.13 \\
16,384 infectious & Reduced & 5.64 & 45.29 & 24.41 \\
16,384 infectious & Infectious only & 5.30 & 34.31 & 5.30 \\
64 paired states & Whitening & 5.20 & 52.56 & 58.18 \\
64 paired states & Infectious only & 4.73 & 34.36 & 4.73 \\
\bottomrule
\end{tabularx}
\caption{Rejection percentages with fitted references. Each cell uses
10,000 complete repetitions and 499 references. The volatility multiplier
is 1.2; the last column uses independent noise added to the susceptible
coordinate as specified above.
Known-parameter reduced weighting duplicates whitening and is omitted.}
\label{tab:sir-refit-all}
\end{table}

With 64 infectious-only observations, reduced weighting lowers null rejection
from 72.64\% to 5.71\%; the latter's pointwise 95\% Wilson interval is
5.27--6.18\%. Fixed two-coordinate weighting still rejects 16.97\% of null
targets. With 16,384 observations, whitening rejects 5.78\%, with interval
5.34--6.25\%. None of these finite-sample comparisons establishes exact
fitted-reference calibration.

With true-null references and 64 infectious training observations,
volatility detection is 29.32\% under whitening and 34.36\% under reduced
weighting; detection of noise added to the susceptible coordinate is
35.01\% and 4.59\%. With 16,384 training observations, the corresponding
rates for this alternative are 57.06\% and 23.44\%.
After joint training on 64 transitions, whitening detects volatility
increases at 51.05\% and noise added to the susceptible coordinate at
57.11\%, compared with 34.72\% and 4.56\%, respectively, for the infectious
coordinate alone. These comparisons separate
the sensitivity cost from differences in fitted-reference false rejection.

The infection-rate root mean squared error falls from 0.006374 to 0.000319
with joint training. To define the reported centring error, evaluate the
zero-noise endpoints \(f_S(\beta),f_Z(\beta)\) and the covariance quantities
\(c(\beta),v(\beta)\) above at the common initial state
\((S_0,\log I_0)\). For fit \(r\), write
\(\widehat c_r=c(\widehat\beta_r)\) and
\(\widehat v_r=v(\widehat\beta_r)\). The standardised error in the fitted
zero-noise centre for the low-variance residual component is
\[
e_r=
\frac{f_S(\beta_0)-f_S(\widehat\beta_r)
-\widehat c_r\{f_Z(\beta_0)-f_Z(\widehat\beta_r)\}}
{\sqrt{\widehat v_r}}.
\]
The numerator compares zero-noise endpoints at the true and fitted
parameters, using the fitted linear coefficient. It vanishes when
\(\widehat\beta_r=\beta_0\). For each training design, the reported
summary is
\[
\left\{\frac1R\sum_{r=1}^{R}e_r^2\right\}^{1/2},
\qquad R=10{,}000,
\]
where the average is over repeated fits at the fixed initial state. This
root mean square is 2.498 with infectious-only training and 0.1245 with
joint training. The variance of this residual component remains small;
the centring error relative to its standard deviation is what changes.
These state-specific summaries are not upper bounds over all future histories.

\subsection{Computation and retained files}

The formal seed is 2026100702. A preliminary run used seed 2026100701,
128 repetitions and 199 references. Joint training preserves the formal
random inputs, adding an independent training-noise stream; its pilot seed
is 2026100703. The infectious-only scoring comparison was added by
rescoring shared random inputs, reproducing the original three-weight
rejection counts. No fits or paths were discarded.

Direct one-day simulations at three states, with 200,000 repetitions each,
give residual variance ratios 0.9952--0.9987 relative to
\eqref{eq:sir-refit-moments}. The whitened energy agrees with direct
covariance inversion to about \(10^{-15}\); analytic centre derivatives
agree with finite differences. Joint-density inverse innovations,
Jacobian, score and curvature were also checked numerically.

Scripts \path{experiments/sir_refit_scale/run.py} and
\path{experiments/sir_refit_scale/joint_training.py} generate the two
training designs. Their \path{results/} and \path{joint_training/}
directories contain settings, rejection counts, pointwise intervals,
paired differences and saved replicate arrays. The independent repetition,
including training and reference generation, is the uncertainty unit.
Intervals are not adjusted over the family of comparisons.

\section[Neural SIR implementation and earlier state diagnostics]{Neural SIR implementation\\and earlier state diagnostics}
\label{app:sir-provenance}

This section records the neural fit, reconstruction and earlier daily-state
comparisons. Their reference paths use the shared fitted parameters and
each country's reconstructed history, which fixes the initial state and
conditioning inputs. These references are simulated directly in state
coordinates, without the reporting model in Section~\ref{app:report-observation}.
Table~\ref{tab:sir-analysis-status} summarises the selection history.
The selection scripts and retained settings are listed in
\path{REPRODUCE.md} in the code directory.
Figures~\ref{fig:sir-country-patterns}--\ref{fig:sir-mda-dependence}
show the country and window comparisons.

\subsection{State reconstruction}
\label{app:sir-reconstruction}
We take country, date, smoothed daily incidence and population from the
archived Our World in Data snapshot \citep{OWIDCOVID}. We replace missing
smoothed incidence with zero and clip negative values to zero.
The retained daily series for the nine training countries contain 337 such
missing values (DEU 90, FRA 134, ITA 3 and ESP 110). Filling them with zero
does not establish that no infections occurred. For CZE and GRC, no smoothed incidence values are missing within the
evaluation windows or the histories of 20 states used to initialise the
reference simulations. For nonnegative preprocessed incidence
\(x_t\) and population \(N\), reconstruction starts with the first recorded
day on or after 15 March 2020, denoted \(t_0\), and \(A_{t_0-1}=0\).
Before clipping, the 28-day reconstruction is
\begin{equation}
\widetilde C_t=\widetilde C_{t-1}+x_t,\qquad
A_t=x_t+\rho A_{t-1},\qquad \rho=\exp(-1/28),
\label{eq:sir-reconstruction}
\end{equation}
To initialise the cumulative series, let \(c_t=\sum_{j=t_0}^t x_j\) and
let \(a\) be the first date with positive \texttt{total\_cases}. We set
\(\widetilde C_t=c_t+\texttt{total\_cases}_a-c_a\), then floor this series
at zero. If no positive anchor exists, preprocessing uses the first
record. The protocol for CZE and GRC requires a positive anchor no later than the
window start. We take the first available population value
on or after \(t_0\). Writing \(\operatorname{clip}(u;a,b)=\min\{b,\max(a,u)\}\),
we construct the unnormalised state proxies as
\begin{equation}
\begin{aligned}
S_t^{obs}&=\operatorname{clip}\{(N-\widetilde C_t)/N;10^{-6},1-10^{-6}\},\\
I_t^{obs}&=\operatorname{clip}\{A_t/N;10^{-8},1-10^{-8}\},\qquad
Z_t^{obs}=\log I_t^{obs}.
\end{aligned}
\label{eq:sir-observation-map}
\end{equation}
We therefore bound \(I\) away from zero before taking its logarithm.
For training, we retain the contiguous block from the first to the last
day on which the untruncated
\(A_t/N>10^{-5}\); intervening days remain in the block. For CZE and GRC, we retain the full series and use the calendar dates
chosen before scoring to locate each evaluation window.
We normalise the states using constants estimated from training data to obtain
\(Y_t^{obs}\). We simulate reference paths conditional on the recorded
reconstructed history, without repeating reporting or reconstruction.
At each simulated substep, the projection imposes bounds
\(S\in[10^{-6},1-10^{-6}]\) and \(Z\in[-20,0]\), distinct from the
preprocessing bounds. These operations are specified in \path{sir_data.py}
and \path{experiments/run_sir_sde_native_country_audit.py}.

We fit the baseline rates by minimising
\[
\sum_k\{(\Delta S_k+\beta S_kI_k)^2+
        (\Delta I_k-(\beta S_k-\gamma)I_k)^2\}
\]
over \(\beta\in[0.01,2]\), \(\gamma\in[0.005,1]\), using pooled consecutive
daily transitions from the nine training countries before the log transform.
The exponential-decay memory used in reconstruction also constrains the
fitted removal rate. With
\(d_T=1-\exp(-1/T)\), the reconstruction before clipping satisfies
\(\Delta I_k=-\Delta S_k-d_T I_k\). No state clipping is active on the
retained training blocks for this snapshot, so the identity also holds there. Put
\(r_{S,k}=\Delta S_k+\widehat\beta S_kI_k\). At an interior solution of the
joint least-squares fit, the normal equation gives
\begin{equation}
\widehat\gamma-d_T=\frac{\sum_k I_k r_{S,k}}{\sum_k I_k^2}.
\label{eq:gamma-reconstruction-normal}
\end{equation}
Thus \(\widehat\gamma\) differs from \(d_T\) by a term determined by the
susceptible-equation residuals.
The baseline \(\widehat\gamma_0=0.0359\) is close to \(d_{28}=0.03508\).
Its reciprocal should therefore not be interpreted as an independently
estimated biological recovery duration.

\subsection{Fitted implementation}
The nonlinear-least-squares baselines are \(\widehat\beta_0=0.0520\) and
\(\widehat\gamma_0=0.0359\). The width-32 tanh network has two outputs;
its first determines the infection rate, while its second does not affect
the fixed removal rate in this fit.
We fit the shared model using 12,661 overlapping 60-state training sequences
from DEU, FRA, ITA, ESP, NLD, BEL, AUT, CHE, and GBR. Preprocessing starts
on 15 March 2020 as above. The last retained days are 7 October 2023 (DEU),
12 November 2023 (FRA), 27 October 2023 (ESP), 4 March 2024 (AUT),
7 August 2024 (ITA), and 4 August 2024 (NLD, BEL, CHE and GBR).
We slide each country's windows by one day, after the initial 19 days
needed for controls. No window crosses a country boundary.

Before constructing overlapping windows, we pool the retained daily
\((S,\log I)\) states, counting each country-day once.
We compute coordinatewise means and sample standard deviations, store them
as float32 and floor standard deviations at \(10^{-8}\). The stored means are approximately \((0.74119,-5.99820)\), and the stored
standard deviations are approximately \((0.20623,1.91037)\). Data preprocessing divides by the stored standard
deviation plus \(10^{-8}\); the simulator uses the stored diagonal scale.
The fitted checkpoint retains these constants and the learned volatility:
\path{models/country_cv/corrected_20260714_fixednorm_matchedsolver/neural_sde_sir_valFIN_baseline_recov28_frzgamma.eqx}.
We construct the predictable control from a 20-state history. At observation-grid time \(t_k\), let
\begin{equation}
\mathcal H_k=(Y_{k-19},\ldots,Y_k),
\qquad
C_k=\operatorname{Sig}_{\le3}
\left\{\operatorname{LL}
\bigl(\operatorname{TimeAug}(\mathcal H_k)\bigr)\right\}.
\label{eq:supp-32}
\end{equation}
Here \(\operatorname{TimeAug}\) appends a channel running from zero to one,
and \(\operatorname{LL}\) replaces consecutive augmented points \(u_j,u_{j+1}\)
by \((u_j,u_j),(u_{j+1},u_j),(u_{j+1},u_{j+1})\), sharing repeated
endpoints. We take the iterated-integral signature of the resulting
piecewise-linear path through levels 1--3, excluding level zero, and flatten
each level in the order used by \texttt{JAXSignatureExtractor} in \path{trajectory.py}.
The resulting
signature has \(6+6^2+6^3=258\) components. We compute controls using only
the current country's history up to the current observation and hold them
fixed over the ten substeps. They enter the width-32 tanh network that determines the infection rate.
The diffusion coefficient is one learned constant scalar.

\paragraph{Training loss.}
We evaluate the loss on normalised one-day transitions. For each current
state--control pair \((Y_{c,k},C_{c,k})\), with \(c\) indexing a training
sequence and \(k\) its within-sequence position, set \(h=0.1\),
\(\widehat Y_{c,k}^{(0)}=Y_{c,k}\), and \(V_{c,k}^{(0)}=0\).  The training code computes the zero-noise path and a diagonal variance
approximation through ten substeps:
\begin{equation}
\begin{aligned}
\widehat Y_{c,k}^{(\ell+1)}
&=\mathsf P\!\left\{\widehat Y_{c,k}^{(\ell)}
+h\,b_\theta\bigl(\widehat Y_{c,k}^{(\ell)},C_{c,k}\bigr)\right\},\\
V_{c,k}^{(\ell+1)}
&=V_{c,k}^{(\ell)}
+h\,g_\theta\bigl(\widehat Y_{c,k}^{(\ell)},C_{c,k}\bigr)
\odot
g_\theta\bigl(\widehat Y_{c,k}^{(\ell)},C_{c,k}\bigr),
\qquad \ell=0,\ldots,9,
\end{aligned}
\label{eq:supp-33}
\end{equation}
where \(\mathsf P\) clips in unnormalised state coordinates and then renormalises, and
\(\odot\) denotes componentwise multiplication.  Write
\(\widehat\mu_{c,k}=\widehat Y_{c,k}^{(10)}\) and
\[
\widehat s_{Z,c,k}
=\max\!\left\{\sqrt{V_{Z,c,k}^{(10)}},10^{-4}\right\}.
\]
For \(\rho_1(r)=r^2/2\) when \(|r|\le1\) and
\(\rho_1(r)=|r|-1/2\) otherwise, the minibatch loss is
\begin{equation}
\begin{aligned}
\mathcal L_{\mathrm{train}}(\theta)
={}&10^4\,\overline{
\rho_1\!\left(\widehat\mu_{S,c,k}-Y_{S,c,k+1}\right)}\\
&+\overline{\left[
\frac12\left(
\frac{Y_{Z,c,k+1}-\widehat\mu_{Z,c,k}}
{\widehat s_{Z,c,k}}
\right)^2
+\log\widehat s_{Z,c,k}
+\frac12\log(2\pi)
\right]},
\end{aligned}
\label{eq:supp-34}
\end{equation}
where each bar averages uniformly over batch elements and the 59 within-window
transitions. We apply a unit-threshold Huber loss to prediction errors in
the normalised \(S\) coordinate and a Gaussian loss to prediction errors in
the normalised \(\log I\) coordinate, using the zero-noise prediction and
approximate variance.
The training variance recursion in \eqref{eq:supp-33} adds substep diffusion
variances without propagating them through the drift. The post-fit tangent
recursion in \eqref{eq:tangent-recursion} includes that propagation.

We train for 60 epochs, each with 24 full batches of
512. We drop incomplete batches after shuffling, giving 1,440 updates
in total. AdamW uses weight
decay \(10^{-4}\), gradient clipping at one, and cosine decay from \(0.005\)
to \(0.0005\) over 1,500 updates; the last applied learning rate is
approximately \(0.00051834\). We recorded validation diagnostics but did not
use them to choose the number of updates, training duration or checkpoint.

\subsection{Windows chosen before scoring}
\label{app:fixed-calendar-holdout}
On 3 September 2026, before calculating the CZE diagnostic from the archived
data, we recorded a 20-state history ending on 1 January 2021 and 60 daily
increments ending on 2 March 2021. The analysis used the common
\texttt{recov28\_frzgamma} fit, 28-day reconstruction, tangent-scaled
normalised \(\log I\) residuals and \(S_{\mathrm{CE}}\), with 2,500 pilot
and 2,500 independent reference paths.

Eligible countries were European, had at least one million residents
according to information available by the window start, and had not been
used for development or diagnostic inspection. All 16 countries in the
pool passed checks on dates, source values and pre-window cumulative
anchors. The draw did not use evaluation incidence magnitudes or
diagnostic results. CZE and GRC had appeared in candidate pools but had
not been drawn or scored. States at each date used incidence only up to
that date; the protocol allowed no replacement for a non-evaluable target.

After seeing the CZE result, we fixed the GRC protocol before drawing it
from the remaining 15 countries. The calendar window, preprocessing,
fitted parameters, score and reference-rank rules were unchanged. We do
not adjust individual ranks across this sequence of analyses.
The dated settings and draws are retained under
\path{cache/summaries/fixed_calendar_holdout/}; the evaluation scripts are
\path{experiments/run_fixed_calendar_holdout.py} and
\path{experiments/run_second_fixed_calendar_holdout.py}.

\subsection{Score choice and analysis history}
\label{app:sir-supporting-results}
Independent pilot paths fix feature medians and sample standard deviations.
The score uses two-sided tails for the signed mean, energy level and two
energy correlations, and upper tails for the two maxima of absolute
cumulative sums. These rules define \(S_{\mathrm{CE}}\) and its centring
and energy groups.

Inspection of Nordic windows informed the score and prompted corrections
to normalisation and alignment of training and simulation substeps. We
retrained on the same preprocessed sequences from the nine training
countries and used the corrected fit for all reported diagnostics. The
feature definitions, tail rules and aggregation rule were fixed before
CZE and GRC were evaluated.

\begin{table}[htbp]
\centering
\small
\begin{tabularx}{\textwidth}{@{}>{\raggedright\arraybackslash}p{0.22\textwidth}Y>{\raggedright\arraybackslash}p{0.22\textwidth}@{}}
\toprule
Analysis & Use of outcomes & Status \\
\midrule
Nordic development & FIN, NOR and SWE windows selected using path outcomes
informed code corrections and the choice of \(S_{\mathrm{CE}}\) & Exploratory development \\
CZE / GRC & Country and window recorded before each diagnostic;
GRC's protocol followed the CZE result & Historical data; choices
recorded before scoring \\
Comparisons in Section~\ref{app:selection-audit} & Incidence determined
eligibility and window starts; additional windows followed the initial
results & Descriptive \\
\bottomrule
\end{tabularx}
\caption{Selection history of the earlier daily-state diagnostics.
The report-model comparisons reuse these windows exploratorily.}
\label{tab:sir-analysis-status}
\end{table}

\subsection{Country selection and window sensitivity}
\label{app:selection-audit}
Here a \emph{reference proportion} is the normalised plus-one upper-tail
rank in \eqref{eq:rank} for the global score. The first draw selected BGR,
HUN and SVK from 22 eligible countries; simulation counts were increased
after inspecting the initial results. Scoring and reporting rules were
fixed before the second draw selected LTU, MDA and SVN, without substitution.
Countries had to be in Europe, have at least one million residents, and
have no previous use in development or diagnostic inspection. The aligned active-period
series starts at its twentieth state, retaining the preceding 19 for controls.
For \(n\) aligned states, the window began at zero-based index
\(\max\{\lfloor n/2\rfloor,20\}\) and required 60 complete increments with
nonzero incidence. All countries passed the completeness check. Eligibility
and window starts thus used the full incidence series, including records
after the conditioning history, and were not repeated in reference simulations.

\begin{figure}[!ht]
\centering
\includegraphics[width=\textwidth]{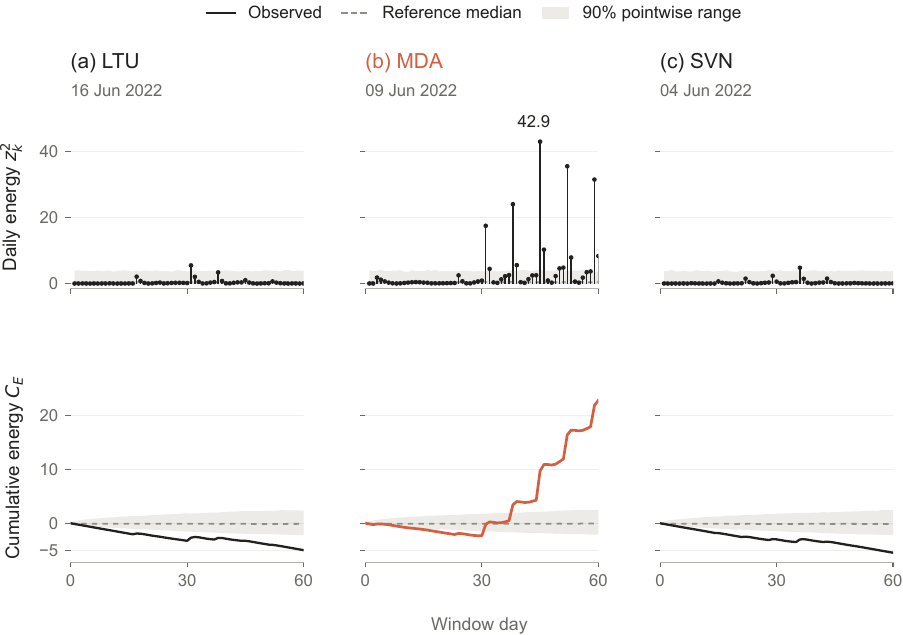}
\caption{Energy patterns in the original windows for LTU, MDA and SVN.
Each column shows daily squared residuals and cumulative energy \(C_E(n)\).
Dates give the start of each 60-transition window; vertical scales are
common within each row. Grey bands and dashed curves show the 90\% pointwise reference ranges
and medians of the quantity in each row, computed from 2,500
country-specific reference paths. MDA has large
squared residuals late in its window; the LTU and SVN curves
\(C_E(n)\) remain negative over extended periods.}
\label{fig:sir-country-patterns}
\end{figure}

With 2,500 pilot and 2,500 independent reference paths per country, BGR's
reference proportion was 0.318; the other five were below 0.05. Removing
lag-seven energy correlation raised HUN's proportion from 0.0064 to 0.0552
and left those for LTU, MDA and SVN unchanged. Other features can still
reflect weekly reporting, so this comparison does not identify the cause.
The shared fit and outcome-based selection also prevent interpreting the
fraction below 0.05 as a population rejection rate.
Figure~\ref{fig:sir-country-patterns} shows the residual energy curves
for the original LTU, MDA and SVN windows.

\begin{figure}[!ht]
\centering
% The lower panels are shown separately in Figure S8.
\includegraphics[width=\textwidth,trim=0 95bp 0 0,clip]{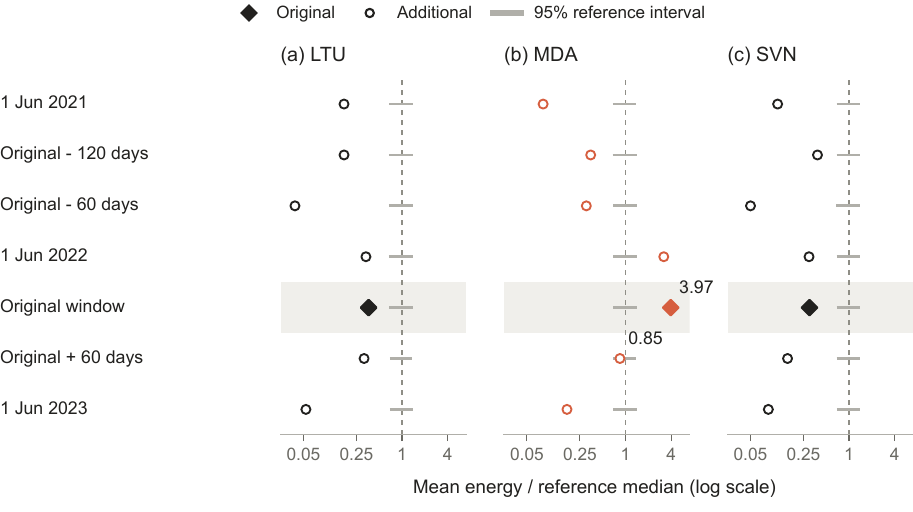}
\caption{The mean of the squared standardised residuals in each of seven
60-transition windows per country, divided by the reference median of that
statistic for the same window. Diamonds mark the original windows in
Figure~\ref{fig:sir-country-patterns}; circles mark additional windows.
Grey segments are 2.5--97.5\% reference intervals from 2,500 reference paths,
describing reference variation rather than uncertainty in the observed
values. The windows can overlap; these comparisons were made after the initial
country analyses.}
\label{fig:sir-window-patterns}
\end{figure}

For LTU, MDA and SVN, six additional windows per country used starts
shifted by \(-120,-60,+60\) days and starts on 1 June 2021, 2022 and 2023.
The settings and seeds
were fixed before these retrospective simulations, with the model and
score unchanged. All 18 reference proportions were below 0.05, but MDA's
original excess of large squared residuals did not persist: mean energy
was below its reference median in five of six additional windows. The
MDA window used in the later report analysis, 8 August--7 October 2022,
was chosen for illustration after these comparisons; it starts 60 days
after the original window. Figure~\ref{fig:sir-window-patterns} compares
mean energy across the seven windows for each country.

\begin{figure}[!ht]
\centering
\includegraphics[width=\textwidth]{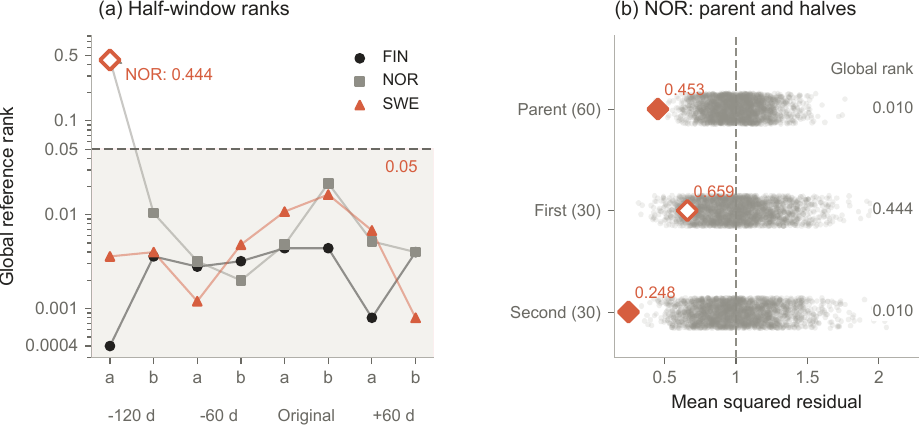}
\caption{Post-hoc analysis of Nordic window sensitivity. (a) Global reference proportions
for 24 half-windows, eight per country. Parent-block starts are relative
to the original window; a and b denote their first and second 30-increment
halves. The 0.05 line is shown as a threshold for this descriptive comparison. (b) NOR's earliest 60-increment block
and its halves, including the half-window with reference proportion 0.444. Grey points in each row show the mean of the squared standardised
residuals for each of 2,500 reference paths matched to that row's window
length. Orange diamonds and their numeric labels show the same statistic
computed from the reconstructed path. Numbers in parentheses give window
lengths, and the right column gives reference proportions. Labels using
``rank'' in the figure denote these normalised plus-one upper-tail ranks.}
\label{fig:nordic-window-sensitivity}
\end{figure}

Separate post-hoc Nordic comparisons used adjacent 60-increment blocks
and their non-overlapping 30-increment halves, with independent pilot and
reference paths matched to each window length. All 12 long windows and
23 of 24 halves had reference proportions below 0.05; the remaining half
had proportion 0.444. Thus the results depend on window location and length
(Figure~\ref{fig:nordic-window-sensitivity}).
The country and window commands, including the lag-seven deletion check,
are retained in \path{REPRODUCE.md}.

\subsection{Tangent-covariance scaling and numerical checks}
\label{app:sir-tangent}
To construct the score, we use the full \(2\times2\) recursion in
\eqref{eq:tangent-recursion} and take its diagonal element for the normalised
\(\log I\) coordinate as the tangent variance in \eqref{eq:sir-coordinate}.
We apply the same variance floor on every path.
For componentwise state clipping, the projection Jacobian is diagonal.
Each diagonal entry is one when the corresponding unnormalised coordinate
lies strictly between its bounds and zero otherwise, including at the bounds.
This Jacobian multiplies both the state Jacobian and the injected loading.
We differentiate the implemented drift automatically, including its
positive-part operation, for which automatic differentiation returns \(1/2\) at zero.
These conventions define the score at branch boundaries;
Proposition~\ref{prop:tangent-first-variation} concerns smooth maps.

The argument of the infection-rate positive-part operation was negative
in 2.18--12.94\% of simulated
substeps across the Nordic and six randomly drawn countries. Zero-noise
paths from observed states entered this branch only in MDA, in 10 of 600
substeps, all on the 11--12 June 2022 transition. For that transition,
20,000 direct one-day simulations, conditional on the reconstructed state
and its fixed 20-state control, gave a tangent-to-sample variance ratio
of 1.012 for normalised \(\log I\).

At eight equally spaced transitions in each Nordic window, we compared
the approximations with 512 direct one-day simulations per history.
At each of these 24 histories, we divided the absolute difference between
the zero-noise prediction and simulated conditional mean of normalised
\(\log I\) by the simulated conditional standard deviation of that coordinate.
The median of these 24 ratios was 0.023. For the same coordinate, the
median ratios to simulated conditional variance were 0.998 for the tangent
variance and 0.999 for the training plug-in variance. The tangent covariance
and diagonal plug-in covariance each had median relative Frobenius error
about 4\% over the 24 selected histories. Replacing plug-in residual scales with tangent
scales left the component ranks unchanged in the paired comparison.
These checks support the approximations at the selected histories and
are computed separately from \(S_{\mathrm{CE}}\), using
\path{experiments/run_sir_sde_native_country_audit.py}.

\subsection{Earlier daily-state results}
\label{app:sir-legacy-results}
The original analysis standardises only the normalised infectious
log-state:
\begin{equation}
z_k=\frac{\Delta Y_{Z,k}-\widehat m_{Z,k}}
{\sqrt{\max\{\widehat\Sigma_{k,ZZ}^{tan},10^{-12}\}}},
\qquad \widehat m_k=\bar Y_{k,L}-Y_k.
\label{eq:sir-coordinate}
\end{equation}
With 2,500 pilot and 2,500 reference paths, the 1 January--2 March 2021
windows gave global ranks 0.0612 for CZE and 0.00040 for GRC. GRC's mean
energy was 0.372 against a reference median of 0.984, with cumulative
energy negative over an extended part of the window.

\begin{figure}[!ht]
\centering
\includegraphics[width=\textwidth]{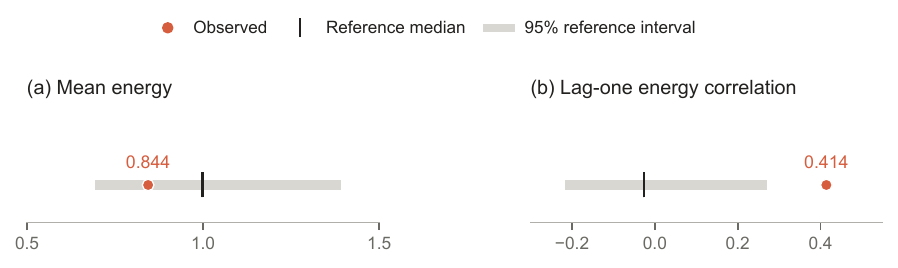}
\caption{Earlier daily-state comparison for Moldova's retrospectively selected
8 August--7 October 2022 window.
The observed mean energy is within its 95\% direct-state reference interval,
whereas energy correlation exceeds the upper endpoint of its own reference
interval. These 2,500 reference paths omit
report generation and reconstruction.}
\label{fig:sir-mda-dependence}
\end{figure}

In MDA's retrospectively selected 8 August--7 October 2022 window, mean
energy was 0.844, within its 95\% reference interval [0.693,1.391], while
energy correlation was 0.414, above its interval's upper endpoint 0.271.
The global rank was 0.0156. Figure~\ref{fig:sir-mda-dependence} shows
this comparison. These results use direct-state references;
Section~\ref{app:report-observation} examines the effect of generating
reports and repeating reconstruction.

\section[Weekly reports, reconstruction and scale comparisons]{Weekly reports, reconstruction\\and scale comparisons}
\label{app:report-observation}

This section specifies the observation model used in
Section~\ref{sec:sir-application}. It reuses six previously examined windows.
The report model, weekly scoring variants and illustrative plots are
exploratory; the earlier country-selection records in
Table~\ref{tab:sir-analysis-status} do not confer a new preselection status.

\subsection{Archived counts and information in reconstructed paths}

The data are the same archived snapshot as in
Section~\ref{app:sir-reconstruction}. We checked the raw \texttt{new\_cases}
and \texttt{new\_cases\_smoothed} fields on every day of the six windows.
No values are missing there, and nonzero counts occur only on Sundays.
The smoothed field equals the trailing seven-day mean after rounding
to two decimal places. This describes the archived allocation of counts
to dates,
not a verified real-time publication schedule.

\begin{table}[!htbp]
\centering\small
\begin{tabular}{@{}llllr@{}}
\toprule
Country & State-window start & First report & Last report & Reports \\
\midrule
CZE & 2021-01-01 & 2021-01-03 & 2021-02-28 & 9 \\
GRC & 2021-01-01 & 2021-01-03 & 2021-02-28 & 9 \\
MDA & 2022-08-08 & 2022-08-14 & 2022-10-02 & 8 \\
FIN & 2022-04-02 & 2022-04-03 & 2022-05-29 & 9 \\
NOR & 2022-06-04 & 2022-06-05 & 2022-07-31 & 9 \\
SWE & 2022-06-04 & 2022-06-05 & 2022-07-31 & 9 \\
\bottomrule
\end{tabular}
\caption{Report dates within the six existing 60-day future windows.
Reference paths start from the 20-day history ending on the Sunday
preceding the first report. Reports within a window need not be independent.}
\label{tab:report-windows}
\end{table}

Write \(q_j=\operatorname{round}(B_j/7,2)\) for the daily equivalent of
the \(j\)th weekly count. Fix the preceding history. Before state clipping,
the full physical \((S,I)\) path can be written as \(y=b+Gq\), where
\(G\) has 120 rows and eight or nine linearly independent columns. It
combines cumulative incidence and the fixed 28-day decay. If \(e=q-q_0\)
and \(D\succ0\) is a covariance used to weight these report equivalents,
\begin{equation}
(Ge)^\top(GDG^\top)^+(Ge)=e^\top D^{-1}e.
\label{eq:report-reconstruction-equivalence}
\end{equation}
Indeed, \(GD^{1/2}\) has full column rank, so its quadratic form with
the pseudoinverse of \(GDG^\top\) is the identity on its column coordinates.
This is a linear algebra identity, not an additional calibration result.
It does not require reports to be independent. Numerical relative errors
are below \(5\times10^{-15}\) in the six windows.

Thus repeated daily state values do not add observed report information.
The claim concerns physical coordinates and rounded report equivalents.
Taking \(\log I\) can introduce covariance directions from curvature even
when the underlying input has one dimension; the rank of an exact
nonlinear covariance need not equal the input dimension.

\subsection{The fitted report and reconstruction model}

Retain the original neural checkpoint, normalisation, daily controls and
ten Euler substeps. From the Sunday reconstructed history, simulate seven
days and take \(L=N[S_{start}-S_{end}]_+\). Draw the weekly count using
\eqref{eq:report-law}. For each of the next six days, the reconstructed
daily incidence is the
preceding Sunday's count divided by seven and rounded to two decimal
places; on the seventh day, it is the new count divided and rounded in
the same way. These values are the rounded trailing seven-day means. Apply
\eqref{eq:sir-reconstruction} with the original cumulative anchor and
unclipped accumulators, imposing the original bounds only on visible
states. Update the 20-day history and repeat from the reconstructed state.
The unobserved infectious endpoint of the seven-day forecast is not kept
as the next observed state.

This is a new reference law built around the fixed predictor. It is not
the original direct-state law or a filtering distribution for latent
epidemic states. The report amplitude and dispersion are fitted by
maximising the conditional mixture likelihood
\[
\sum_i\log\left\{\frac1R\sum_{r=1}^R
p_{\mathrm{NB}}(B_i;aL_{ir},\phi)\right\},
\]
using \(R=4096\) fixed simulated intensities per training history.
We fit \(\log a\in[-5,5]\) and \(\log\phi\in[-12,4]\).
The retained fit uses all 332 eligible reports before 1 January 2021
from the original nine training countries, including zero reports;
histories unavailable on the preceding Sunday are omitted.
It gives \(a=1.04873184\), \(\phi=0.09669523\).
The network itself remains trained over its original, longer period.
The dispersion can absorb mean-model and reporting errors and is not
identified as a pure measurement-noise parameter.

At 512 intensity simulations per history, the earlier-period estimates
are \(a=1.04877\), \(\phi=0.09637\); two halves of the 4096-draw bank
give dispersion 0.096689 and 0.096726. A fit to all 1832 available
training reports is unstable: dispersion changes from 0.16367 at 512
draws to 0.13474 at 4096, with half-bank estimates 0.134795 and 0.155762.
The problem is concentrated in an extreme Dutch report with very low
predicted intensity. We retain that fit as a numerical sensitivity record,
not as the coefficient source for this application. Integration checks
do not measure training uncertainty.

For fixed history, the exact report moments satisfy
\begin{align*}
\E[B\mid\mathcal H]&=a\E[L\mid\mathcal H],\\
\operatorname{Var}(B\mid\mathcal H)
&=a\E[L\mid\mathcal H]+\phi a^2\E[L\mid\mathcal H]^2
+(1+\phi)a^2\operatorname{Var}(L\mid\mathcal H).
\end{align*}
Equation~\eqref{eq:report-scales} replaces \(\E[L\mid\mathcal H]\)
by \(L(0)\) and \(\operatorname{Var}(L\mid\mathcal H)\) by
\(\|D_\varepsilon L(0)\|^2\). The derivative covers all 70 Gaussian
substep inputs and daily feedback. The centre uses zero latent inputs
and the conditional report mean; it is not a zero-noise convention for
the discrete negative-binomial draw itself.

Nonnegative infection-rate truncation can give zero local centre and
variance despite positive flow on some stochastic paths. We therefore
fix variance floors of one and 100 squared cases before the full-path
comparisons. Neither variance floor is active at any of the 53 observed
report histories. Variance approximations along reference paths can fall
below these floors, so changing the floor can still alter the reference
score distribution. State projection and reconstruction clipping
were inactive in the reported full-path application samples.

Reapplying the full reconstruction reproduces the actual paths to
\(1.3\times10^{-15}\), and checked simulated paths to
\(2.3\times10^{-15}\). Zero-input centres agree with the separate
conditional forecast calculation to relative error below
\(2.9\times10^{-10}\). Target and reference updates use the same rules.

\subsection{Feature transformations and observed report ranks}

We generate 5000 report paths per country. Alternating paths form
independent 2500-path pilot and reference samples. For both scales, use
the same counts and the five features specified in
Section~\ref{sec:six-residual-statistics}. The primary score takes the
maximum pilot-standardised distance from each feature's median, with
upper tails for cumulative absolute maxima and two sides for the other
three features.

A second, exploratory transformation uses empirical pilot tails. For
pilot size \(P\) and feature value \(t\), define
\[
\ell_j(t)=\frac{1+\sum_{p=1}^P\1\{T_{jp}\le t\}}{P+1},\qquad
u_j(t)=\frac{1+\sum_{p=1}^P\1\{T_{jp}\ge t\}}{P+1}.
\]
For cumulative absolute maxima, use the transformed feature
\(-\log u_j(t)\). For the other features, use
\[
-\log\min\{1,2\min(\ell_j(t),u_j(t))\}.
\]
We rank their maximum against independently generated reference maxima.
The tail values are feature transformations, not the final global ranks.
Pilot ties and reference ties use the displayed weak inequalities.

\begin{table}[!htbp]
\centering\small
\begin{tabular}{@{}lrrrr@{}}
\toprule
& \multicolumn{2}{c}{Median distance} & \multicolumn{2}{c}{Empirical tails} \\
\cmidrule(lr){2-3}\cmidrule(l){4-5}
Country & Report & Propagated & Report & Propagated \\
\midrule
CZE & 0.0116 & 0.0164 & 0.0780 & 0.1004 \\
GRC & 0.1727 & 0.1467 & 0.5970 & 0.5470 \\
MDA & 0.1483 & 0.1767 & 0.1140 & 0.1076 \\
FIN & 0.0412 & 0.0676 & 0.0360 & 0.0268 \\
NOR & 0.2583 & 0.2391 & 0.3055 & 0.2095 \\
SWE & 0.6917 & 0.5418 & 0.1899 & 0.0936 \\
\bottomrule
\end{tabular}
\caption{Global reference rank proportions for the actual reports,
with a variance floor of one. Both transformations and both scales use
the same report paths. These exploratory ranks have no cross-country
adjustment and do not estimate detection power.}
\label{tab:report-ranks}
\end{table}

Changing the scale does not uniformly decrease observed ranks. In Finland,
the propagated residual mean is \(-0.7643\), compared with \(-0.9252\)
using the report-only scale. In Czechia the corresponding energy
correlations are 0.6466 and 0.6659. Both scales retain the direction of
these descriptive findings, while crossing a 0.05 threshold depends on
the feature transformation. Czechia and Finland were chosen for the
cumulative plots after inspecting all six windows.

\paragraph{A separate weekly-report reference.}
For this comparison, weekly counts follow a negative-binomial persistence
model with conditional mean \(\mu_j=c(B_{j-1}+1)\) and conditional variance
\(\mu_j+\phi_p\mu_j^2\), where \(c=1.19926550\) and
\(\phi_p=0.38895910\). We fitted both coefficients by maximum likelihood
to 1,840 available weekly reports with observed predecessors from the
original nine training countries, covering 5 April 2020--4 August 2024.
Generated counts undergo the same trailing seven-day averaging, rounding
to two decimal places and daily state reconstruction as the observations.
We retain the original daily
infectious-coordinate residual construction and its observed values,
changing only the reference generation. This persistence model differs
from the seven-day SIR report model used in
Section~\ref{sec:sir-application}.

Moldova's observed daily energy correlation, 0.414, lies outside the
direct-state reference interval [-0.216,0.254] but inside the persistence
model's reconstruction interval [0.184,0.446]. The intervals use the
2,500 reference paths from each 5,000-path sample, with alternating paths
reserved for pilot calculations. The direct-state simulation batch
differs from the earlier one in Section~\ref{app:sir-legacy-results}.
This retrospective comparison illustrates how the interpretation depends
on the reference law. Since the reference means and laws also change,
it does not isolate the effect of smoothing.

\subsection{Scale contributions and conditional sensitivity}
\label{app:report-scale-checks}

A controlled two-substep SIR report experiment separates observation
variation from propagation. For latent intensity mean \(\mu\), compare
\(\mu+\phi\mu^2\) with
\(\mu+\phi\mu^2+(1+\phi)\operatorname{Var}(L)\), fixing the same
nine-week paths and upper-tail mean-energy score. With known parameters,
5000 repetitions and 499 new references each, multiplying SIR volatility
by 1.25 gives detection rates 13.66\% and 17.12\% when \(\phi=0\),
9.26\% and 9.26\% when \(\phi=0.096\), and 8.18\% and 8.16\% when
\(\phi=0.164\). In this setting \(\operatorname{Var}(L)\) is close to
a constant times \(\mu^2\), so adding substantial report dispersion
makes the propagation correction nearly a common multiplier. The
ordering of an upper-tail homogeneous energy score is unchanged under
an exact common positive multiplier. This argument does not apply
unchanged to every cumulative feature.

We next fix the first-report history of each of the six windows and use
the fitted neural predictor with report amplitude \(a=1\).
At history \(h\), the residual centre is \(m_{0,h}=L_h(0)\).
A separate pilot of 32,768 latent forecasts per history gives estimates
\(\widehat\mu_h\) and \(\widehat v_h\) of the conditional mean and variance
of \(L_h\). We fix the two variance weights
\[
d_h^{report}=\widehat\mu_h+\phi\widehat\mu_h^2,\qquad
d_h^{prop}=d_h^{report}+(1+\phi)v_h^{tan},
\]
where \(v_h^{tan}=\|D_\varepsilon L_h(0)\|^2\).
Unlike \eqref{eq:report-scales}, this experiment evaluates the reporting
variance at the pilot mean while retaining the zero-input residual centre.
Each repetition generates one independent target report at each history
and, separately, 499 independent null reference vectors, each containing
one independent report from each history. We compare upper-tail mean
energy across the six reports,
using the same generated reports for both weights and assigning no temporal
order to the countries.

For the constructed common-variance departure, the latent intensity law
is unchanged. At history \(h\), we replace the negative-binomial dispersion by
\[
\phi'_h=
\frac{1.5625\,\widehat V_h-\widehat\mu_h-\widehat v_h}
     {\widehat\mu_h^2+\widehat v_h},
\qquad
\widehat V_h=d_h^{report}+(1+\phi)\widehat v_h.
\]
This preserves the conditional report mean and targets a common multiplier
of 1.5625 for the total report variance.
With baseline \(\phi=0.096\), 10,000 independent repetitions give report-only
and propagated detection rates of 14.08\% and 16.15\%.
The paired gain is 2.07 percentage points, with pointwise 95\% interval
[1.25,2.89]. Fresh reference simulations give estimated variance ratios
of 1.5613--1.5643 across histories. For SIR volatility multiplied by 1.25,
the rates are 9.11\% and 8.40\%; for an infection-rate increase of
0.005 per day, they are 62.09\% and 60.13\%.

These target laws differ: raising latent volatility alone need not multiply
all report variances by the same factor. Reports are non-Gaussian and
their zero-input centres need not equal their means. The comparison
therefore illustrates conditional weighting effects without asserting
that Proposition~\ref{prop:common-scale-efficiency} applies to these counts.
It is distinct from the complete report-path experiment below.

\subsection{Complete-path null rejection and detection}
\label{app:report-power}

The complete-path experiment fixes the earlier-period report fit and the
network. Each country has 100 independent groups. Each group contains
one bank of 499 new null reference paths, shared by 50 new target paths
from each generating law: the null and four alternatives. Targets
start from the same recorded history as their references and thereafter
use their own generated and reconstructed histories. The feature pilots
come from the separate full-path application simulation. Scoring always
uses the original fitted centre, scale rule and report parameters at the
current history, including for paths generated under alternatives.

Alternatives multiply SIR volatility by 1.25, add 0.001 or 0.005 per day
to the infection rate in both state equations, or multiply report
dispersion by 1.5625. Volatility changes include their It\^o correction.
The report-dispersion alternative multiplies the single \(\phi\), unlike
the constructed common-total-variance alternative in the preceding
subsection. The smaller infection-rate change was included before
examining complete-path power; both rates are retained.

For each path and weight, we calculate upper-tail mean energy, the
five-feature median-distance maximum and the empirical-tail maximum.
Methods share reports. The main scale comparison fixes variance floor
one; floor 100 is a separate sensitivity comparison. Tables
\ref{tab:report-power-countries-a}--\ref{tab:report-power-countries-b}
give all countries and targets for the main score. The complete saved
table also retains all scores and floors.

Table~\ref{tab:report-power} gives descriptive means over the six
specified histories. These histories are not a random sample of
countries; the individual results below retain the differences between
them.

\begin{table}[!htbp]
\centering
\small
\begin{tabularx}{\textwidth}{@{}Yrr@{}}
\toprule
Target change & Report-only scale & Propagated scale \\
\midrule
None & 5.06 & 5.05 \\
SIR volatility multiplied by 1.25 & 7.13 & 6.67 \\
Infection rate increased by 0.001 per day & 5.85 & 6.75 \\
Infection rate increased by 0.005 per day & 26.95 & 29.75 \\
Report dispersion multiplied by 1.5625 & 10.76 & 11.33 \\
\bottomrule
\end{tabularx}
\caption{Rejection percentages for the five-feature median-distance score,
averaged descriptively over the six specified initial histories. Each
history has 5,000 targets per row. The same target and reference reports
are used for both scales. These are fixed-model simulations, not measured
power on the observed country records or averages over randomly selected countries.}
\label{tab:report-power}
\end{table}

\begin{table}[!htbp]
\centering\small
\begin{tabular}{@{}llrrr@{}}
\toprule
Country & Target & Report & Propagated & Difference interval \\
\midrule
CZE & None & 5.14 & 5.26 & [-0.22, 0.46] \\
CZE & \(\sigma\times1.25\) & 5.12 & 4.90 & [-0.57, 0.13] \\
CZE & \(\beta+0.001\) & 4.72 & 4.80 & [-0.19, 0.35] \\
CZE & \(\beta+0.005\) & 3.78 & 3.98 & [0.00, 0.40] \\
CZE & \(\phi\times1.5625\) & 5.58 & 5.68 & [-0.36, 0.56] \\
\addlinespace
GRC & None & 5.26 & 5.26 & [-0.56, 0.56] \\
GRC & \(\sigma\times1.25\) & 6.62 & 6.58 & [-0.48, 0.40] \\
GRC & \(\beta+0.001\) & 4.14 & 3.96 & [-0.63, 0.27] \\
GRC & \(\beta+0.005\) & 7.08 & 4.78 & [-2.82, -1.78] \\
GRC & \(\phi\times1.5625\) & 6.26 & 6.92 & [0.20, 1.12] \\
\addlinespace
MDA & None & 4.78 & 4.86 & [-0.33, 0.49] \\
MDA & \(\sigma\times1.25\) & 6.36 & 5.96 & [-0.82, 0.02] \\
MDA & \(\beta+0.001\) & 6.40 & 6.58 & [-0.18, 0.54] \\
MDA & \(\beta+0.005\) & 21.06 & 23.48 & [1.83, 3.01] \\
MDA & \(\phi\times1.5625\) & 17.34 & 16.16 & [-1.75, -0.61] \\
\bottomrule
\end{tabular}
\caption{Complete-path five-feature median-distance comparison: CZE, GRC, MDA. Rates are percentages; the last column is a pointwise approximate
95\% interval for propagated minus report-only rejection, in percentage
points. Each country and target law uses 100 simulation groups, each with
499 reference paths and 50 target paths.}
\label{tab:report-power-countries-a}
\end{table}

\begin{table}[!htbp]
\centering\small
\begin{tabular}{@{}llrrr@{}}
\toprule
Country & Target & Report & Propagated & Difference interval \\
\midrule
FIN & None & 4.88 & 4.64 & [-0.73, 0.25] \\
FIN & \(\sigma\times1.25\) & 6.48 & 6.80 & [-0.18, 0.82] \\
FIN & \(\beta+0.001\) & 6.34 & 7.94 & [1.06, 2.14] \\
FIN & \(\beta+0.005\) & 36.38 & 39.94 & [2.77, 4.35] \\
FIN & \(\phi\times1.5625\) & 13.68 & 13.94 & [-0.51, 1.03] \\
\addlinespace
NOR & None & 5.20 & 5.10 & [-0.82, 0.62] \\
NOR & \(\sigma\times1.25\) & 9.20 & 8.10 & [-1.92, -0.28] \\
NOR & \(\beta+0.001\) & 6.54 & 8.86 & [1.57, 3.07] \\
NOR & \(\beta+0.005\) & 49.40 & 55.56 & [4.92, 7.40] \\
NOR & \(\phi\times1.5625\) & 9.28 & 12.18 & [1.85, 3.95] \\
\addlinespace
SWE & None & 5.10 & 5.20 & [-0.44, 0.64] \\
SWE & \(\sigma\times1.25\) & 8.98 & 7.70 & [-1.94, -0.62] \\
SWE & \(\beta+0.001\) & 6.96 & 8.38 & [0.81, 2.03] \\
SWE & \(\beta+0.005\) & 44.02 & 50.78 & [5.67, 7.85] \\
SWE & \(\phi\times1.5625\) & 12.40 & 13.10 & [-0.05, 1.45] \\
\bottomrule
\end{tabular}
\caption{Complete-path five-feature median-distance comparison: FIN, NOR, SWE. Rates are percentages; the last column is a pointwise approximate
95\% interval for propagated minus report-only rejection, in percentage
points. Each country and target law uses 100 simulation groups, each with
499 reference paths and 50 target paths.}
\label{tab:report-power-countries-b}
\end{table}

We estimate Monte Carlo uncertainty from the rejection proportions in
the 100 independent groups. Variation between groups includes variation
in their reference banks, each shared by the targets within that group.
Targets sharing a bank are not treated as independent unconditional
decisions. Paired intervals use differences between the two rejection
proportions within each group and the 97.5th Student \(t\) quantile with 99
degrees of freedom. They are conditional on the fixed feature pilots and
do not include training uncertainty or simultaneous adjustment. When
empirical group variance is zero, the first group's value is used as an
anchor and the other 99 groups bound the probability of leaving it,
giving a conservative nonzero uncertainty range rather than a zero-width
interval. Saved path ranks were used to recalculate the group summaries.

The alternative score and floor comparisons limit the interpretation of
energy gains. For Czechia's larger infection-rate change, empirical-tail
aggregation gives detection 27.76\% with the report-only scale and 37.40\%
with propagation, whereas median-distance aggregation gives 3.78\% and
3.98\%. For Greece, mean-energy detection rises from 0.04\% to 16.02\%,
but median-distance detection falls from 7.08\% to 4.78\% and empirical-tail
detection from 3.58\% to 2.50\%. Increasing the propagation variance
floor from one to 100 raises mean-energy detection to 26.08\%, a paired
gain of 10.06 percentage points [8.44,11.68]. Its gains for the two
composite scores are only 0.12 and 0.16 points, with intervals spanning
zero. Changing a weight can therefore help one statistic while doing
little for the diagnostic maximum. These fixed-parameter comparisons
do not assess whether regularisation controls fitting error.

\subsection{Conditional report moments and their effect on the score}
\label{app:report-moment-comparison}

The local report construction uses the centre \(m_0=aL(0)\) and the
propagated variance \(d^{prop}\) in \eqref{eq:report-scales}. Its scale can
be small because the forecast at zero latent noise inputs stays in a
branch with no infection flow, even though finite Gaussian inputs can
leave that branch. We first examine five fixed histories from the original
application pilot, then compare two residual rules on complete paths.
Both checks retain the fitted network and report coefficients. They
concern approximation of moments under that fitted law, not whether the
law agrees with the data-generating process.

\paragraph{Conditional moments at five selected histories.}
The three unusual histories are Czechia path 4702 before the report on
28 February 2021, Czechia path 264 before 31 January 2021, and Greece
path 2796 before 31 January 2021. Indices are zero-based in the original
5000-path application arrays; all are even and belong to the feature
pilot. The first and third have zero local centre and variance. The
second has large path energy without reaching the variance floor at any
report. They were selected after inspecting the pilot extremes.
For comparison, Czechia path 2250 and Greece path 3454, both before
31 January 2021, minimise distance of full-path mean energy from their
country's pilot median and do not reach the floor. These two controls
were fixed before the conditional simulations. Reconstructing each
history from its preceding saved reports reproduces the saved local
centre and variance to relative error below \(10^{-8}\).

Let \(\mathcal H\) denote the reconstructed history immediately before a
report. At each selected history, we generate 32,768 fresh seven-day latent
forecasts and their negative-binomial reports. For intensity draw \(L_r\),
write \(u_r=aL_r\) and \(w_r=u_r+\phi u_r^2\). Averaging \(u_r\) estimates
the conditional report mean, while
\(\overline w+\overline{(u-\overline u)^2}\) estimates the conditional
report variance.
These calculations integrate the conditional negative-binomial moments
rather than estimating both moments from the report draws alone. For the
selected saved report \(b_*\), we also average the exact conditional
negative-binomial probabilities \(\Pr(B\ge b_*\mid L_r)\).

\begin{table}[!htbp]
\centering\small
\begin{tabular}{@{}lrrrrrr@{}}
\toprule
History & \(b_*\) & \(m_0\) & \(\widehat m\)
& \(v_{\rm local}/10^6\) & \(\widehat v/10^6\)
& \(\widehat{\Pr}(B\ge b_*\mid\mathcal H)\) \\
\midrule
CZE 4702 & 1462 & 0 & 573.36 & 0 & 0.82490 & 0.12850 \\
CZE 264 & 9526 & 76.57 & 807.99 & 0.00064 & 1.84227 & 0.00191 \\
GRC 2796 & 965 & 0 & 118.85 & 0 & 0.04928 & 0.01336 \\
CZE 2250 & 8028 & 11422.65 & 11622.14 & 30.45857 & 29.88687 & 0.71841 \\
GRC 3454 & 3024 & 3241.13 & 3251.60 & 1.67765 & 1.68436 & 0.51404 \\
\bottomrule
\end{tabular}
\caption{Conditional report moments at five selected simulated histories.
The local centre is \(m_0=aL(0)\), and \(v_{\rm local}=d^{prop}\) is
the propagated variance in \eqref{eq:report-scales} before flooring.
The estimates \(\widehat m\), \(\widehat v\) and the tail probability
use 32,768 latent draws per history. Variance columns are in
millions of squared cases. The final
column is a one-report upper-tail probability for a selected simulated
report, not a global test rank or a probability for an observed report.}
\label{tab:report-conditional-moments}
\end{table}

The first and third rows show that a zero derivative at the zero-input
forecast does not describe all finite-input variation. The selected
Czech report 1462 has conditional upper-tail probability about 0.13,
despite its very large locally scaled residual. The Czech report 9526
remains unusual under the full conditional distribution: its estimated
upper-tail probability is 0.00191. Thus replacing the local centre and
variance approximations need not make every extreme report ordinary.
At the two controls, the estimated
moments are close to the local values.

Thirty-two consecutive blocks of 1024 draws give Monte Carlo standard
errors for the estimated means of 4.05, 6.57, 1.14, 21.60 and 4.92,
respectively, in table order. The corresponding standard errors for the
tail probabilities are 0.00130, 0.00012, 0.00038, 0.00168 and 0.00158.
For the second moment about \(m_0\), the largest single draw contributes
at most 0.324\% of the empirical sum, and the largest block at most
3.79\%. The retained half-sample estimates reveal remaining variation;
for CZE 264, the variance estimate changes from \(1.783\times10^6\)
with 16,384 draws to \(1.842\times10^6\) with all draws. These checks
describe simulation precision at the five selected histories; they do
not establish that rarer events have been resolved at every history.

\paragraph{A fixed simulation rule for complete paths.}
To assess whether the local approximation affects path scores, we replace
both its centre and variance by moments computed from \(N_{\mathrm{in}}=256\) ordinary
Gaussian inputs. One bank, generated with seed 2026101003 independently
of the outer paths, is shared by all histories, countries, pilots,
references and targets. At each pre-report history, all bank inputs pass
through the original seven-day forecast, using the original fitted
coefficients. With \(u_r(\mathcal H)=aL_r(\mathcal H)\), define
\begin{align}
\widehat m_{\mathrm{in}}(\mathcal H)
&=\frac1{N_{\mathrm{in}}}\sum_{r=1}^{N_{\mathrm{in}}} u_r(\mathcal H),\notag\\
\widehat v_{\mathrm{in}}(\mathcal H)
&=\frac1{N_{\mathrm{in}}}\sum_{r=1}^{N_{\mathrm{in}}}
\left\{u_r(\mathcal H)+\phi u_r(\mathcal H)^2
+[u_r(\mathcal H)-\widehat m_{\mathrm{in}}(\mathcal H)]^2\right\}.
\label{eq:report-finite-moments}
\end{align}
The new residual is
\((B-\widehat m_{\mathrm{in}})/\sqrt{\max(\widehat v_{\mathrm{in}},1)}\).
These are moments of a finite simulation bank, not exact conditional
moments. A finite bank can miss positive-flow events, so the floor
remains part of both rules. The comparison changes the centre and
variance together and re-estimates the pilot feature transforms; it does
not isolate the contribution of each change.

The comparison uses Czechia and Greece because the earlier checks
identified approximation problems there. It does not assess the other
four country windows. Both residual rules use the same generated reports,
five features and maximum aggregation. Each rule recomputes its own
feature medians and interquartile ranges from the same original 2500
pilot paths. For feature \(T_j\), with pilot median \(c_j\) and positive
interquartile range \(d_j\), the transformed value is
\begin{equation}
\begin{aligned}
R_j(t)&=
\begin{cases}
\max\{t-c_j,0\}/d_j,&\text{for the two cumulative absolute maxima},\\
|t-c_j|/d_j,&\text{for the other three features},
\end{cases}\\
S&=\max_jR_j(T_j).
\end{aligned}
\label{eq:report-iqr-transform}
\end{equation}
All pilot interquartile ranges are positive. Using them prevents isolated
pilot extremes from inflating a sample-standard-deviation denominator,
but does not ensure that every component can cross the global threshold.

Each country has ten independent groups. Each group contains one bank
of 199 null reference paths, shared by 50 target paths from each of the
same five generating laws as Section~\ref{app:report-power}.
There are 500 targets per law and country.
The nominal level is 0.05. With the plus-one rank and reference ties
counted as exceedances, rejection requires a score strictly above the
tenth-largest reference score in its group. Approximate pointwise 95\%
intervals use the ten group means, or paired group differences, and the
97.5th Student \(t\) quantile with nine degrees of freedom. The same
nonzero interval convention as in Section~\ref{app:report-power} applies
when group variance is zero. Intervals condition on the fixed pilot paths
and inner bank; they do not include fitting or inner-bank uncertainty.

\begin{table}[!htbp]
\centering\small
\begin{tabular}{@{}llrrr@{}}
\toprule
Country & Generating law & \shortstack{Local\\approximation} & \shortstack{Simulation\\moments}
& \shortstack{Difference and\\95\% interval} \\
\midrule
CZE & Null & 6.4 & 6.0 & \(-0.4\;[-1.88,1.08]\) \\
CZE & \(\sigma\times1.25\) & 6.0 & 6.6 & \(0.6\;[-0.37,1.57]\) \\
CZE & \(\beta+0.001\) & 8.8 & 8.6 & \(-0.2\;[-1.01,0.61]\) \\
CZE & \(\beta+0.005\) & 43.0 & 42.4 & \(-0.6\;[-2.26,1.06]\) \\
CZE & \(\phi\times1.5625\) & 21.8 & 20.6 & \(-1.2\;[-2.58,0.18]\) \\
\addlinespace
GRC & Null & 3.4 & 5.6 & \(2.2\;[0.24,4.16]\) \\
GRC & \(\sigma\times1.25\) & 8.0 & 8.4 & \(0.4\;[-3.85,4.65]\) \\
GRC & \(\beta+0.001\) & 0.4 & 9.2 & \(8.8\;[6.26,11.34]\) \\
GRC & \(\beta+0.005\) & 10.6 & 59.4 & \(48.8\;[40.70,56.90]\) \\
GRC & \(\phi\times1.5625\) & 9.8 & 14.8 & \(5.0\;[0.88,9.12]\) \\
\bottomrule
\end{tabular}
\caption{Complete-path rejection rates for the same five-feature
interquartile-range score, using local centre and variance approximations
or finite-bank estimates. Rates are percentages; differences are rejection
rates under the simulation estimates minus those under the local
approximations, in percentage points. All four alternatives and the null
are retained.
The comparison is conditional on the fitted model, feature pilots and
256-input bank. Infection-rate increments are per day.}
\label{tab:report-moment-power}
\end{table}

In the first-bank comparison, Czechia has no detected improvement for
this IQR score: each paired
interval includes zero, including the report-dispersion alternative
where detection decreases. Greece gains sensitivity to both infection-rate
changes and the report-dispersion change, but not clearly to the volatility
change. Its realised null rejection also rises by 2.2 percentage points,
with paired interval [0.24,4.16]. Both individual null-rate intervals
include 5\%, but the paired difference means that the observed null rejection rates
should not be described as identical. The separate upper-tail mean-energy
benchmark gives 39.6\% versus 40.8\% in Czechia and 11.2\% versus 54.2\%
in Greece for the larger infection-rate change. All its null and alternative
rates and paired intervals are retained with those of the composite score.

\paragraph{A subsequent decomposition using the original standard deviations.}
After seeing nearly equal IQR-score detection in Czechia, we applied the
median-distance transformation rule from Section~\ref{sec:score-construction}
to the same saved features. We call the resulting score the SD score here
to distinguish it from the IQR score. Each residual rule computes its own
feature medians and sample standard deviations from the same 2,500 pilot
paths, using divisor \(2499\) for the sample variance. Tail directions and
reference calibration stay unchanged. This is a post hoc
decomposition of the original standard-deviation problem, not a further
primary comparison or a rule for selecting a preferred score.

For Czechia, the pilot mean-energy standard deviation decreases from
4880.148 to 0.570 when simulation estimates replace the local centre and
variance approximations; for Greece it decreases from 4230.687 to 0.615.
The corresponding pilot mean
energies decrease from 104.676 and 415.118 to 1.006 and 1.031, and maxima
from 243,526.9 and 103,470.5 to 7.327 and 7.491. These are finite-sample
descriptions. Their values near one do not establish exact moment
standardisation for every history.

Table~\ref{tab:report-moment-scaling} compares two residual constructions
and two pilot feature scales for the infection-rate increase of 0.005 per
day. Its entries use the same outer paths and first inner bank as
Table~\ref{tab:report-moment-power}, not independent replications.

\begin{table}[!htbp]
\centering
\small
\begin{tabularx}{\textwidth}{@{}lYrr@{}}
\toprule
Country & Residual centre and variance & SD scaling & IQR scaling \\
\midrule
Czechia & Zero-input centre and propagated variance & 3.4 & 43.0 \\
        & Conditional simulation & 46.8 & 42.4 \\
Greece  & Zero-input centre and propagated variance & 4.2 & 10.6 \\
        & Conditional simulation & 66.8 & 59.4 \\
\bottomrule
\end{tabularx}
\caption{Detection rates (\%) for an infection-rate increase of 0.005
per day, at nominal level 5\%. Each country has 10 independent groups,
with 199 references and 50 targets per group, shared across all four
rules. Each rule uses its own transformation of the same 2,500 pilot
paths. The IQR comparison was specified before this experiment;
the SD comparison was added after examining the IQR results.}
\label{tab:report-moment-scaling}
\end{table}

For the SD score, replacing the local centre and variance approximations
increases detection by 43.4 percentage points [35.19,51.61] in Czechia
and 62.6 points [56.80,68.40] in Greece. This connects the original
composite score's low detection rates to the extreme locally scaled
pilot energies. Gains are not
uniform: for Greece's volatility alternative, SD-score detection falls
from 7.0\% to 5.6\%, with paired difference \(-1.4\) points
[-3.31,0.51]. All five generating laws, both residual rules and both
feature transformations, including their null rates, remain in the
saved follow-up tables. Choosing the better score from these outcomes
would use the evaluation results for selection.

Table~\ref{tab:report-moment-sd-all} gives the SD-score comparison for
every generating law. Table~\ref{tab:report-moment-null} gives individual
null-rate intervals for both feature transformations, so the calibration
comparisons can be read alongside the paired power differences.

\begin{table}[!htbp]
\centering\small
\begin{tabular}{@{}llrrr@{}}
\toprule
Country & Generating law & \shortstack{Local\\approximation}
& \shortstack{Simulation\\moments} & \shortstack{Difference and\\95\% interval} \\
\midrule
CZE & Null & 6.0 & 6.4 & \(0.4\;[-2.45,3.25]\) \\
CZE & \(\sigma\times1.25\) & 4.4 & 8.0 & \(3.6\;[0.75,6.45]\) \\
CZE & \(\beta+0.001\) & 5.8 & 8.8 & \(3.0\;[0.20,5.80]\) \\
CZE & \(\beta+0.005\) & 3.4 & 46.8 & \(43.4\;[35.19,51.61]\) \\
CZE & \(\phi\times1.5625\) & 6.8 & 18.6 & \(11.8\;[7.46,16.14]\) \\
\addlinespace
GRC & Null & 4.4 & 5.8 & \(1.4\;[-1.84,4.64]\) \\
GRC & \(\sigma\times1.25\) & 7.0 & 5.6 & \(-1.4\;[-3.31,0.51]\) \\
GRC & \(\beta+0.001\) & 2.0 & 9.2 & \(7.2\;[4.57,9.83]\) \\
GRC & \(\beta+0.005\) & 4.2 & 66.8 & \(62.6\;[56.80,68.40]\) \\
GRC & \(\phi\times1.5625\) & 6.0 & 13.6 & \(7.6\;[3.24,11.96]\) \\
\bottomrule
\end{tabular}
\caption{Post hoc SD-score rejection rates on the same paths as
Table~\ref{tab:report-moment-power}. Rates are percentages; differences
are rejection rates under the first-bank simulation estimates minus
those under the local approximations, in percentage points. The intervals
use paired group differences. All four alternatives
and the null are shown; the primary comparison remains the IQR score.}
\label{tab:report-moment-sd-all}
\end{table}

\begin{table}[!htbp]
\centering\small
\begin{tabular}{@{}llrr@{}}
\toprule
Country & Centre and variance & \shortstack{SD null rate\\and 95\% interval}
& \shortstack{IQR null rate\\and 95\% interval} \\
\midrule
CZE & Local & \(6.0\;[2.84,9.16]\) & \(6.4\;[2.97,9.83]\) \\
CZE & Simulation, first bank & \(6.4\;[2.97,9.83]\) & \(6.0\;[3.39,8.61]\) \\
CZE & Simulation, second bank & \(7.4\;[4.53,10.27]\) & \(6.4\;[3.89,8.91]\) \\
\addlinespace
GRC & Local & \(4.4\;[1.99,6.81]\) & \(3.4\;[1.37,5.43]\) \\
GRC & Simulation, first bank & \(5.8\;[2.83,8.77]\) & \(5.6\;[2.68,8.52]\) \\
GRC & Simulation, second bank & \(6.4\;[2.78,10.02]\) & \(5.6\;[2.68,8.52]\) \\
\bottomrule
\end{tabular}
\caption{Individual null rejection rates and pointwise 95\% intervals,
in percent. Within each country, all rows use the same ten groups and
500 null target paths; the two inner banks are not independent outer
repetitions.
The intervals condition on the pilot paths and the specified bank.
The SD-score analysis was added after the IQR comparison.
}
\label{tab:report-moment-null}
\end{table}

\paragraph{What the correlation component can contribute.}
For the interquartile-range score in \eqref{eq:report-iqr-transform}, let
\(c_\rho\) and \(d_\rho\) be the pilot median and interquartile range of
energy correlation. Since \(\rho\in[-1,1]\), its transformed distance
satisfies the bound
\begin{equation}
\frac{|\rho-c_\rho|}{d_\rho}
\le \frac{1+|c_\rho|}{d_\rho}.
\label{eq:report-correlation-bound}
\end{equation}
Under the local approximation, Greece's bound is 3.352, while its ten
global critical values range from 4.705 to 432.646. The correlation component cannot
trigger rejection in any group. Removing it changes neither a critical
value nor any decision among the saved null and alternative targets.
With first-bank simulation moments, the bound is 3.174 and critical
values range
from 2.746 to 3.864; five groups still have critical values at least as
large as the bound. Removing the component changes one critical value
but no saved target decision. In Czechia, three of ten critical values
exceed the corresponding bound for each residual rule. Thus reducing
extreme pilot energies does not ensure that every component can
independently exceed the global reference threshold. The saved component comparisons use
exceedance of each group's global threshold; overlapping component
events are not added, and a feature attaining the maximum is not by
itself treated as an explanation of rejection.

\paragraph{A second inner bank.}
To check dependence on the finite integration bank, the calculation is
repeated with 256 ordinary Gaussian inputs from seed 2026101005. The
same pilot, reference and target report paths are retained, including
their grouping; only their conditional moments and the resulting feature
transforms and ranks are recomputed. This is a paired sensitivity check
of numerical integration. The two banks do not supply independent
outer experiments or double the effective number of groups.
For the infection-rate increase of 0.005 per day, second-bank detection
is 47.2\% with SD scaling and 43.4\% with IQR scaling in Czechia, and
66.2\% and 60.6\% in Greece. The corresponding first-bank rates are
46.8\%, 42.4\%, 66.8\% and 59.4\%. Across both countries, all five
generating laws and both composite scores, the largest absolute change
between banks is 1.4 percentage points, for Czechia's IQR score under
the report-dispersion alternative (20.6\% to 22.0\%). All individual
null-rate intervals still include 5\%; the Greek IQR null-rate difference
from the local rule remains 2.2 percentage points. The original-rule
features, scores and ranks are unchanged on every saved path.

No saved pilot, reference, target or observed history reaches the variance
floor under the second-bank rule. For this interquartile-range score, the
correlation component cannot alone exceed the global threshold in two Czech
and three Greek reference groups,
compared with three and five groups for the first bank. Thus the main
two-by-two comparison is preserved in this second calculation, while
component contributions can change. Two banks do not establish uniform
integration accuracy or estimate general bank-to-bank uncertainty.

\begin{table}[!htbp]
\centering\small
\begin{tabular}{@{}llrrrr@{}}
\toprule
Country & Law & IQR & SD & \shortstack{IQR difference\\and 95\% interval}
& \shortstack{SD difference\\and 95\% interval} \\
\midrule
CZE & Null & 6.4 & 7.4 & \(0.0\;[-1.51,1.51]\) & \(1.4\;[-0.51,3.31]\) \\
CZE & \(\sigma\times1.25\) & 7.8 & 7.8 & \(1.8\;[-0.67,4.27]\) & \(3.4\;[0.31,6.49]\) \\
CZE & \(\beta+0.001\) & 8.4 & 8.8 & \(-0.4\;[-1.30,0.50]\) & \(3.0\;[0.54,5.46]\) \\
CZE & \(\beta+0.005\) & 43.4 & 47.2 & \(0.4\;[-1.82,2.62]\) & \(43.8\;[34.89,52.71]\) \\
CZE & \(\phi\times1.5625\) & 22.0 & 18.8 & \(0.2\;[-1.98,2.38]\) & \(12.0\;[7.90,16.10]\) \\
\addlinespace
GRC & Null & 5.6 & 6.4 & \(2.2\;[0.24,4.16]\) & \(2.0\;[-1.63,5.63]\) \\
GRC & \(\sigma\times1.25\) & 7.4 & 5.8 & \(-0.6\;[-4.48,3.28]\) & \(-1.2\;[-3.83,1.43]\) \\
GRC & \(\beta+0.001\) & 9.6 & 9.2 & \(9.2\;[6.84,11.56]\) & \(7.2\;[4.84,9.56]\) \\
GRC & \(\beta+0.005\) & 60.6 & 66.2 & \(50.0\;[41.99,58.01]\) & \(62.0\;[56.48,67.52]\) \\
GRC & \(\phi\times1.5625\) & 14.8 & 14.0 & \(5.0\;[1.05,8.95]\) & \(8.0\;[3.09,12.91]\) \\
\bottomrule
\end{tabular}
\caption{Rejection percentages
using simulation moments from inner seed 2026101005 and the original
outer paths. Differences are against the unchanged local approximation,
in percentage points, with the same group-based intervals as
Table~\ref{tab:report-moment-power}. The SD columns retain their post hoc
status.}
\label{tab:report-moment-bank-check}
\end{table}

The inner bank and feature pilot are part of the fixed analysis
information \(\mathcal A\). Conditional on \(\mathcal A\), independent
target and reference paths use the same deterministic scoring rule,
so reusing the inner bank does not itself violate the independence
conditions of the rank result; its calibration conclusion still requires
the target and reference score laws to match.
For the infection-rate increase of 0.005 per day, replacing the local
centre and variance approximations with simulation estimates increases
detection under both feature scales in Greece. In Czechia, it gives a large increase under standard-deviation
scaling, but no clear gain under interquartile-range scaling.

\subsection{Predictive benchmarks and interpretation}
\label{app:report-benchmarks}

As a separate check of the predictor, compare next-report distributions
at the same 53 observed histories. A negative-binomial persistence model
has mean \(a(B_{k-1}+1)\); a recent-growth model has mean
\[
a(B_{k-1}+1)\exp\left\{b\log\frac{B_{k-1}+1}{B_{k-2}+1}\right\}.
\]
Their coefficients and dispersions are fitted to earlier training-country
reports. For predictive draws \(X_1,\ldots,X_M\) and observed count \(y\),
estimate the continuous ranked probability score by
\[
\frac1M\sum_i|X_i-y|
-\frac1{2M(M-1)}\sum_{i\ne j}|X_i-X_j|.
\]
Divide each score by the preceding report plus one, a positive common
scale known before that report. The mean scaled losses are 0.1430 for
recent growth, 0.1586 for persistence, 0.2084 for neural SIR with the
earlier-period report fit, and 0.2220 for neural SIR with Poisson reports.
Recent growth has lower loss in each of the six windows. This is a small
retrospective comparison; it supports model criticism rather than a
claim of predictive superiority for the neural model.

The observed report ranks, fixed-model power comparisons and controlled
SIR repeated fits answer different questions. The fitted report procedure
has not been evaluated after repeatedly fitting the network and report
layer together. Reporting dispersion and local covariance floors should
not be interpreted as having removed that uncertainty.

\subsection{Reproduction and retained settings}
\label{app:report-reproduction}

The scripts, saved tables and settings are in
\path{experiments/sir_case_observation/} in the GitHub code repository:
\url{https://github.com/clemmentin/Residual-Model-Criticism-for-Fitted-SDE-Implementations}.
\path{docs/REPRODUCE_20261008.md} gives the required archived inputs,
dependencies, command order and result locations; the experiment
\path{README.md} describes the report and reconstruction comparisons.
The commands regenerate the large simulation arrays from the recorded
seeds. The supplied tables retain adverse comparisons as well as gains.

For Section~\ref{app:report-moment-comparison},
\path{conditional_scale_check.py} runs the five-history check with
32,768 forecasts per history and seed 2026101002.
\path{conditional_moment_power.py} runs the complete-path comparison
with outer seed 2026101004 and inner seed 2026101003;
\path{analyze_conditional_moment_sd.py} applies the post hoc SD
transformation to the saved features.
\path{conditional_moment_bank_check.py} reuses those outer paths with
inner seed 2026101005. The experiment's
\path{conditional_moment_power.md} gives the comparison design,
commands and interpretation. Detailed numerical checks and the complete
second-bank comparison are in Section~\ref{app:report-moment-comparison};
script and output-file descriptions are in the reproduction instructions.

\end{document}